\documentclass[journal=jpcbfk,manuscript=article]{achemso}

\usepackage[T1]{fontenc}
\usepackage{lmodern}

\usepackage{longtable}
\usepackage{multirow}
\usepackage{amsmath}
\usepackage[usenames,dvipsnames]{color}
\usepackage{soul}
\usepackage{threeparttable}
\usepackage{graphicx}
\usepackage{amsmath,amssymb}
\usepackage{caption}
\usepackage{color}
\usepackage{xcolor}
\usepackage{dcolumn}
\usepackage{bm}
\usepackage{float}
\usepackage{subcaption}
\usepackage{textcomp}
\usepackage{url}
\usepackage[normalem]{ulem}
\usepackage{booktabs}

\usepackage{microtype}

\makeatletter

\author{Kai T\"opfer} \affiliation[University of Basel]{Department of
  Chemistry, University of Basel, Klingelbergstrasse 80 , CH-4056
  Basel, Switzerland}

\author{Shyamsunder Erramilli} \affiliation[Boston University]{
  Department of Physics and the Photonics Center, Boston University,
  590 Commonwealth Ave, MA 02215, Boston}

\author{Lawrence D. Ziegler} \affiliation[Boston University]{
  Department of Chemistry and the Photonics Center, Boston University,
  8 St Mary's St, MA 02215, Boston}

\author{Markus Meuwly} \affiliation[University of Basel]{Department of
  Chemistry, University of Basel, Klingelbergstrasse 80 , CH-4056
  Basel, Switzerland} \altaffiliation{Department of Chemistry, Brown
  University, Providence, RI 02912, USA} \email{m.meuwly@unibas.ch}

\title{Energy Relaxation of N$_2$O in Gaseous, Supercritical and
  Liquid Xenon and SF$_6$}

\begin{document}
\date{\today}

\begin{abstract}
Rotational and vibrational energy relaxation (RER and VER) of N$_2$O
embedded in xenon and SF$_6$ environments ranging from the gas phase
to the liquid, including the supercritical regime, is studied at a
molecular level. Calibrated intermolecular interactions from
high-level electronic structure calculations, validated against
experiments for the pure solvents were used to carry out classical
molecular dynamics simulations corresponding to experimental state
points for near-critical isotherms. Computed RER rates in low-density
solvent of $k_{\rm rot}^{\rm Xe} =
(3.67\pm0.25)\cdot10^{10}$\,s$^{-1}$M$^{-1}$ and $k_{\rm rot}^{\rm
  SF_6} = (1.25\pm0.12)\cdot10^{11}$\,s$^{-1}$M$^{-1}$ compare well
with rates determined by analysis of 2-dimensional infrared
experiments. Simulations find that an isolated binary collision (IBC)
description is successful up to solvent concentrations of $\sim 4$
M. For higher densities, including the supercritical regime, the
simulations do not correctly describe RER, probably due to neglect of
solvent-solute coupling in the analysis of the rotational motion. For
VER, the near-quantitative agreement between simulations and
pump-probe experiments captures the solvent density-dependent trends.
\end{abstract}

\section{Introduction}
Understanding the precise molecular level details of rotational and
vibrational energy relaxation dynamics in high density gas and
supercritical fluid (SCF) solutions is of fundamental importance for
the description, optimization and control of chemical reactivity in
inherently dense media. For example, dense fluids at high temperatures
and pressures, often in supercritical (sc) regimes, are solvent
environments where many combustion reactions
occur.\cite{fostermiller:2010,bares:2017,peters:1986} Furthermore,
SCFs offer the possibility for selective control of chemical processes
including energy relaxation/transfer dynamics, and have already been
successfully employed in a range of
applications.\cite{kajimoto:1999,skerget:2014,brunner:2010,knez:2018,perrut:2000,udupa:2011,smith:2011}
Aside from offering the ``green'' potential to replace organic
solvents (e.g. scH$_2$O or scCO$_2$), local density augmentation and
the dynamics associated with the long-spatial correlations that
develop in near critical state points contribute to the special
solvation properties of near critical
fluids.\cite{kajimoto:1999,tucker:1999,nishikawa:2003} These density
correlation length increases result in a corresponding correlation
relaxation time increase at the critical point, often described as
critical slowing and evident in scattering
experiments\cite{nishikawa:2000,nishikawa:2002,nishikawa:2005} and
simulations.\cite{tucker:1999a,tucker:2000c} \\

\noindent
In prior ultrafast studies of dense gas and supercritical fluid
solutions, two-dimensional infrared (2DIR) experiments have revealed
rates of rotational energy relaxation (RER) as a function of solvent
density for the N$_2$O asymmetric stretching mode ($\nu_{\rm as}$) in
SF$_6$ and Xe even in the absence of any rotational features in the
corresponding $\nu_{\rm as}$ rovibrational
spectrum.\cite{ziegler:2018,ziegler:2019,ziegler:2022} Pump-probe
measurements provide vibrational energy relaxation (VER) rates for
these same high density state points. A simple isolated binary
collision (IBC) model was sufficient to describe RER of N$_2$O
$\nu_{\rm as}$ up to SF$_6$ and Xe solvent densities of $\approx 4$\,M
for $T \leq 1.01 T_{\rm c}$ isotherms derived from 2DIR
measurements.\cite{ziegler:2022} However, the RER dynamics of N$_2$O
in these two similar non-dipolar solvents show different concentration
dependencies as the critical density is approached and where many body
effects begin to play a larger role for solvation.\cite{ziegler:2022}
The RER of N$_2$O is slower than the IBC predicted rates in the
density region ($\approx 4$\,M - $\approx 6$\,M) of the SF$_6$
critical point ($\rho_{\rm c}$(SF$_6$) = 5.09\,M) and attributed to
the longer length scale, slower density fluctuations coupling to the
closest N$_2$O solvation shells. However, no such slowing effect was
observed for N$_2$O RER in xenon.  For xenon densities $> \sim 4$\,M,
the RER rate monotonically increased through the critical point region
($\rho_{\rm c}$(Xe) = 8.40\,M).  It was hypothesized that the many
body interactions at the higher critical density for the xenon solvent
shielded the N$_2$O rotors from coupling to the long correlation
length fluctuations.  N$_2$O rotational relaxation in both solvents is
a highly efficient process.  Following $\nu_{\rm as}$ excitation
rotational equilibrium is re-established after 1.7 and 2.4 collisions
in SF$_6$ and Xe respectively in the IBC density
region.\cite{ziegler:2022} \\

\noindent
Rotational and vibrational equilibrium are established on very
different timescales following N$_2$O $\nu_{\rm as}$ excitation in
these high density gas and SCF SF$_6$ and Xe
solvents.\cite{ziegler:2022} While only a couple of collisions are
required for N$_2$O rotations to thermalize in both solvents, the VER
of N$_2$O may take hundreds of collisions or more to return the
molecule fully to thermal equilibrium.  The density dependence of
N$_2$O VER is also strikingly different in xenon and SF6, and
underscores that VER has a highly variable rate due to the inherently
quantum nature ($\hbar \omega_{\rm vib} = kT$) of this relaxation
mechanism.  The N$_2$O $\nu_{\rm as}$ lifetime is $\approx 300$ times
shorter in SF$_6$ than in xenon than because xenon has no possible
resonant V$\rightarrow$V, solute$\rightarrow$solvent, energy transfer
relaxation pathway as found for relatively efficient VER from N$_2$O
$\nu_{\rm as}$ in SF$_6$.  The first step in N$_2$O $\nu_{\rm as}$ VER
in SF$_6$ is the collision enabled intramolecular relaxation to the
N$_2$O symmetric stretching mode before a much slower return to the
ground vibrational state. Furthermore, VER is nearly an order of
magnitude slower for this initial VER component in SF$_6$ as compared
to RER, and a critical slowing effect is not as clearly
evident.\cite{ziegler:2022} The density dependence of the N$_2$O
$\nu_{\rm as}$ rovibrational absorption line shapes in SF$_6$ and
xenon were well-captured in a previous classical MD simulation
study.\cite{MM.n2ovib:2023} Even the rovibrational spectral contours,
inherently determined by $J,M$ quantum transition selection rules, were
accurately duplicated by these classical simulation results. The goal
of this current report is to test the ability of this same simulation
approach to capture the previously reported N$_2$O $\nu_{\rm as}$
rotational and vibrational relaxation dynamics determined by these
2DIR and pump-probe measurements spanning the dense gas to
supercritical density regime in SF$_6$ and Xe.  \\

Relaxation phenomena in liquids provide valuable information about
intermolecular interactions and coupling between solvent and solute
modes. In the present work RER and VER of N$_2$O as the solute
immersed in Xe and SF$_6$ as atomic and molecular solvents at a wide
range of densities is investigated from atomistic simulations. The
environments are under conditions which range from gaseous to the
supercritical and regular liquids.\\

\noindent
The present work is structured as follows. First, the methods are
presented, followed by the validation of the intermolecular
interactions. Next, results on the RER and VER of N$_2$O are 
presented and discussed. Finally, conclusions are drawn.\\

\section{Methods}

\subsection{Inter- and Intramolecular Interactions}
{\it The N$_2$O PES:} The intramolecular potential energy surface (PES)
of N$_2$O in its
electronic ground state ($^1$A$'$) is provided by a machine-learned
representation using the reproducing kernel Hilbert space (RKHS)
method.\cite{MM.rkhs:2017,koner:2020} Reference energies at the
CCSD(T)-F12/aug-cc-pVQZ level of theory were determined on a grid of
Jacobi coordinates $(R,r,\theta)$ with $r$ the N-N separation, $R$ the
distance between the center of mass of the diatom and the oxygen atom,
and $\theta$ the angle between the two distance vectors.  All
calculations were carried out using the MOLPRO
package.\cite{werner:2020} The full-dimensional RKHS model for a
N$_2$O potential energy surface was originally developed for the
investigation of N+NO collision reaction dynamics.\cite{koner:2020}
The Pearson coefficient $R^2$ of the RKHS representation and the full
set of reference values is $0.99983$ and the root mean squared error
(RMSE) between RKHS and reference energies up to 20\,kcal/mol above
the equilibrium structure (78 reference energies) is
$0.13$\,kcal/mol.\\

\noindent
{\it The Solvent PES:} Intra- and intermolecular force field
parameters for SF$_6$ were those from the work of Samios {\it et
  al.}.\cite{samios:2010} Intermolecular interactions were based on
Lennard-Jones potentials only and the parameters were optimized such
that MD simulations of pure SF$_6$ reproduce the experimentally
observed $pVT$ state points for liquid and gas SF$_6$, as well as the
states of liquid-vapor coexistence below and supercritical fluid above
the critical temperature $T_c(\mathrm{SF_6})=318.76$\,K,
respectively.\cite{ambrose:1987,haynes:2014crc} For Xe the
Lennard-Jones parameters were optimized to reproduce the $pVT$ state
points of gaseous, supercritical and liquid Xe at different densities
and temperatures below and above the critical temperature of
$T_c(\mathrm{Xe})=289.74$\,K.\cite{ambrose:1987,haynes:2014crc} \\

\noindent
The Xe--Xe Lennard-Jones (LJ) potential parameters for the
van-der-Waals (vdW) potential energy contribution were fit to match
experimental $pVT$ state points of xenon, which were computed with a
similar setup as used by Samios {\it et al.} to fit the force field
parameters for SF$_6$.\cite{samios:2010} A set of $N_{pVT}=44$
experimentally measured $pVT$ state points of xenon at temperatures
(number of state points) $273.15$ (5), $286.65$ (3), $298.15$ (12),
$323.15$ (12) and $373.15$\,K (12) was
selected.\cite{louwerse:1954xenon} For each state point a separate
$NVT$ molecular dynamics (MD) simulation was run using a Langevin
thermostat with a friction coefficient of $0.1$\,ps$^{-1}$. The system
contained 1200 Xe atoms, the structure was minimized, heated to the
target temperature and equilibrated for a total of 100\,ps with a time
step of 1\,fs. Next, the average pressure was determined from a 300 ps
simulation and compared with the experimental
reference.\cite{louwerse:1954xenon} The xenon LJ parameters
$\epsilon_\mathrm{Xe}$ and $R_\mathrm{min, Xe}$ were then optimized by
the Nelder–Mead method\cite{neldermead:1965} to minimize the relative
root mean square error (rRMSE) between sampled average and
experimental pressures at respective volume and temperature
\begin{equation}
    \mathrm{rRMSE} = \sqrt{ \dfrac{1}{N_{pVT}} \sum_i^{N_\mathrm{PVT}}
    \left( \dfrac{p_{\mathrm{sim.}, i}(V,T) - p_{\mathrm{ref.}, i}(V,T)}
    {p_{\mathrm{ref.}, i}(V,T)} \right)^2}
\end{equation}
The LJ parameter set with the lowest rRMSE were chosen after
the rRMSE arguably reached a convergence limit for several tens of
optimization steps.  \\

\noindent
Electrostatic solvent-solute interactions were computed based on a
minimally distributed charge model (MDCM) for N$_2$O that correctly
describes higher-order multipole
moments.\cite{MM.dcm:2014,MM.mdcm:2017,MM.mdcm:2020} For
parametrization, a reference electrostatic potential (ESP) of N$_2$O
in its linear equilibrium conformation was computed at the
CCSD/aug-cc-pVTZ level using the Gaussian program
package.\cite{gaussian16} The optimized MDCM fit reproduces the ESP
with a RMSE of $0.31$\,kcal/mol. For SF$_6$ in its octahedral
equilibrium conformation the ESP was computed at the MP2/aug-cc-pVTZ
level of theory using the Gaussian program. The RMSE between the
fitted ESP from MDCM and the reference ESP was
$0.11$\,kcal/mol. Recently,\cite{MM.mdcm:2020} non-iterative
polarization was also included in MDCM, and this is also used here for
N$_2$O, SF$_6$ and Xe. The polarizability of linear N$_2$O at the
CCSD/aug-cc-pVTZ level is $2.85$\,\AA$^3$ (with each atom contributing
$\sim 0.95$\,\AA$^3$ per atom), compared with $2.998$\,\AA$^3$ from
experiment.\cite{olney:1997} For Xe at the CCSD/aug-cc-pVTZ the
computed value of $2.96$\,\AA$^3$ compares with $4.005$\,\AA$^3$ from
experiment\cite{olney:1997} and for SF$_6$ the experimentally measured
polarizability of $4.49$\,\AA$^3$ was used and evenly distributed over
the fluorine atoms ($0.74$\,\AA$^3$ per fluorine
atom).\cite{gussoni:1998}\\

\noindent
The LJ parameters for N$_2$O ($\epsilon_\alpha, R_{{\rm min}, \alpha}$
with $\alpha = \{{\rm N, O}\}$) were individually optimized for each
N$_2$O atom by least-squares fitting using the trust region reflective
algorithm\cite{trf:1999} to best reproduce the N$_2$O--Xe and
N$_2$O--SF$_6$ interaction energies, respectively. Reference energies
were computed for a single N$_2$O molecule immersed within differently
dense Xe or SF$_6$ clusters to sample gaseous, supercritical and
liquid solvent environments. For this, 50 cluster conformations for
each solvent state were extracted from earlier MD simulations of
difference solvent concentration.\cite{MM.n2ovib:2023} The clusters
contained 3, 7, and 10 xenon atoms and 2, 6, and 10 SF$_6$ molecules
for gaseous, supercritical and liquid samples within the radius of the
first solvation shell around the N$_2$O center of mass of $6.5$ and
$7.5$\,\AA, respectively. Cluster conformations were randomly
extracted from the trajectories with the respective number of solvent
atoms or molecules within the cluster radius is maintained.\\

\noindent
Counterpoise corrected reference interaction energies between N$_2$O
and the solvent shell of the cluster were computed at the
M06-2X/aug-cc-pVTZ level of theory including D3 dispersion corrections
with the Gaussian program package.\cite{cpc,grimmeD3:2011,gaussian16}
Computations at the CCSD(T) level of theory, as done for the
intramolecular N$_2$O potential, are not feasible for the interaction
energies between N$_2$O and up to 10 Xe atoms or 10 SF$_6$
molecules.\\

\subsection{Molecular Dynamics Simulations}
Molecular dynamics simulations were performed with the CHARMM program
package\cite{Charmm-Brooks-2009} including provisions for RKHS and
MDCM.\cite{MM.rkhs:2017,MM.dcm:2014,MM.mdcm:2020} Each system (N$_2$O
in Xe and N$_2$O in SF$_6$ at given temperature and solvent
concentration) was sampled from 5 independent MD simulations with
initially random solute and solvent arrangements generated using the
packmol code.\cite{martinez:2009} Each simulation was initially heated
and equilibrated for $100$\,ps each, followed by 10\,ns production
simulations in the $NVT$ ensemble using a time step $\Delta t = 1$\,fs
for the leapfrog integration scheme. In total, a total of 50\,ns was
sampled for each system condition.\\

\noindent
The N$_2$O/Xe systems were simulated at a temperature of 291.2\,K and
for N$_2$O/SF$_6$ the temperature was 321.9\,K which both are slightly
above the experimental critical temperatures for condensation of xenon
and SF$_6$, respectively ($T_c(\mathrm{Xe}) = 289.74$\,K,
$T_c(\mathrm{SF_6}) =
318.76$\,K).\cite{ziegler:2019,ambrose:1987,haynes:2014crc} A Langevin
thermostat (coupling $0.1$\,ps$^{-1}$) was used to maintain the
temperature constant but was applied only to the solvent (Xe and
SF$_6$) atoms. Positions and velocities of snapshots of the
simulations were stored every $1$\,fs for analysis. As intermolecular
vibrational energy transfer is slow,\cite{ziegler:2022} the structure
of N$_2$O was optimized and new velocities from a Boltzmann
distribution at the simulation temperature were assigned to N$_2$O
after the heating step. This ensures that the kinetic energies along
the asymmetric, symmetric and bending modes match the thermal energy
with respect to the target simulation temperature.\\

\noindent
The different simulation systems were prepared according to the
conditions used in the
experiments.\cite{ziegler:2018,ziegler:2019,ziegler:2022} Table
S1 summarizes the N$_2$O concentrations $c$(N$_2$O),
molar volumes $V_m$ and critical density ratios $\rho^* = \rho/\rho_c$
used in the simulations. The experimentally determined critical
densities are $\rho_c = 1.11$\,g/ml for xenon and $\rho_c =
0.74$\,g/ml for SF$_6$ from which critical concentrations of $8.45$\,M
and $5.06$\,M for xenon and SF$_6$ are obtained,
respectively.\cite{ambrose:1987,haynes:2014crc} In all setups, the
simulation box contained one N$_2$O molecule and 600 Xe atoms or 343
SF$_6$ molecules which corresponds to similar simulation box volumes
for similar relative density ratios of the two solvents. In the
original parametrization study a simulation box containing 343 SF$_6$
molecules was used to fit temperature-pressure
properties.\cite{samios:2010} \\

\noindent
In the MD simulations for N$_2$O in SF$_6$, electrostatic and
polarization interactions were only computed between the N$_2$O solute
and SF$_6$ solvent. Electrostatic and polarization contributions to
the SF$_6$ solvent-solvent interactions were neglected. Such a
procedure ensures that the pure (liquid, gas) properties of the
solvent are unaltered. \\

\subsection{Analysis}
{\it Rotational Relaxation:} RER times of the N$_2$O solute were
determined by fitting single- or bi-exponential functions to the
autocorrelation functions involving angular momentum-dependent
quantities. The normalized correlation functions $C(t)$ for 
time-dependent scalar $(A(t))$ 
\begin{equation}
\label{eq:lcorr}
C(t) = \dfrac{ \left \langle A(0) \cdot A(t) \right \rangle } {\left
  \langle A(0)^2 \right \rangle}
  \end{equation}
or vectorial $(\vec{A}(t) = \{ A_x(t), A_y(t), A_z(t) \})$
\begin{equation}
    C(t) = \sqrt{ \dfrac{1}{3}
  \sum_{i=\{x,y,z\}} \dfrac{ \ \left \langle A_i(0) \cdot A_i(t) \right
    \rangle ^2 } {\left \langle A_i(0)^2 \right \rangle} }
\end{equation}
observables were computed for time series of the rotational energy of N$_2$O where
$E_\mathrm{rot}(t) = \left| \vec{L}(t) \right|^2 / 2I(t)$ and $I(t)$
is the moment of inertia, the squared angular momentum $\left|
\vec{L}(t) \right|^2$ and angular momentum $\vec{L}(t)$.\\

\noindent
The amplitude $A$, RER rates $1/\tau_i$ and 
offset $\Delta$ of a single exponential function $C_1(t)$
\begin{align}
C_1(t)  = A \mathrm{e}^{-t/\tau} + \Delta \label{eq:lfit1}
\end{align}
were optimized to fit the sampled rotational energy correlation
function from simulations. For the fit, the lower limit for the time
was $t = 0.2$\,ps, as had also been done for fitting the experimental
results to avoid any pulse overlap effects,\cite{ziegler:2022} and was
restricted to $C(t) \geq 0$. It is noted that using a double
exponential function does not provide improved representations of the
data even near the critical point.\\

\noindent
{\it Vibrational Relaxation:} 
Transition rates $k_{i \rightarrow j} = \tau_{ij}^{-1}$ for the 
vibrational state transition ${i \rightarrow j}$
of the N$_2$O solute in solution were computed from a Landau-Teller
model\cite{Hynes:1992,Hynes:1998,kato:1998,anfinrud:1999,skinner:1996,skinner:1998,skinner:1999}
\begin{equation}
    k_{i \rightarrow j} = \dfrac{1}{\tau_{ij}(\omega_{ij})} =
    \gamma_{ij} \int_{-\infty}^\infty dt \exp(i \omega_{ij} t) \zeta
    (t)
\end{equation}
with a proportionality factor $\gamma_{ij}$ defined as
\begin{align}
    \gamma_{ij} &=
        \dfrac{2 \hbar^{-2}}{1 + \exp(-\beta\hbar\omega_{ij})}
        \cdot Q(\omega_{ij}) \\
    Q(\omega_{ij}) &=
        \dfrac{\beta \hbar \omega_{ij}}{2}
        \coth{\left ( \dfrac{\beta \hbar \omega_{ij}}{2} \right )}
\end{align}
which includes $\beta = 1/(k_\mathrm{B}T)$ and a quantum correction 
factor $Q(\omega)$ for the classically
obtained correlation function $\zeta(t)$.\cite{kato:1998,berne:1994}
Here, $\zeta(t)$ is the time-dependent friction acting on the
oscillation of the solute and is derived from the correlation function
\begin{equation}
    \zeta (t) = \left \langle V_{ij}(t) \cdot V_{ji}(0) \right \rangle
\end{equation}
of the classical solute-solvent interaction friction potential $V_{ij}(t)$.
The solute-solvent interaction is obtained from the solute-solvent force 
$\vec{F}_\mathrm{int}(t)$ projected along the normal mode vector $\vec{q}_i$ by
the sum over the solute atoms $\alpha$ according to
\begin{equation} \label{eq:fproj}
    V_{ij}(t) = \mu_i \sum_\alpha \dfrac{1}{m_\alpha} 
    \dfrac{\partial V_\mathrm{int}(t)}{\partial \vec{q}_{i,\alpha}(t)}
    \cdot \vec{q}_{i,\alpha}(t) 
    = - \sum_\alpha \vec{F}_{\mathrm{int},\alpha}(t) \cdot \vec{q}_{i,\alpha}(t),
\end{equation}
where $m_\alpha$ is the atom mass of $\alpha$ and 
$\mu_i$ is the reduced mass of normal mode $i$.\cite{kato:1998} Because
a quantum mechanical correlation function was replaced by its classical analogue, quantum correction factors $Q$ have been introduced.\cite{berne:1994,skinner:1998,skinner:2001} The typical behaviour of $Q(\omega)$ is 
$\approx 1$ for $\hbar \omega \ll k_\mathrm{B}T$
($\lim_{\omega \rightarrow 0} Q(\omega) = 1$),
whereas $Q(\omega)$ can become significantly larger than 1 for 
$\hbar \omega \gtrsim k_\mathrm{B}T$.\\

\section{Results}

\subsection{Validation of Inter- and Intramolecular Interactions}
For investigating the RER and VER of N$_2$O in gaseous, supercritical,
or liquid Xe or SF$_6$ environments an accurate representation of the
solute-solvent interaction is
crucial.\cite{ziegler:2022,ziegler:2019,ziegler:2018} The results of
the parameter optimization for the intermolecular interactions are
summarized next.\\

\noindent
{\it Solute Potential:} To establish the quality of the N$_2$O PES,
the vibrational modes were determined by solving the 3d nuclear
Schr\"odinger equation using the DVR3D\cite{dvr3d:2004} package. The
computed fundamental asymmetric, symmetric, and bending vibrations
were at $\nu_{\rm as} = 2229$\,cm$^{-1}$, $\nu_{\rm s} =
1291$\,cm$^{-1}$, and $\nu_{\rm b} = 598$\,cm$^{-1}$ and agree well
with 2224\,cm$^{-1}$, 1285\,cm$^{-1}$, and 589\,cm$^{-1}$ from
experiments in the gas
phase.\cite{herzberg:1945,herzberg:1950,kagann:1982} For the bending
overtone $(2 \nu_{\rm b})$ the computations yield 1184 cm$^{-1}$
compared with 1168 cm$^{-1}$ from
experiments.\cite{herzberg:1945,herzberg:1950,kagann:1982}\\

\begin{figure}[htb!]
\begin{center}
\includegraphics[width=0.5\textwidth]{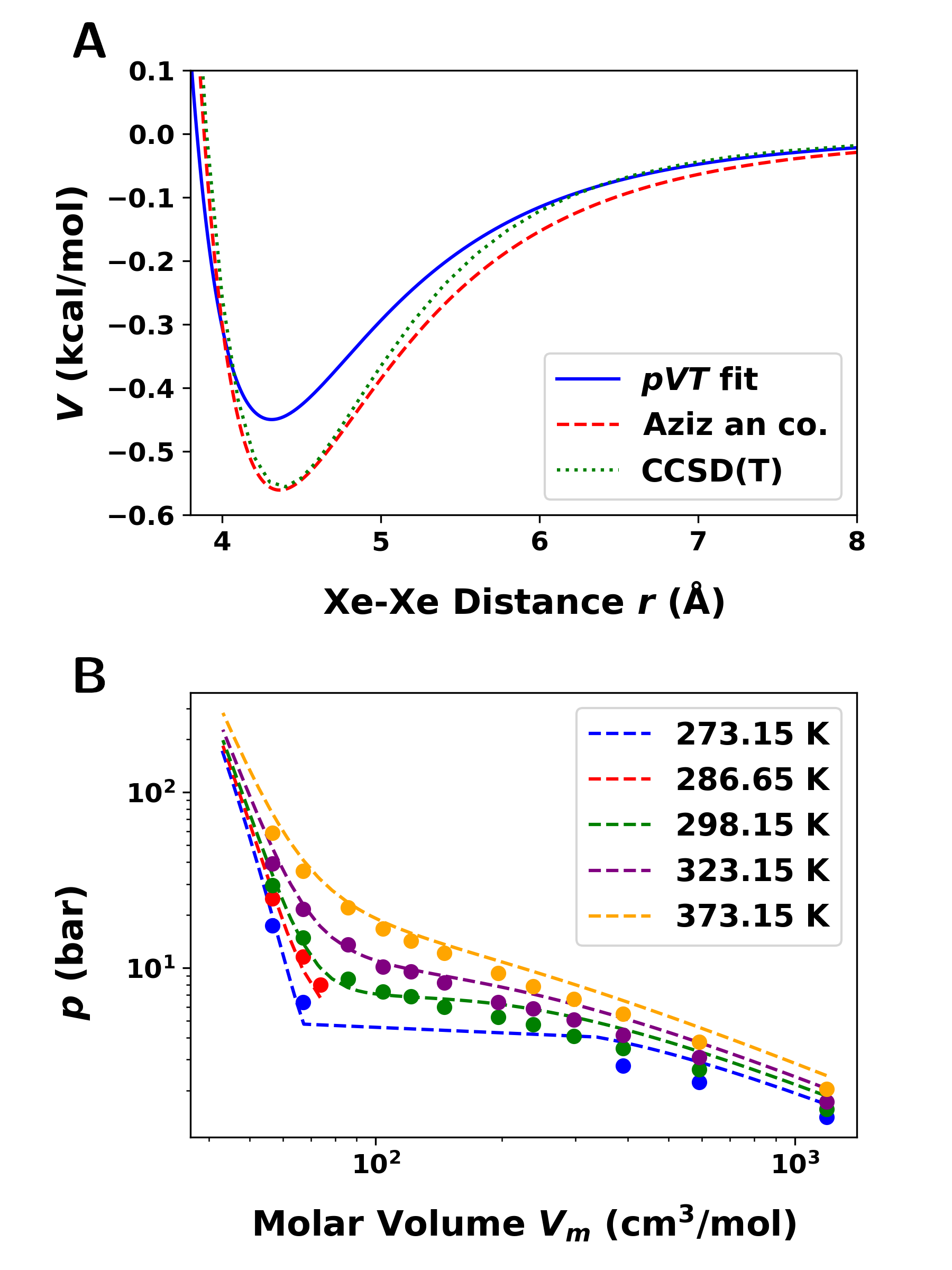}
\caption{Panel A: Two-body interaction potential $V$ between Xe atoms
  computed by Lennard-Jones potential function with the optimized LJ
  parameters (solid blue line) and from Aziz {\it et al.} (dashed red
  line).\cite{aziz:1986} The dotted green line shows {\it ab initio}
  reference potential for a Xe-Xe atom pair at CCSD(T)/cc-pV6Z level
  of theory corrected by higher coupled-cluster level
  contributions.\cite{bich:2017} Panel B: Experimentally observed
  isotherms in the $pVT$ diagram for pure xenon (dashed
  lines)\cite{louwerse:1954xenon} and computed $pVT$ state points
  (full circles) with optimized LJ parameters. }
\label{fig:vdw_fit}
\end{center}
\end{figure}

\noindent
{\it Solvent Potential:} The Xe and SF$_6$ solvent potential models
aim at accurately reproducing the experimentally measured $pVT$ state
points.  The atomic LJ parameters for SF$_6$ were already fitted to
match reference $pVT$ state points of pure gaseous, supercritical and
liquid SF$_6$ systems.\cite{samios:2010} In contrast, for xenon the
previously used LJ parameters were fitted to match dilute gas
macroscopic properties such as virial coefficient, viscosity and
thermal conductivity over a wide temperature range but not
specifically for correct phase transition and supercritical fluid
properties.\cite{aziz:1986} As described above, the LJ parameters were
refitted to reproduce the reference system pressure of pure xenon at
44 different density and temperature
conditions.\cite{louwerse:1954xenon} Figure \ref{fig:vdw_fit}A shows
the measured reference isotherms in the $pVT$ diagram (dashed lines)
and the computed $pVT$ state points from simulations using the
optimized LJ parameters with the lowest relative RMSE of $15.1$\%
between reference and computed system pressures (absolute RMSE of
$32.7$\,Pa). This is an improvement by a factor of $\sim 5$ (relative
RMSE of $75.1$\%; absolute RMSE of $124.3$\,Pa) when using the
original parameters.\cite{aziz:1986} \\

\noindent
Figure \ref{fig:vdw_fit}B compares the Xe--Xe interaction based on a
LJ potential using the optimized LJ parameters and those from Aziz
{\it et al.}. The dissociation energies are $0.450$\,kcal/mol and
$0.561$\,kcal/mol, respectively, at equilibrium separations of $r_{\rm
  e} = 4.31$\,\AA\/ and 4.36\,\AA\/. This compares with a dissociation
energy of 0.556\,kcal/mol and $r_{\rm e} = 4.40$\,\AA\/ from
high-level CCSD(T)/cc-pV6Z calculations.\cite{bich:2017} The zero
point vibrational energy for a xenon pair is found to be
$10.4$\,cm$^{-1}$ ($0.030$\,kcal/mol) experimentally\cite{tanaka:1974}
and $0.029$\,kcal/mol computationally using the CCSD(T)/cc-pV6Z
potential data and a 1-dimensional DVR method.\cite{bich:2017}\\

\begin{figure}[htb!]
\begin{center}
\includegraphics[width=0.35\textwidth]{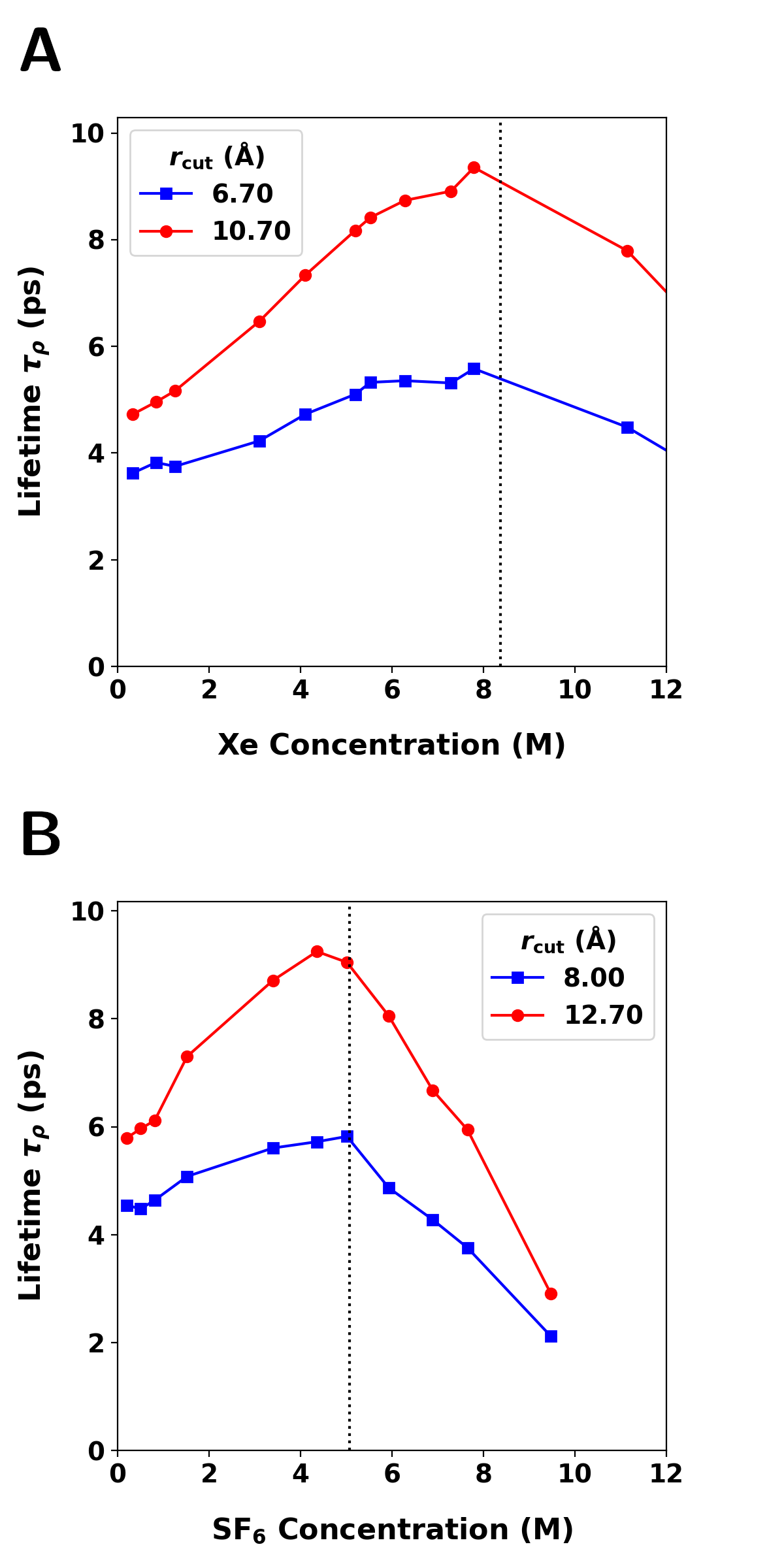}
\caption{Local solvent reorganization lifetime $\tau_\rho$ from MD
  simulations of pure (A) Xenon at $T = 291.2$\,K and (B) SF$_6$
  system at $T = 321.9$\,K. The cutoff radii of the local solvent
  clusters are determined from the local minima in the respective
  radial distribution functions. The vertical dotted line marks the
  experimentally determined critical concentration at the critical
  temperatures of $c({\rm Xe}) = 8.45$\,M and $c({\rm SF_6}) =
  5.02$\,M, respectively.\cite{haynes:2014crc}}
\label{fig:ldtau_solv}
\end{center}
\end{figure}

\noindent
Following previous work,\cite{MM.n2ovib:2023} the local solvent
reorganization lifetime $\tau_\rho$ was used to probe for distinct SCF
properties and as a proxy to determine the computationally predicted
critical density at the given temperature, following established
routes.\cite{tucker:2000a,tucker:2000b} Figures \ref{fig:ldtau_solv}A
and B show the computed $\tau_\rho$ for liquid xenon and SF$_6$,
respectively. For the analysis, the cutoff radii for the first and
second solvation shells of a local solvent residue cluster were
estimated from the local minima of the respective radial distribution
function $g(r)$ determined from the same trajectory. For both
solvents, the maxima of $\tau_\rho$ for the simulated solvent
concentration are close to the experimentally observed critical
concentrations at the critical temperature. As was found in previous
work,\cite{tucker:2000a,tucker:2000b} the solvent fluctuation
lifetime, $\tau_\rho$, in LJ gas systems is a maximum for
supercritical fluids at the critical concentration. This indicates
that the supercritical dynamical properties are well captured by the
solvent potential model and are one more manifestation of critical
slowing, at least in pure xenon and SF$_6$.\\

\begin{figure}[htb!]
\begin{center}
\includegraphics[width=0.35\textwidth]{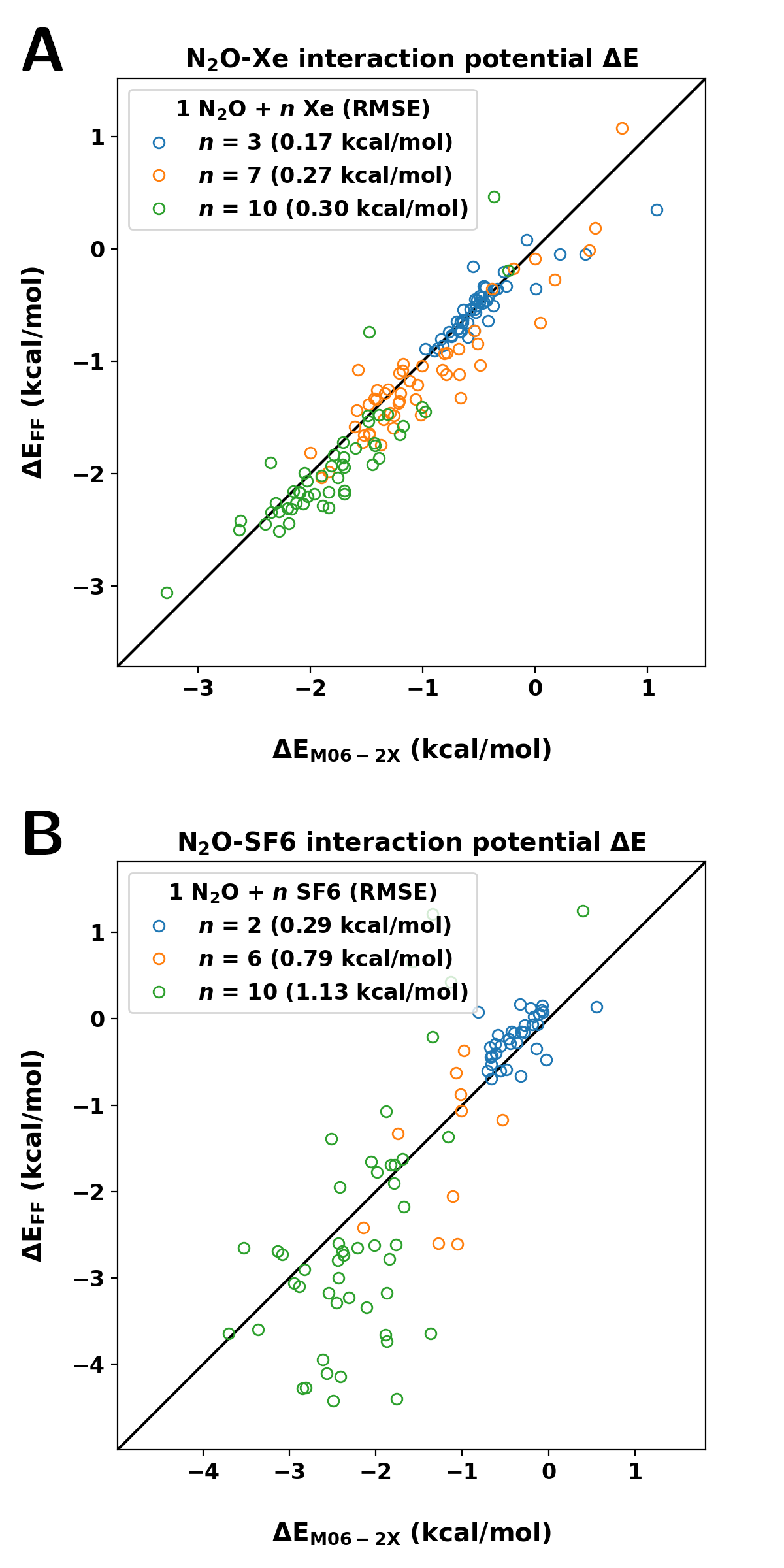}
\caption{Correlation plot of the optimized interaction potential model
  against the {\it ab initio} reference interaction energies between
  N$_2$O within (A) xenon and (B) SF$_6$ solvent clusters of different
  densities. The number of solvent particles in the cluster are given
  in the respective legend, respectively.}
\label{fig:fit_inter}
\end{center}
\end{figure}

\noindent
{\it Intermolecular Interactions:} The intermolecular interactions
between the N$_2$O solute and the Xe or SF$_6$ solvent were fit to
reproduce counterpoise corrected {\it ab initio} reference interaction
energies for various solute-solvent clusters. Figures
\ref{fig:fit_inter}A and B show the correlation between the modelled
and reference interactions for optimized LJ parameter of the N$_2$O
atoms, respectively. The solute-solvent clusters with N$_2$O at their
center have a maximum radius of $6.5$ \AA\/ for Xe and $7.5$ \AA\/ for
SF$_6$, respectively. The number of solvent atoms or molecules within
the cluster are representative for the local density of gaseous,
supercritical and liquid solvent environments, respectively, and the
selection of cluster conformations are described in the Method
section.\\

\noindent
The N$_2$O--Xe interaction in Figure \ref{fig:fit_inter}A with
optimized LJ parameter for N$_2$O reproduces the reference interaction
energies with RMSEs from gaseous to liquid-like solvent clusters with
$0.17$, $0.27$ and $0.30$\,kcal/mol, respectively. In contrast, for
the N$_2$O--SF$_6$ interactions (see Figure \ref{fig:fit_inter}B) the
RMSEs increase to $0.29$, $0.79$ and $1.13$\,kcal/mol,
respectively. The interactions in supercritical and liquid-like
cluster environments with $n=6$ and 10 SF$_6$ molecules are
overestimated due to stronger electrostatic interactions compared with
Xe.\\

\subsection{Rotational Energy Relaxation (RER)}
\begin{figure}[htb!]
\begin{center}
\includegraphics[width=0.90\textwidth]{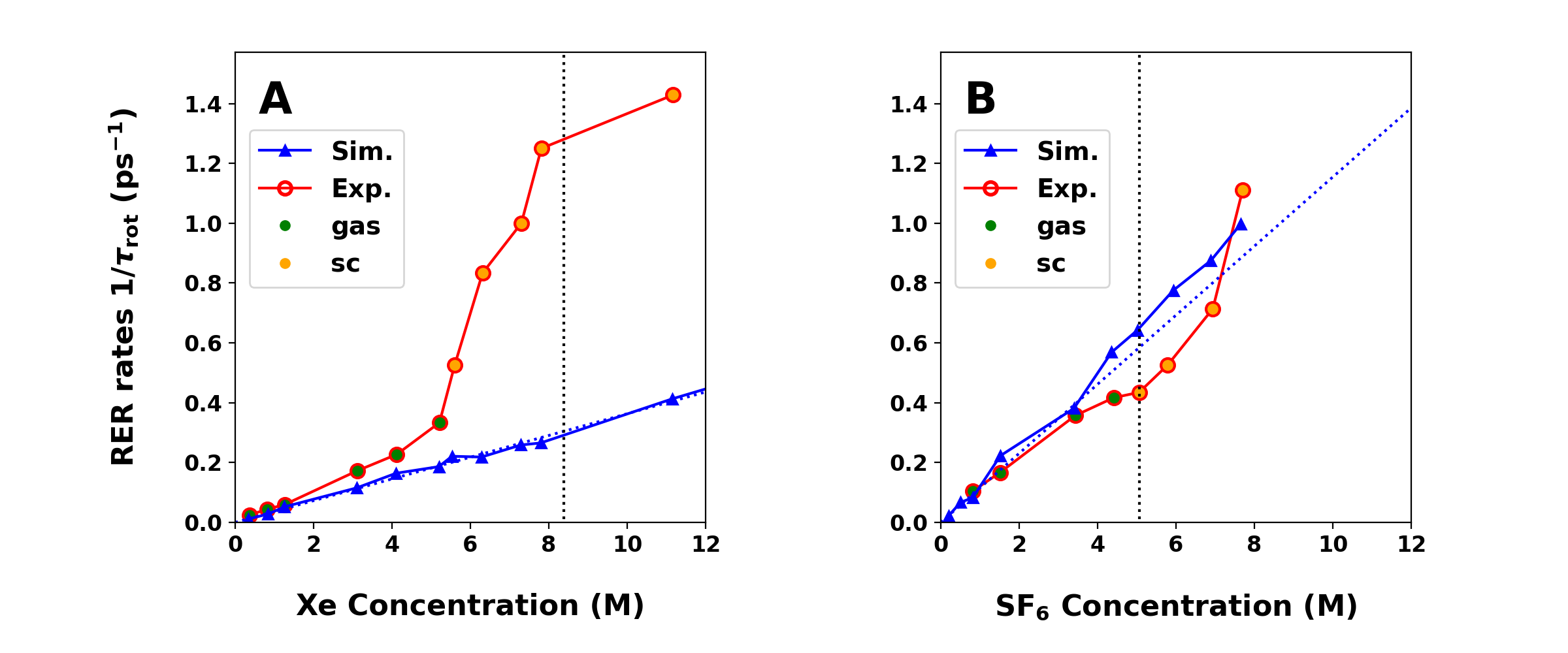}
\caption{RER rates from the RER times $\tau_\mathrm{rot}$ of a fitted
  single-exponential function to correlation function of the
  rotational energy $E_\mathrm{rot}(t)$ of N$_2$O in (A) xenon and (B)
  SF$_6$ at different solvent concentrations (solid blue lines).  The
  solid red line with colored circle markers shows the experimentally
  measured RER rates and indicate the solvent state gaseous or
  supercritical (sc).\cite{ziegler:2022} The dotted blue line is an
  linear extrapolation of the first 3 simulated RER rates. The
  vertical dotted lines mark the experimentally observed solvent
  concentration at the respective critical density of xenon and
  SF$_6$, respectively.}
\label{fig:rotrates}
\end{center}
\end{figure}

From the simulations the N$_2$O atom positions and velocities were
recorded every 50\,fs from the aggregate of 50\,ns ($5 \times 10$ ns)
for each system and composition. The RER rates $1/\tau_{\rm rot}$ were
then determined from the $E_\mathrm{rot}(t)$ autocorrelation function
by fitting to a single--exponential function (see Equation
\ref{eq:lfit1}), yielding a single RER time $\tau_{\rm rot}$.  Figure
\ref{fig:rotrates} shows RER rates in xenon and SF$_6$ retrieved from
the corresponding calculated rotational energy correlation functions,
see Figures S1 and S2.

\noindent
As an extension, a bi-exponential function was used to fit the
rotational energy correlation functions. This yields two time scales:
fast and a slow RER times $\tau_\mathrm{rot,fast}$ and
$\tau_\mathrm{rot,slow}$ shown in Figure S3. In
general, $\tau_{\rm rot}$ from the single-exponential fit matches the
slow component of the RER time $\tau_{\rm rot,slow}$ from which the
respective rates are closest to the experimental results. Thus, only
the fit parameters of the single-exponential function will be
discussed.\\

\noindent
Analysis of the 2D IR experiments provided RER rates for N$_2$O
molecules in gaseous, supercritical and liquid solvent
regions.\cite{ziegler:2018, ziegler:2019, ziegler:2022} For xenon a
characteristic increase in $1/\tau_{\rm rot}$ for $c{\rm [Xe]} > 4$ M
was found whereas for SF$_6$ critical slowing for $c{\rm [SF_6]} > 4$
M was observed (red traces in Figures \ref{fig:rotrates}A and B).
Both measurements deviate from results predicted by the simplest IBC
model which predicts a linear dependence between $c{\rm [solvent]}$
and $1/\tau_{\rm rot}$. From the simulations in xenon, however, the
results follow that expected from IBC (blue symbols in Figure
\ref{fig:rotrates}A) whereas for SF$_6$ the computed results deviate
around the predictions from IBC, see Figure \ref{fig:rotrates}B.\\

\noindent
The agreement between computed RER rates and the IBC model rather than
the experimentally observed RER rates at supercritical solvent
concentrations becomes apparent when scaling the rates with the
concentration to yield the RER rates $k_\mathrm{rot}$. From the slopes
of the RER rate dependence of the IBC region in the Figure
\ref{fig:rotrates}, the simulated N$_2$O RER rates are $k_{\rm
  rot}^{\rm Xe} = (3.67\pm0.25)\cdot10^{10}$\,s$^{-1}$M$^{-1}$ and
$k_{\rm rot}^{\rm SF_6} = (1.25\pm0.12)\cdot10^{11}$\,s$^{-1}$M$^{-1}$
for N$_2$O in Xe and SF$_6$, respectively. 2DIR derived measured RER
rates for N$_2$O in gaseous xenon and SF$_6$ solvent within the IBC
region are $5.36\cdot10^{10}$\,s$^{-1}$M$^{-1}$ and
$1.02\cdot10^{11}$\,s$^{-1}$M$^{-1}$, respectively, which are rather
close to the simulation results.\cite{ziegler:2022} The ratio $k_{\rm
  rot}^{\rm SF_6} / k_{\rm rot}^{\rm Xe} \sim 3$ also qualitatively
agrees with a factor of $\sim 2$ from the experiments.\\

\noindent
Up to this point the rotational energy was used to determine the
correlation function and to obtain RER rates. To probe whether other
angular-momentum-dependent quantities lead to similar conclusions,
correlation functions for N$_2$O angular momentum $\vec{L}(t)$, and
the squared angular momentum $|L(t)|^2$ were considered, see Figures
S4 and S5. Using
$\vec{L}(t)$ yields higher rates than those from the squared angular
momentum $|L(t)|^2$ which in turn closely matches the rates based on
the rotational energy correlation function. This is not unexpected as
the rotational energy is proportional to the squared angular momentum:
$E_\mathrm{rot}(t) = |L(t)|^2 / (2I)$ with $I$ the N$_2$O moment of
inertia. The rates from $\vec{L}(t)$ are higher as they also include
the reorientation of the rotational axis of N$_2$O known as rotational
Brownian motion.\cite{berne:2000}\\

\noindent
The respective correlation functions for $\vec{L}(t)$ and $|L(t)|^2$
of N$_2$O in xenon and SF$_6$ are shown in Figures S6
to S9. The correlation functions are the average of
the correlation functions computed from each single sample run at the
respective solvent concentration. RER rates from fits to the
correlation functions of the angular momentum $\vec{L}(t)$, squared
angular momentum $| \vec{L}(t) |^2 $ and rotational energy
$E_\mathrm{rot}$ are shown in Figures S10
and S11.  \\

\noindent
To conclude, for low solvent densities ($c{\rm [Xe]} < 4$ M and $c{\rm
  [SF_6]} < 4$ M) the computed RER rates agree very favourably with
experiments whereas for higher densities the deviations from the IBC
model observed in the experiments are not correctly captured,
specifically for xenon. Rather, the simulations follow the predicted
behaviour from the IBC model. As the deviations are systematic with
increasing solvent concentration it is conjectured that as the solvent
concentration increases the property probed by the experiment is not
only the solute rotational energy. This is discussed further below. In
addition, many-body effects may be insufficiently accounted for in the
computational model, thus contributing to the lack of quantitative
agreement particularly for N$_2$O in xenon at higher density.\\

\subsection{Collision Analysis}
The computed RER of N$_2$O in Xe and SF$_6$ followed the simple IBC
model and is in disagreement with the experimentally observed rapid
increase around $c{\rm [Xe] > 5}$ M and critical slowing at the
critical density for SF$_6$. To gain deeper insight into the
solute-solvent interactions, the change in kinetic energy before and
after N$_2$O--solvent collisions was further analyzed and decomposed
into translational, rotational, and vibrational contributions of
N$_2$O at different solvent densities.  \\

\begin{figure}[htb!]
\begin{center}
\includegraphics[width=0.90\textwidth]{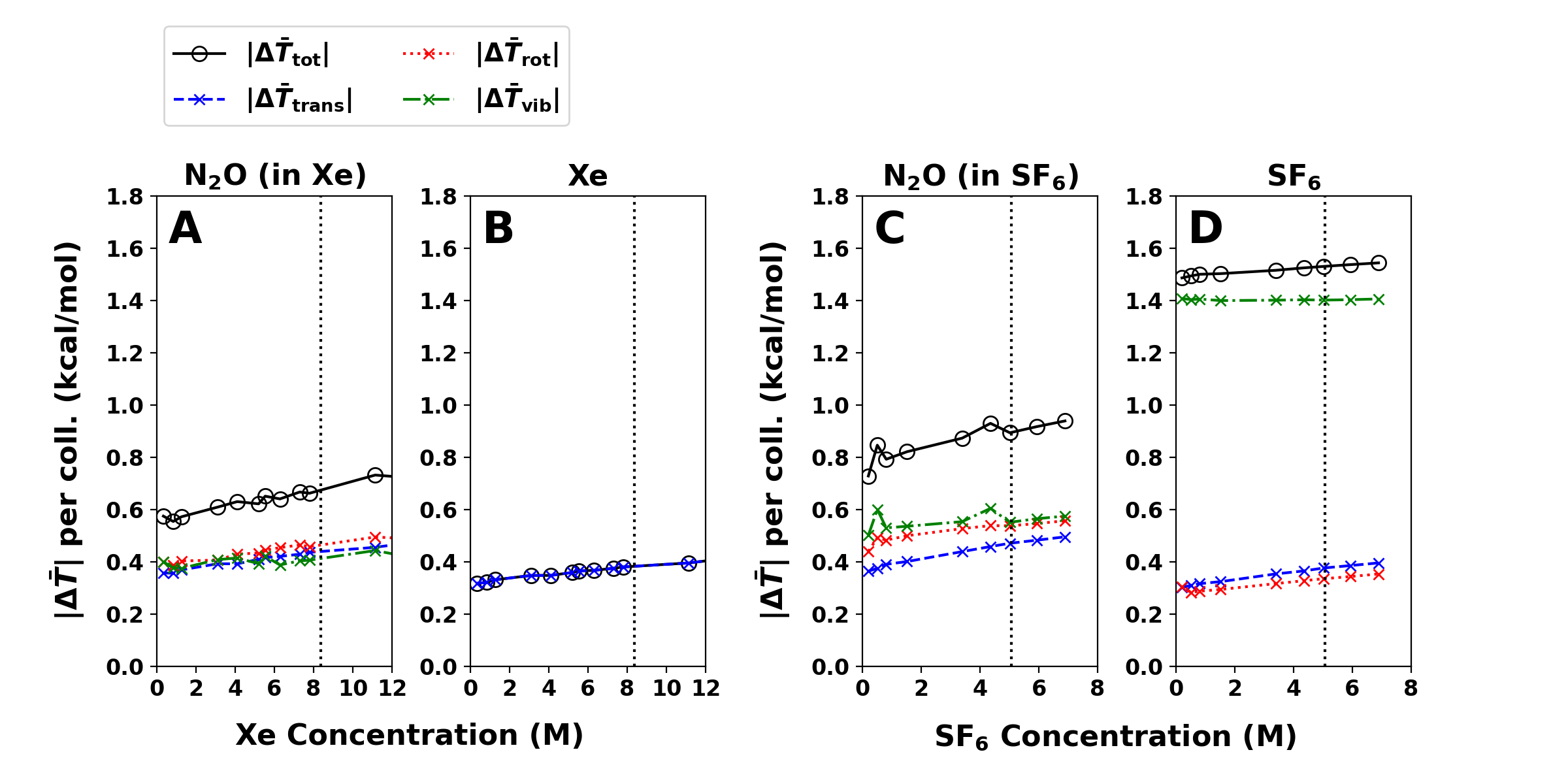}
\caption{Averaged absolute energy difference $| \Delta \bar{T} |$ of
  the total, translational, rotational and vibrational kinetic energy
  after a collision between (A-B) N$_2$O and xenon, and (C-D) N$_2$O
  and SF$_6$, respectively, at different solvent concentrations.  The
  vertical dotted lines mark the experimentally observed solvent
  concentration at the respective critical density. Note the different
  system temperatures of 291.2\,K and 321.9\,K for N$_2$O in xenon and
  SF$_6$, respectively.}
\label{fig:ekin_diff_all}
\end{center}
\end{figure}

\noindent
Figure \ref{fig:ekin_diff_all} shows the average absolute change in
the kinetic energies of N$_2$O, xenon atoms and SF$_6$ molecules upon
collision. A collision between N$_2$O and solvent was considered to
have occurred when any of the N$_2$O--Xe or N$_2$O--SF$_6$ atom
separations is smaller than the sum of their atomic van-der-Waals
radii ($1.55, 1.52, 2.16, 1.80, 1.47$\,\AA\/ for N, O, Xe, S and F,
respectively).\cite{vdwradii:1964} The initial and final kinetic
energies of N$_2$O and solvent were extracted from the simulation at
the frame after crossing this van-der-Waals radii threshold from which
the changes $|\Delta \bar{T}_{\alpha}|$ for $\alpha = {\rm total, vib,
  rot, trans}$ and corresponding distributions $P(|\Delta
\bar{T}_{\alpha}|)$ were determined. Due to intramolecular energy flow
and intermolecular exchange of kinetic energy between the solute and
the surrounding solvent molecules during the contact time, the sum of
changes in the kinetic energy contributions does not necessarily add
up to the total kinetic energy difference.\\

\noindent
Related to RER are changes in the N$_2$O rotational kinetic energy,
$|\Delta \bar{T}_\mathrm{rot}|$, see Figures \ref{fig:ekin_diff_all}A
and C (red dotted line). For N$_2$O in Xe, $|\Delta
\bar{T}_\mathrm{rot}|$ is lower than in SF$_6$
and the computed ratio $|\Delta \bar{T}_{\rm rot}^{\rm SF_6} / \Delta \bar{T}_{\rm rot}^{\rm Xe}| \sim 1.2$ compares with 1.9 from the experiments.\cite{ziegler:2022} 
This indicates that in simulations with SF$_6$ as the solvent the change in the N$_2$O rotational energy
occurs not only due to collisions as for xenon. Rather, RER is also
influenced by the longer-ranging, anisotropic intermolecular
interactions between N$_2$O and and the nonvanishing higher-order
multipoles of SF$_6$. As the average change in vibrational kinetic
energy in $|\Delta \bar{T}_\mathrm{vib}|$ (dash-dotted green line) of
N$_2$O is larger in SF$_6$ than in xenon, the overall change in total
kinetic energy $|\Delta \bar{T}_\mathrm{tot}|$ (solid black line) is
considerably larger, whereas the changes in translational kinetic
energy $|\Delta \bar{T}_\mathrm{trans}|$ (dashed blue line) for the
two solvents are comparable because the masses of xenon ($131.3$\,u)
and SF$_6$ ($146.1$\,u) are similar.\\

\noindent
The change in the kinetic energy contributions of the solvent
particles, xenon and SF$_6$, after a collision with N$_2$O are shown
in Figures \ref{fig:ekin_diff_all}B and D. At an atomistic level the
kinetic energy of a single xenon atom only consists of the
translational contribution. Again, the translational contributions for
xenon and SF$_6$ are comparable (blue symbols in Figures
\ref{fig:ekin_diff_all}B and D) due to their similar masses. However,
because SF$_6$ has internal degrees of freedom, additional rotational
and vibrational kinetic energy contributions arise and the total
$|\Delta \bar{T}_\mathrm{tot}|$ (open circles) for collisions with
SF$_6$ are considerably larger than those of xenon. For SF$_6$ the
rotational and translational contributions are similar in magnitude
$|\Delta \bar{T}_\mathrm{rot}| \sim |\Delta \bar{T}_\mathrm{trans}|$
whereas $|\Delta \bar{T}_\mathrm{vib}|$ is significantly larger. This
is due to the larger number of vibrational degrees of freedom (15)
compared with rotation (3) and translation (3). It is also interesting
to note that the translational and rotational contributions for SF$_6$
depend on solvent-concentration whereas the vibrational contribution
does not. This is consistent with the analysis of the experimental
data which indicates that vibrational relaxation through
intramolecular energy relaxation upon collision is IBC-like as opposed
to rotational relaxation.\cite{ziegler:2022}\\

\begin{figure}[htb!]
\begin{center}
\includegraphics[width=0.90\textwidth]{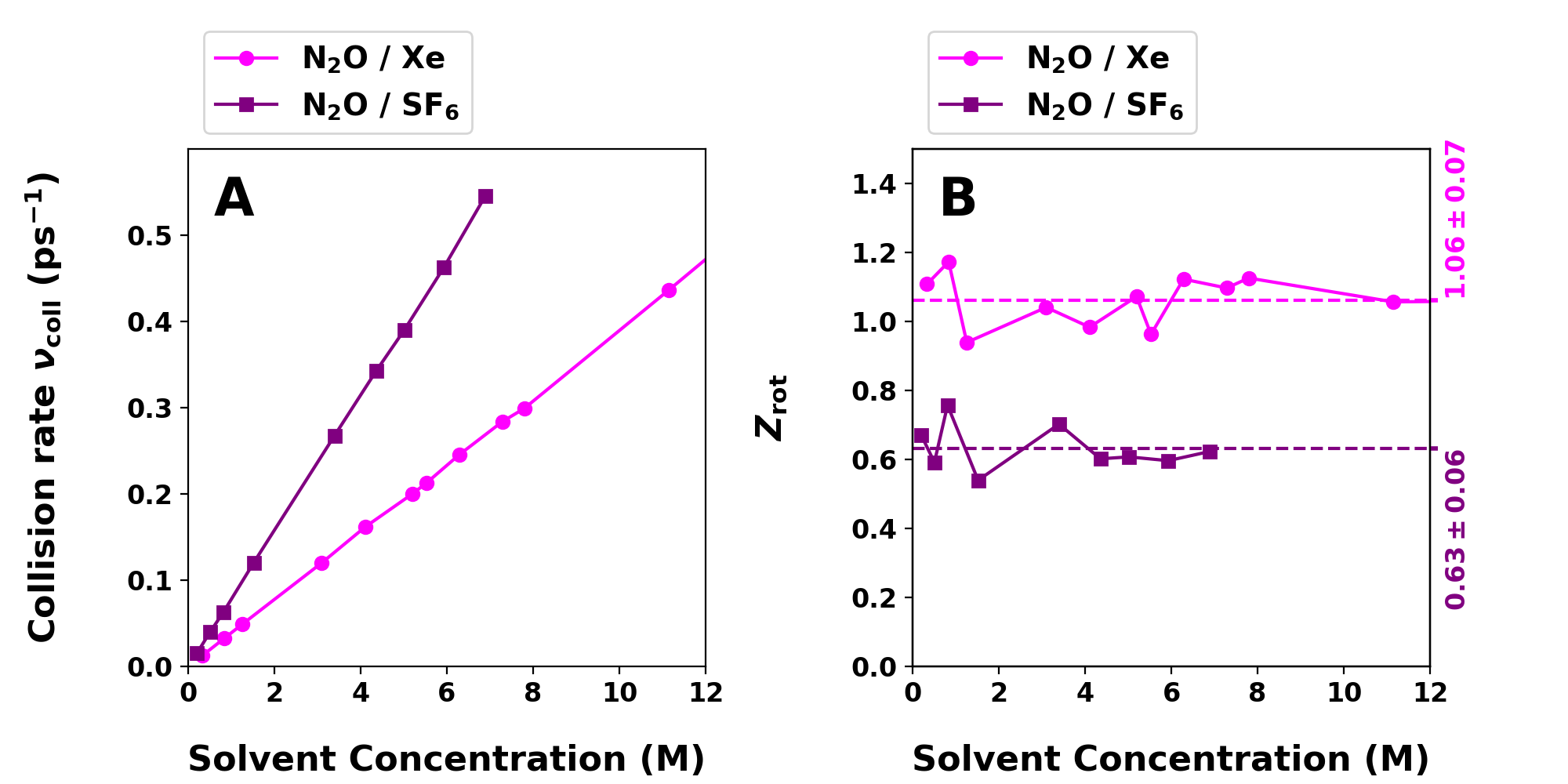}
\caption{Panel A: Average collision rates $\nu_\mathrm{coll}$ between
  N$_2$O and xenon (magenta) and SF$_6$ (purple) according to the
  collision conditions using atom distances and van-der-Waals radii
  thresholds defined for this work. Panel B: Average Number of
  collisions $Z_\mathrm{rot}$ for rotational relaxation of N$_2$O in
  xenon and SF$_6$ in the average and standard deviation displayed on
  the right axis.}
\label{fig:zrot}
\end{center}
\end{figure}

\noindent
Based on the RER rates $1/\tau_\mathrm{rot}$ determined in the
previous subsection and the average collision frequency
$\nu_\mathrm{coll} = N_\mathrm{coll} / t_\mathrm{sim}$
($t_\mathrm{sim} = 50$\,ns) it is also possible to compute the average
number of collisions $Z_\mathrm{rot} = \tau_\mathrm{rot} \cdot
\nu_\mathrm{coll}$ required to rotationally relax the solute
(N$_2$O). Here, $N_\mathrm{coll}$ is the number of collisions between
N$_2$O and either xenon or SF$_6$, and $1/\tau_\mathrm{rot}$ was
already obtained from fitting a single-exponential decay to the
rotational energy correlation function of N$_2$O in both
solvents. Figure \ref{fig:zrot}A and B shows the computed average
collision rate $\nu_\mathrm{coll}$ and number of collision
$Z_\mathrm{rot}$ for RER of N$_2$O in xenon and SF$_6$,
respectively. The collision rate of N$_2$O with SF$_6$ (purple line in
Figure \ref{fig:zrot}A) is higher than with xenon (magenta line)
because of the higher simulation temperature ($321.9$\,K
vs. $291.2$\,K), the larger molecular volume occupied by SF$_6$
compared to a single xenon atom (based on atomic van der Waals radii),
and the anisotropic electrostatic intermolecular interactions between
N$_2$O and SF$_6$.\\

\noindent
The experimentally reported $Z_\mathrm{rot}^{\rm expt}$ are $2.4$ and
$1.7$ for RER of N$_2$O in xenon and SF$_6$, respectively, based on a
hard sphere model for both types of solvent
molecules.\cite{yardley1980,ziegler:2022} Within the IBC regime
($c{\rm [Xe]} < 4$ M and $c{\rm [SF_6]} < 4$ M) the
$Z_\mathrm{rot}^{\rm expt}$ are larger than those from
simulations. This is mostly due to the higher computed collision rate
$\nu_\mathrm{coll}$, estimated from the inverse of the mean free
collision time. On the other hand, the ratio $Z_\mathrm{rot}^{\rm
  expt, Xe} /Z_\mathrm{rot}^{\rm expt, SF_6}$ is comparable but
somewhat smaller ($\sim1.4$) than that from the simulations
($\sim1.7$). This is also related to lower simulated RER rates for
N$_2$O in xenon within the IBC region compared with experiments for
which the simulated RER rates for N$_2$O in SF$_6$ matches well. Thus,
the simulations reasonably well agree with the experiments for RER of
N$_2$O over solvent concentrations within the IBC regime. However, the
simulations do not reproduce the steep increase or critical slowing in
the RER rates in xenon and SF$_6$, respectively (see Figure
\ref{fig:rotrates}).\\

\subsection{Vibrational Energy Relaxation}
VER lifetimes and rates were determined from the Landau-Teller
formalism described in the Methods section. VER of a particular N$_2$O
mode can either involve pure intramolecular relaxation or a
combination of inter- and intramolecular processes, both of which can
be obtained from MD
simulations.\cite{Hynes:1992,Hynes:1998,kato:1998,anfinrud:1999,skinner:1999}
For this, the forces acting on the solute atoms derived only from the
non-bonding solute-solvent interaction potential were extracted and
projected along the normal mode vector of the asymmetric stretch
vibration $\nu_{\rm as}$ of N$_2$O, see Eq. \ref{eq:fproj}.\\

\noindent
For N$_2$O in SF$_6$ and from the experimental
analysis,\cite{ziegler:2022} the most effective vibrational relaxation
channel of $\nu_{\rm as}^{\rm N_2O} \sim 2220$ cm$^{-1}$ occurs
through coupling with the symmetric N$_2$O stretch vibration $\nu_{\rm
  s}^{\rm N_2O} \sim 1280$ cm$^{-1}$ and the three-fold degenerate
S--F stretch vibration $\nu_3^{\rm SF_6} \sim 948$ cm$^{-1}$.
Following the procedure outlined in the methods section, the VER rate
for $(\nu_\mathrm{s}, \nu_\mathrm{as}) = (0,1) \rightarrow (1,0)$ were
computed.  The forces of N$_2$O are projected along the normal mode
vectors of $\nu_\mathrm{as}$ obtained from instantaneous normal mode
analysis at the respective frames, which can be interpreted as solvent
friction.  The Fourier transform of the correlation function of this
quantity yields the frequency dependent friction function, see Figures
S16 and S17.\\

\begin{figure}[htb!]
\begin{center}
\includegraphics[width=0.90\textwidth]{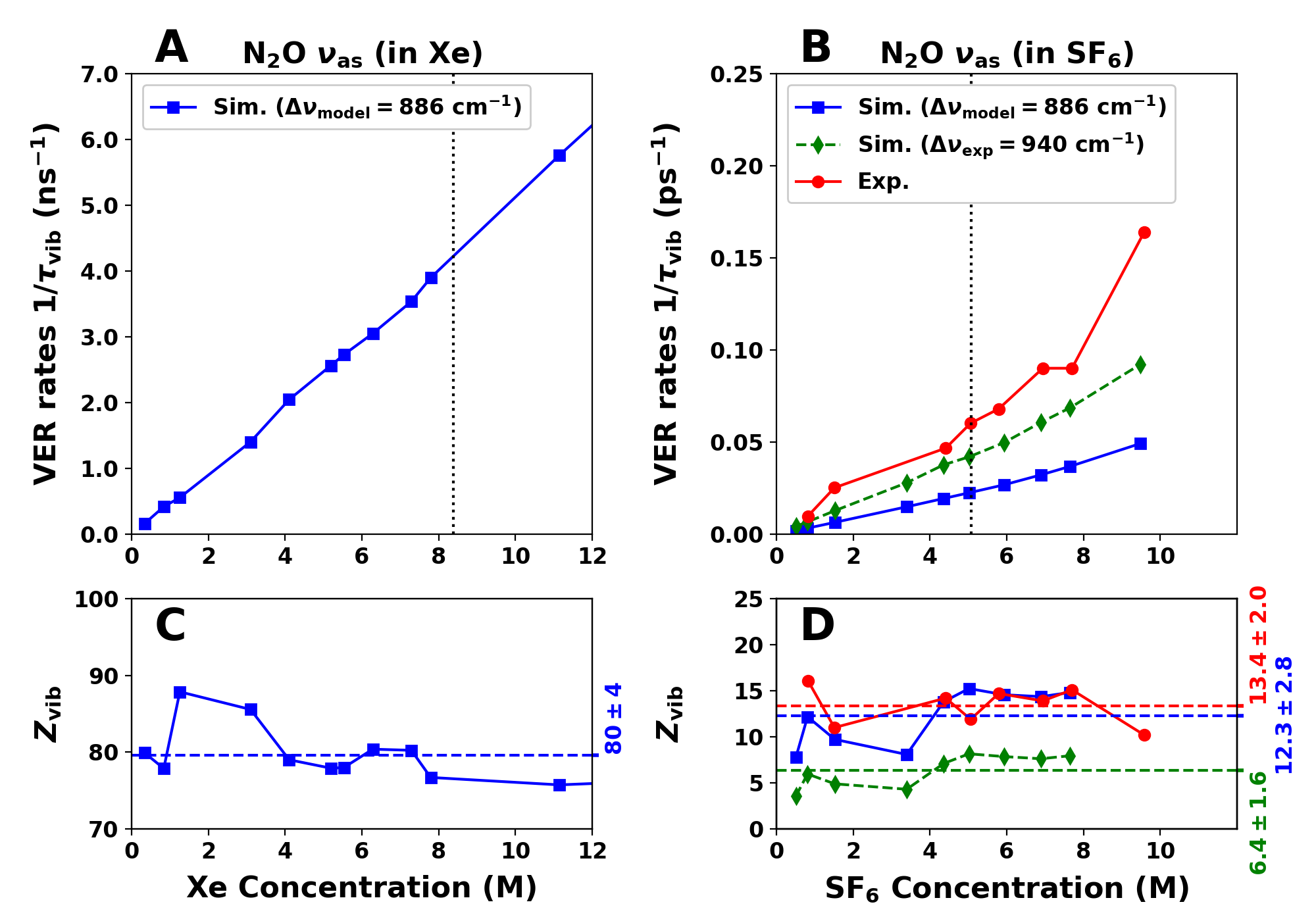}
\caption{Computed VER rates (blue squares) for $(\nu_\mathrm{s},
  \nu_\mathrm{as}) = (0,1) \rightarrow (1,0)$ with frequency
  difference $\Delta \nu = \nu_\mathrm{as} - \nu\mathrm{s}$ computed
  from the frequency dependent friction function of the asymmetric
  stretch vibration of N$_2$O in (A) xenon and (B) SF$_6$ solvent at
  different solvent densities.\cite{Hynes:1992, Hynes:1998} The
  frequency difference $\Delta \nu$ is given in the legend in
  brackets. For N$_2$O in SF$_6$ (B), experimental values are
  available from Ref. \citenum{ziegler:2022} and VER rates are shown
  in assumption that $\Delta \nu$ matches the experimentally observed
  transition frequency of N$_2$O in SF$_6$ (green diamonds). Note the
  different unit representation of the VER rates in panels A and
  B. The vertical dotted lines mark the experimentally observed solvent
  concentration at the respective critical density of xenon and
  SF$_6$, respectively.
  Panels C and D show the computed (and experimentally observed)
  number of collision $Z_\mathrm{vib}$ for the vibrational relaxation
  of N$_2$O with average and standard deviation displayed on the right
  axis.}
\label{fig:vibtimes_conc}
\end{center}
\end{figure}

\noindent
The VER lifetimes $\tau_{\rm vib}$ for the relaxation $(0,1)
\rightarrow (1,0)$ in N$_2$O were determined by the average of the
friction function weighted by the frequency difference distribution
$P(\nu_\mathrm{as} - \nu_\mathrm{s})$.  The resulting rates
$1/\tau_{\rm vib}$ from simulation and experiments\cite{ziegler:2022}
are shown in Figure \ref{fig:vibtimes_conc}.  The average frequency
difference in the simulation were determined to be
$\Delta\nu_\mathrm{model} = 886$\,cm$^{-1}$ but
$\Delta\nu_\mathrm{exp} = 940$\,cm$^{-1}$ in
experiments.\cite{ziegler:2022} For consistency, the same distribution
width centered around the experimentally observed frequency difference
$\Delta\nu_\mathrm{exp}$ was used to compute the $\tau_{\rm vib}$ (see
green dashed line in Figure \ref{fig:vibtimes_conc}B).
At the given conditions the quantum correction factor
increases the VER lifetimes $\tau_{\rm vib}$ by about 4\%
($Q(T_\mathrm{Xe}, \Delta\nu_\mathrm{model}) = 1.040$,
$Q(T_\mathrm{SF_6}, \Delta\nu_\mathrm{model}) = 1.034$ or
$Q(T_\mathrm{SF_6}, \Delta\nu_\mathrm{exp}) = 1.038$).\\

\noindent
For N$_2$O in xenon, the computed N$_2$O VER rates for
$\nu_\mathrm{as}$ in Figure \ref{fig:vibtimes_conc}A are on the
ns$^{-1}$ time scale for supercritical xenon and indicate a rather
slow VER relaxation process, as expected. 
The average solvent concentration-weighted VER rate is
$k_\mathrm{VER}=1/(\tau_{\rm vib} \cdot c[\mathrm{Xe}]) = 
(4.89 \pm 0.22) \cdot 10^{8}$\,s$^{-1}$M$^{-1}$.
Experimentally, $k_{\rm VER}^{\rm exp} \approx
5.1 \cdot 10^{7}$\,s$^{-1}$M$^{-1}$ ($\approx 8.5 \cdot
10^{-14}$\,cm$^{3}$s$^{-1}$), which corresponds to $\tau_\mathrm{vib} \approx 2.5$\,ns,
was indirectly obtained at one solvent concentration ($[c(\rm Xe)] = 7.8$M) from a small
$(\sim 5 \%)$ reduction of the pump-probe decay signal within 200 ps.\cite{ziegler:2022}
At this concentration ($[c(\rm Xe)] = 7.8$M), the computed rate $k_{\rm VER}^{\rm sim} = 5.00 \cdot 10^{8}$ s$^{-1}$M$^{-1}$ is $\sim 10$ times faster compared with the estimate
from the experiments. It should be noted that the Landau-Teller model
was successfully applied to rather long VER times ($\sim 600$ ps) for
the case of CO in myoglobin.\cite{anfinrud:1999} Hence it is likely
that the Landau-Teller approach for VER is valid but the limitations
for N$_2$O in xenon arise due to neglect of many-body interactions, in
particular at high solvent densities. \\

\noindent
On the
other hand, for N$_2$O in SF$_6$ (panel B), the computed VER rates
obtained with the model frequency difference distribution around
$\Delta\nu_\mathrm{model}$ (blue line) are on the ps$^{-1}$ time scale
for gaseous and supercritical SF$_6$. Assuming a frequency difference
distribution around the experimental value $\Delta\nu_\mathrm{exp}$
the VER rates are higher by a factor of two (green dashed line).  The
VER of $\nu_\mathrm{as}^{\rm N_2O}$ is faster in SF$_6$ than in xenon
due to the coupling with three-times degenerated the S-F stretch modes
at $\nu_3 = 948$\,cm$^{-1}$, which is well visible by a peak around
this wavenumber in the friction function in Figure
S17.\\

\noindent
The available experimental data of the VER rates for N$_2$O
$\nu_\mathrm{as}$ in SF$_6$ (red circles in Figure
\ref{fig:vibtimes_conc}B) are still about four or two times faster
than VER rates computed with the frequency difference distribution
from the model simulation or experimental observations,
respectively.
The computed VER rates, however, feature a linear raise
with increasing SF$_6$ concentration which corresponds to VER rates of
$k_\mathrm{VER}(\Delta\nu_\mathrm{model}) =
(4.46\pm0.36)\cdot10^{9}$\,s$^{-1}$M$^{-1}$ and
$k_\mathrm{VER}(\Delta\nu_\mathrm{exp}) =
(8.62\pm0.44)\cdot10^{9}$\,s$^{-1}$M$^{-1}$.
The experimentally
reported VER rate for N$_2$O $\nu_\mathrm{as}$ in gaseous SF$_6$ or
the IBC regime is $k_\mathrm{VER,IBC} =
1.56\cdot10^{10}$\,s$^{-1}$M$^{-1}$
($2.6\cdot10^{-11}$\,cm$^{3}$s$^{-1}$).\cite{ziegler:2022}\\

\noindent
Scaling the computed VER lifetimes $\tau_\mathrm{vib}$ with the
estimated collision rate $\tau_\mathrm{coll}$ reveals that for N$_2$O
in xenon an average of $Z_\mathrm{vib}=\tau_\mathrm{vib} \cdot
\tau_\mathrm{coll} = 80 \pm 4$ collisions between N$_2$O and xenon
atoms are required to relax the asymmetric stretch vibration in
N$_2$O.
Using the experimentally estimated $\tau_{\rm vib}$ would
yield a yet larger value for $Z_{\rm vib}$ by one order of
magnitude.
For N$_2$O in SF$_6$, $Z_\mathrm{vib}($SF$_6) = 12.3 \pm
2.8$ or $6.4 \pm 1.6$ is about one order of magnitude smaller
depending on whether $\Delta\nu_\mathrm{model}$ or
$\Delta\nu_\mathrm{exp}$ are used for estimating the VER lifetime,
respectively.
From experiments, $Z_\mathrm{vib}^{\rm expt.} = 13.4 \pm
2.0$ collisions between N$_2$O and SF$_6$ are required. 
Due to differences in the estimated collision rate,
$Z_\mathrm{vib}^{\rm expt.}$ is closer to the computed value 
from the VER rate estimation at $\Delta\nu_\mathrm{model}$
which deviate more strongly from the experimental
VER rates than the estimation at $\Delta\nu_\mathrm{exp}$.
The collision rate from the estimated mean free
collision time - which is larger than the one obtained from counting
collision events - together with the underestimated VER rate lead to
closer agreement of computed and experimentally estimated
$Z_\mathrm{vib}$ than comparison of the VER.\\

\noindent
If VER rates from simulations are computed using the
experimentally reported frequency difference $\Delta\nu_\mathrm{exp}$,
the VER rates are $\sim 0.8$ times as fast
compared with the experimentally
measured VER rates for N$_2$O in SF$_6$. The first-order Landau-Teller
model to compute the VER lifetimes used here only includes the term
describing solvent friction, i.e. the influence of the solute-solvent
interaction potential $V_\mathrm{int}$ on the normal mode
$\nu_\mathrm{as}^{\rm N_2O}$. Shorter lifetimes might be obtained by
including higher-order correction terms such as solvent-induced
coupling to the solute-solvent interaction potential expression in
Eq. \ref{eq:fproj}. Such terms include the second derivative of
$V_\mathrm{int}$ along the solute normal mode of $\nu_\mathrm{as}^{\rm
  N_2O}$ and $\nu_\mathrm{s}^{\rm N_2O}$ describing the relaxation
pathway of intramolecular VER assisted by the
solvent.\cite{kato:1998}\\

\section{Discussion and Conclusions}
The present work provides a atom-resolved picture of rotational and
vibrational energy relaxation of N$_2$O in xenon and SF$_6$ covering a
wide range of thermodynamic states ranging from the gas phase to the
liquid. In an effort to improve the solvent-solute interactions, the
electrostatic and other nonbonded interactions were fit to reference
electronic structure calculations at different levels of theory. While
for VER satisfactory results compared with experiment were found, RER
agrees with experiment only up to $c{\rm [solvent]} \sim 4$ M. For
higher concentrations the simulation follow the IBC model whereas
experimentally different behaviours are found for xenon and SF$_6$.\\

\noindent
For N$_2$O in xenon the experiments find a pronounced increase in the
RER time as the concentration increases from 6\,M to 8\,M, see Figure
S4. The experiments are expected to report on the
solvent rotational energy normalized autocorrelation function $\langle
E_{\rm rot}(0) E_{\rm rot}(t) \rangle$. For low-density solvent the
experimental result could be reproduced almost quantitatively, whereas
for increasing density no discontinuous change of the RER time was
found. Rather, the linear behaviour - consistent with the IBC model -
continued up to the highest simulated densities yield RER rates of
$(3.67\pm0.25)\cdot10^{10}$\,s$^{-1}$M$^{-1}$ and
$(1.25\pm0.12)\cdot10^{11}$\,s$^{-1}$M$^{-1}$ for N$_2$O in xenon and
SF$_6$, respectively. These computed RER rates agree well with those
from experiment within the IBC region of
$5.36\cdot10^{10}$\,s$^{-1}$M$^{-1}$ and
$1.02\cdot10^{11}$\,s$^{-1}$M$^{-1}$, respectively.\\

\noindent
In search for a reason for the disagreement, additional correlation
functions, including $\langle \vec{L}_{\rm N_2O}(0) \vec{L}_{\rm
  N_2O}(t) \rangle$, $\langle E_{\rm rot,N_2O}(0) E_{\rm rot,N_2O}(t)
\rangle$, and $\langle \mu_{\rm N_2O}(0) \mu_{\rm N_2O}(t) \rangle$
were analyzed. However, all of them confirm the behaviour of the
computed $\langle \vec{L}_{\rm N_2O}^2(0) \vec{L}_{\rm N_2O}^2(t)
\rangle$ correlation function. 
One possible explanation for the disagreement is the observation 
that with increasing density the solvent packing around the solute
becomes so dense, that isolated N$_2$O
the properties ($E_{\rm rot,N_2O}$ or $\vec{L}_{\rm N_2O}$) cannot be directly
compared with the experimentally observed signals. In other words,
analysis of the 
N$_2$O rotational motion without accounting for the coupling to the solvent shell(s) - as is usually done and was done here, too - becomes unrealistic. Rather, for sufficiently high solvent densities the
rotational motion of N$_2$O and the solvent molecules are coupled and
solvent molecules are ``slaved'' to the N$_2$O rotation.
Quantum-mechanically, the rotational quantum number correlating with
$\vec{L}_{\rm N_2O}$ is a ``good quantum number'' for gas phase and
low solvent densities whereas this property is lost for increasing
solvent density.\\

\noindent
Analysis of the trajectories with regards to changes in kinetic energy
(squared momentum transfer) between the collision partners (N$_2$O and
xenon; N$_2$O and SF$_6$) revealed a larger average change for the
rotational energy of N$_2$O when colliding with SF$_6$ than with
xenon. The translational kinetic energy change of the solvent
molecules xenon and SF$_6$ are comparable, whereas the additional
rotational and vibrational degrees of freedom in SF$_6$ provide
further energy transfer pathways compared with xenon.\\

\noindent
Considering RER, the average number of collisions to rotationally
relax N$_2$O over the concentration range considered are
$Z_\mathrm{rot}^{\rm Xe} = 1.1$ and $Z_\mathrm{rot}^{\rm SF6} = 0.6$,
respectively. This shows that RER for N$_2$O in SF$_6$ is about
$\sim1.7$ times more effective than in xenon, which is consistent with
experiments. There, the values of 2.4 and 1.7 collision are about 2 to
3 times larger for SF$_6$ compared with xenon, indicating a less
efficient rotational relaxation per collision in the simulations. The
largest deviations between experiments and the simulations are in the
collision rates which are twice as high when estimated from the
inverse of the mean free collision times. Together with the simulated
RER rates this yields $Z'_\mathrm{rot}$ of 2.5 and 1.4 for N$_2$O in
xenon and SF$_6$, respectively, which is much closer to the
experiments. The relative change $|\Delta \bar{T}_\mathrm{rot}^{\rm
  SF_6}| / |\Delta \bar{T}_\mathrm{rot}^{\rm Xe}|$ is somewhat smaller
but comparable to $|\Delta \bar{T}_\mathrm{vib}^{\rm SF_6}| / |\Delta
\bar{T}_\mathrm{vib}^{\rm Xe}|$ depending on solvent concentration
which indicates that angular anisotropy of the intermolecular
interactions influence RER. \\

\noindent
VER rates for the asymmetric N$_2$O stretch vibration $\nu_\mathrm{as}$
in xenon were computed following the Landau-Teller approach and yielded
$k_{\rm VER} = (4.89 \pm 0.22) \cdot 10^{8}$\,s$^{-1}$M$^{-1}$ which is
one order of magnitude faster compared with experiment. To put this
comparison into context it is useful to recall earlier experiments for
O$_2$ vibrational relaxation in argon but at considerably lower
temperatures.\cite{skinner:1998} The reported relaxation rates were
5 orders of magnitude smaller than experiments from entirely classical
simulations which changed to 2 orders smaller or 1 order larger than
experiments depending on the quantum correction factor employed.\cite{skinner:1998}
This work concluded that "Relaxation rates on long
time scales are extremely sensitive to details in the
potential, and errors in the potential model could
easily account for an order of magnitude discrepancy
between theory and experiment." In SF$_6$, the computed
rates are considerably faster (by one order of magnitude) with
$(4.46\pm0.36)\cdot10^{9}$\,s$^{-1}$M$^{-1}$ or
$(8.62\pm0.44)\cdot10^{9}$\,s$^{-1}$M$^{-1}$
depending on the applied
frequency difference distribution $P(\nu_\mathrm{as} -
\nu_\mathrm{s})$ for the relaxation process $(\nu_\mathrm{s},
\nu_\mathrm{as}) = (0,1) \rightarrow (1,0)$. These rates are about
$\sim 0.4$ to $\sim 0.8$ times 
the experimentally observed VER rate in
the IBC regime at low solvent densities of
$1.56\cdot10^{10}$\,s$^{-1}$M$^{-1}$, but also do the computed VER
rates of N$_2$O $\nu_\mathrm{as}$ in SF$_6$ around its critical density
not show the effect of critical slowing as in the experiments (see
Figure \ref{fig:vibtimes_conc}B). This deficiency of the simulation,
however, is consistent with the missing effect of critical slowing in
the computed RER rates for N$_2$O in SF$_6$.  \\

\noindent
In summary, this work presented a molecular-level study of rotational
and vibrational energy relaxation of N$_2$O in xenon and SF$_6$
ranging from highly dilute to liquid phases of the solvent, including
the supercritical regime. For regimes in which the IBC model is valid
almost quantitative agreement with experiment was found whereas near
the supercritical point simulations and experimental findings
disagree, in particular for RER. 
The most likely reason is that the
effect of solvent packing is neglected if only the
rotational motion of the solute (N$_2$O) is considered in the analysis - as is usually done - without accounting for solvent-solute coupling. The average
number of collisions for rotational and vibrational relaxation from
the simulations agree rather well with those obtained from
experiments. In conclusion, VER of N$_2$O across different solvent
density regimes and RER in the low-density region can be characterized
well at a molecular level whereas for the higher-density regimes,
analysis of the RER requires additional refinements.

\section*{Acknowledgment}
This work has been financially supported by the Swiss National Science
Foundation (NCCR-MUST, grants 200021-117810, 200020-188724), the
AFOSR, the University of Basel, and by the European Union's Horizon 2020 
research and innovation program under the Marie Sk{\l}odowska-Curie grant
agreement No 801459 -FP-RESOMUS. The support of the National
Science Foundation, Grant No. CHE-2102427 (L.D.Z.), and the Boston
University Photonics Center is gratefully acknowledged.\\

\section*{Supporting Information}
The supporting material includes system setups (Table
S1), force field parameters (Table S2),
various correlation functions (Figures S1, 
S2, S6-S9)
with further evaluation results  (Figures
S3-S5,
S10-S15)
and the frequency-dependent friction functions (Figures
S16, S17).\\

\section*{Data Availability}
Relevant data for the present study are available at
\url{https://github.com/MMunibas/SCF_N2O/}.

\bibliography{refs}

\providecommand{\latin}[1]{#1}
\makeatletter
\providecommand{\doi}
  {\begingroup\let\do\@makeother\dospecials
  \catcode`\{=1 \catcode`\}=2 \doi@aux}
\providecommand{\doi@aux}[1]{\endgroup\texttt{#1}}
\makeatother
\providecommand*\mcitethebibliography{\thebibliography}
\csname @ifundefined\endcsname{endmcitethebibliography}
  {\let\endmcitethebibliography\endthebibliography}{}
\begin{mcitethebibliography}{62}
\providecommand*\natexlab[1]{#1}
\providecommand*\mciteSetBstSublistMode[1]{}
\providecommand*\mciteSetBstMaxWidthForm[2]{}
\providecommand*\mciteBstWouldAddEndPuncttrue
  {\def\EndOfBibitem{\unskip.}}
\providecommand*\mciteBstWouldAddEndPunctfalse
  {\let\EndOfBibitem\relax}
\providecommand*\mciteSetBstMidEndSepPunct[3]{}
\providecommand*\mciteSetBstSublistLabelBeginEnd[3]{}
\providecommand*\EndOfBibitem{}
\mciteSetBstSublistMode{f}
\mciteSetBstMaxWidthForm{subitem}{(\alph{mcitesubitemcount})}
\mciteSetBstSublistLabelBeginEnd
  {\mcitemaxwidthsubitemform\space}
  {\relax}
  {\relax}

\bibitem[Foster and Miller()Foster, and Miller]{fostermiller:2010}
Foster,~J.; Miller,~R.~S. \emph{High Pressure Processes in Chemical
  Engineering}; pp 53--75\relax
\mciteBstWouldAddEndPuncttrue
\mciteSetBstMidEndSepPunct{\mcitedefaultmidpunct}
{\mcitedefaultendpunct}{\mcitedefaultseppunct}\relax
\EndOfBibitem
\bibitem[Guardiola \latin{et~al.}(2017)Guardiola, Olmeda, Pla, and
  Bares]{bares:2017}
Guardiola,~C.; Olmeda,~P.; Pla,~B.; Bares,~P. In-cylinder pressure based model
  for exhaust temperature estimation in internal combustion engines.
  \emph{Appl. Therm. Eng.} \textbf{2017}, \emph{115}, 212--220\relax
\mciteBstWouldAddEndPuncttrue
\mciteSetBstMidEndSepPunct{\mcitedefaultmidpunct}
{\mcitedefaultendpunct}{\mcitedefaultseppunct}\relax
\EndOfBibitem
\bibitem[Farrell and Peters(1986)Farrell, and Peters]{peters:1986}
Farrell,~P.~V.; Peters,~B.~D. Droplet vaporization in supercritical pressure
  environments. \emph{Acta Astronaut.} \textbf{1986}, \emph{13}, 673--680,
  Congress of the International Academy of Astronautics\relax
\mciteBstWouldAddEndPuncttrue
\mciteSetBstMidEndSepPunct{\mcitedefaultmidpunct}
{\mcitedefaultendpunct}{\mcitedefaultseppunct}\relax
\EndOfBibitem
\bibitem[Kajimoto(1999)]{kajimoto:1999}
Kajimoto,~O. Solvation in Supercritical Fluids: Its Effects on Energy Transfer
  and Chemical Reactions. \emph{Chem. Rev.} \textbf{1999}, \emph{99},
  355--*390\relax
\mciteBstWouldAddEndPuncttrue
\mciteSetBstMidEndSepPunct{\mcitedefaultmidpunct}
{\mcitedefaultendpunct}{\mcitedefaultseppunct}\relax
\EndOfBibitem
\bibitem[Knez \latin{et~al.}(2014)Knez, Marko\v{c}i\v{c}, Leitgeb,
  Primo\v{z}i\v{c}, {Knez Hrn\v{c}i\v{c}}, and {\v{S}}kerget]{skerget:2014}
Knez,~{\v{Z}}.; Marko\v{c}i\v{c},~E.; Leitgeb,~M.; Primo\v{z}i\v{c},~M.; {Knez
  Hrn\v{c}i\v{c}},~M.; {\v{S}}kerget,~M. Industrial applications of
  supercritical fluids: A review. \emph{Energy} \textbf{2014}, \emph{77},
  235--243\relax
\mciteBstWouldAddEndPuncttrue
\mciteSetBstMidEndSepPunct{\mcitedefaultmidpunct}
{\mcitedefaultendpunct}{\mcitedefaultseppunct}\relax
\EndOfBibitem
\bibitem[Brunner(2010)]{brunner:2010}
Brunner,~G. Applications of Supercritical Fluids. \emph{Annu. Rev. Biomed.
  Eng.} \textbf{2010}, \emph{1}, 321--342\relax
\mciteBstWouldAddEndPuncttrue
\mciteSetBstMidEndSepPunct{\mcitedefaultmidpunct}
{\mcitedefaultendpunct}{\mcitedefaultseppunct}\relax
\EndOfBibitem
\bibitem[Hrn\v{c}i\v{c} \latin{et~al.}(2018)Hrn\v{c}i\v{c}, C\"{o}r, Verboten,
  and Knez]{knez:2018}
Hrn\v{c}i\v{c},~M.~K.; C\"{o}r,~D.; Verboten,~M.~T.; Knez,~{\v{Z}}. Application
  of supercritical and subcritical fluids in food processing. \emph{Food Qual.
  Saf.} \textbf{2018}, \emph{2}, 59--67\relax
\mciteBstWouldAddEndPuncttrue
\mciteSetBstMidEndSepPunct{\mcitedefaultmidpunct}
{\mcitedefaultendpunct}{\mcitedefaultseppunct}\relax
\EndOfBibitem
\bibitem[Perrut(2000)]{perrut:2000}
Perrut,~M. Supercritical Fluid Applications: Industrial Developments and
  Economic Issues. \emph{Industrial \& Engineering Chemistry Research}
  \textbf{2000}, \emph{39}, 4531--4535\relax
\mciteBstWouldAddEndPuncttrue
\mciteSetBstMidEndSepPunct{\mcitedefaultmidpunct}
{\mcitedefaultendpunct}{\mcitedefaultseppunct}\relax
\EndOfBibitem
\bibitem[Deshpande \latin{et~al.}(2011)Deshpande, Kumar, Kumar, Shavi, Karthik,
  Reddy, and Udupa]{udupa:2011}
Deshpande,~P.~B.; Kumar,~G.~A.; Kumar,~A.~R.; Shavi,~G.~V.; Karthik,~A.;
  Reddy,~M.~S.; Udupa,~N. Supercritical Fluid Technology: Concepts and
  Pharmaceutical Applications. \emph{PDA Journal of Pharmaceutical Science and
  Technology} \textbf{2011}, \emph{65}, 333--344\relax
\mciteBstWouldAddEndPuncttrue
\mciteSetBstMidEndSepPunct{\mcitedefaultmidpunct}
{\mcitedefaultendpunct}{\mcitedefaultseppunct}\relax
\EndOfBibitem
\bibitem[Machida \latin{et~al.}(2011)Machida, Takesue, and Smith]{smith:2011}
Machida,~H.; Takesue,~M.; Smith,~R.~L. Green chemical processes with
  supercritical fluids: Properties, materials, separations and energy.
  \emph{The Journal of Supercritical Fluids} \textbf{2011}, \emph{60},
  2--15\relax
\mciteBstWouldAddEndPuncttrue
\mciteSetBstMidEndSepPunct{\mcitedefaultmidpunct}
{\mcitedefaultendpunct}{\mcitedefaultseppunct}\relax
\EndOfBibitem
\bibitem[Tucker(1999)]{tucker:1999}
Tucker,~S.~C. Solvent Density Inhomogeneities in Supercritical Fluids.
  \emph{Chem. Rev.} \textbf{1999}, \emph{99}, 391--418\relax
\mciteBstWouldAddEndPuncttrue
\mciteSetBstMidEndSepPunct{\mcitedefaultmidpunct}
{\mcitedefaultendpunct}{\mcitedefaultseppunct}\relax
\EndOfBibitem
\bibitem[Saitow \latin{et~al.}(2003)Saitow, Otake, Nakayama, Ishii, and
  Nishikawa]{nishikawa:2003}
Saitow,~K.; Otake,~K.; Nakayama,~H.; Ishii,~K.; Nishikawa,~K. Local density
  enhancement in neat supercritical fluid due to attractive intermolecular
  interactions. \emph{Chem. Phys. Lett.} \textbf{2003}, \emph{368},
  209--214\relax
\mciteBstWouldAddEndPuncttrue
\mciteSetBstMidEndSepPunct{\mcitedefaultmidpunct}
{\mcitedefaultendpunct}{\mcitedefaultseppunct}\relax
\EndOfBibitem
\bibitem[Nishikawa and Morita(2000)Nishikawa, and Morita]{nishikawa:2000}
Nishikawa,~K.; Morita,~T. Inhomogeneity of molecular distribution in
  supercritical fluids. \emph{Chem. Phys. Lett.} \textbf{2000}, \emph{316},
  238--242\relax
\mciteBstWouldAddEndPuncttrue
\mciteSetBstMidEndSepPunct{\mcitedefaultmidpunct}
{\mcitedefaultendpunct}{\mcitedefaultseppunct}\relax
\EndOfBibitem
\bibitem[Saitow \latin{et~al.}(2002)Saitow, Ochiai, Kato, and
  Nishikawa]{nishikawa:2002}
Saitow,~K.-i.; Ochiai,~H.; Kato,~T.; Nishikawa,~K. Correlation time of density
  fluctuation for supercritical ethylene studied by dynamic light scattering.
  \emph{J. Chem. Phys.} \textbf{2002}, \emph{116}, 4985--4992\relax
\mciteBstWouldAddEndPuncttrue
\mciteSetBstMidEndSepPunct{\mcitedefaultmidpunct}
{\mcitedefaultendpunct}{\mcitedefaultseppunct}\relax
\EndOfBibitem
\bibitem[Saitow \latin{et~al.}(2005)Saitow, Kajiya, and
  Nishikawa]{nishikawa:2005}
Saitow,~K.-i.; Kajiya,~D.; Nishikawa,~K. Time Evolution of Density Fluctuation
  in Supercritical Region. I. Non-hydrogen-bonded Fluids Studied by Dynamic
  Light Scattering. \emph{J. Phys. Chem. A} \textbf{2005}, \emph{109},
  83--91\relax
\mciteBstWouldAddEndPuncttrue
\mciteSetBstMidEndSepPunct{\mcitedefaultmidpunct}
{\mcitedefaultendpunct}{\mcitedefaultseppunct}\relax
\EndOfBibitem
\bibitem[Goodyear and Tucker(1999)Goodyear, and Tucker]{tucker:1999a}
Goodyear,~G.; Tucker,~S.~C. Glass-like behavior in supercritical fluids: The
  effect of critical slowing down on solute dynamics. \emph{J. Chem. Phys.}
  \textbf{1999}, \emph{111}, 9673--9677\relax
\mciteBstWouldAddEndPuncttrue
\mciteSetBstMidEndSepPunct{\mcitedefaultmidpunct}
{\mcitedefaultendpunct}{\mcitedefaultseppunct}\relax
\EndOfBibitem
\bibitem[Goodyear \latin{et~al.}(2000)Goodyear, Maddox, and
  Tucker]{tucker:2000c}
Goodyear,~G.; Maddox,~M.~W.; Tucker,~S.~C. The correlation between local and
  long-range structure in compressible supercritical fluids. \emph{J. Chem.
  Phys.} \textbf{2000}, \emph{112}, 10327--10339\relax
\mciteBstWouldAddEndPuncttrue
\mciteSetBstMidEndSepPunct{\mcitedefaultmidpunct}
{\mcitedefaultendpunct}{\mcitedefaultseppunct}\relax
\EndOfBibitem
\bibitem[Mandal \latin{et~al.}(2018)Mandal, Ng~Pack, Shah, Erramilli, and
  Ziegler]{ziegler:2018}
Mandal,~A.; Ng~Pack,~G.; Shah,~P.~P.; Erramilli,~S.; Ziegler,~L.~D. Ultrafast
  Two-Dimensional Infrared Spectroscopy of a Quasifree Rotor: $J$ Scrambling
  and Perfectly Anticorrelated Cross Peaks. \emph{Phys. Rev. Lett.}
  \textbf{2018}, \emph{120}, 103401\relax
\mciteBstWouldAddEndPuncttrue
\mciteSetBstMidEndSepPunct{\mcitedefaultmidpunct}
{\mcitedefaultendpunct}{\mcitedefaultseppunct}\relax
\EndOfBibitem
\bibitem[Ng~Pack \latin{et~al.}(2019)Ng~Pack, Rotondaro, Shah, Mandal,
  Erramilli, and Ziegler]{ziegler:2019}
Ng~Pack,~G.; Rotondaro,~M.~C.; Shah,~P.~P.; Mandal,~A.; Erramilli,~S.;
  Ziegler,~L.~D. Two-dimensional infrared spectroscopy from the gas to liquid
  phase: density dependent $J-$scrambling{,} vibrational relaxation{,} and the
  onset of liquid character. \emph{Phys. Chem. Chem. Phys.} \textbf{2019},
  \emph{21}, 21249--21261\relax
\mciteBstWouldAddEndPuncttrue
\mciteSetBstMidEndSepPunct{\mcitedefaultmidpunct}
{\mcitedefaultendpunct}{\mcitedefaultseppunct}\relax
\EndOfBibitem
\bibitem[Rotondaro \latin{et~al.}(2022)Rotondaro, Jain, Erramilli, and
  Ziegler]{ziegler:2022}
Rotondaro,~M.~C.; Jain,~A.; Erramilli,~S.; Ziegler,~L.~D. Ultrafast 2DIR
  comparison of rotational energy transfer, isolated binary collision
  breakdown, and near critical fluctuations in Xe and SF$_6$ solutions.
  \emph{J. Chem. Phys.} \textbf{2022}, \emph{157}, 174305\relax
\mciteBstWouldAddEndPuncttrue
\mciteSetBstMidEndSepPunct{\mcitedefaultmidpunct}
{\mcitedefaultendpunct}{\mcitedefaultseppunct}\relax
\EndOfBibitem
\bibitem[Töpfer \latin{et~al.}(2023)Töpfer, Erramilli, Ziegler, and
  Meuwly]{MM.n2ovib:2023}
Töpfer,~K.; Erramilli,~D.~K.; Ziegler,~L.~D.; Meuwly,~M. Molecular-level
  understanding of the rovibrational spectra of N$_2$O in gaseous,
  supercritical, and liquid SF$_6$ and Xe. \emph{J. Chem. Phys.} \textbf{2023},
  \emph{158}, 144302\relax
\mciteBstWouldAddEndPuncttrue
\mciteSetBstMidEndSepPunct{\mcitedefaultmidpunct}
{\mcitedefaultendpunct}{\mcitedefaultseppunct}\relax
\EndOfBibitem
\bibitem[Unke and Meuwly(2017)Unke, and Meuwly]{MM.rkhs:2017}
Unke,~O.~T.; Meuwly,~M. Toolkit for the Construction of Reproducing
  Kernel-based Representations of Data: Application to Multidimensional
  Potential Energy Surfaces. \emph{J. Chem. Inf. and Mod.} \textbf{2017},
  \emph{57}, 1923--1931\relax
\mciteBstWouldAddEndPuncttrue
\mciteSetBstMidEndSepPunct{\mcitedefaultmidpunct}
{\mcitedefaultendpunct}{\mcitedefaultseppunct}\relax
\EndOfBibitem
\bibitem[Koner \latin{et~al.}(2020)Koner, San Vicente~Veliz, Bemish, and
  Meuwly]{koner:2020}
Koner,~D.; San Vicente~Veliz,~J.~C.; Bemish,~R.~J.; Meuwly,~M. Accurate
  reproducing kernel-based potential energy surfaces for the triplet ground
  states of N$_2$O and dynamics for the N + NO $\leftrightarrow$ O + N$_2$ and
  N$_2$ + O $\rightarrow$ 2N + O reactions. \emph{Phys. Chem. Chem. Phys.}
  \textbf{2020}, \emph{22}, 18488--18498\relax
\mciteBstWouldAddEndPuncttrue
\mciteSetBstMidEndSepPunct{\mcitedefaultmidpunct}
{\mcitedefaultendpunct}{\mcitedefaultseppunct}\relax
\EndOfBibitem
\bibitem[Werner \latin{et~al.}(2020)Werner, Knowles, Manby, Black, Doll,
  Heßelmann, Kats, Köhn, Korona, Kreplin, Ma, Miller, Mitrushchenkov,
  Peterson, Polyak, Rauhut, and Sibaev]{werner:2020}
Werner,~H.-J.; Knowles,~P.~J.; Manby,~F.~R.; Black,~J.~A.; Doll,~K.;
  Heßelmann,~A.; Kats,~D.; Köhn,~A.; Korona,~T.; Kreplin,~D.~A.
  \latin{et~al.}  The Molpro quantum chemistry package. \emph{J. Chem. Phys.}
  \textbf{2020}, \emph{152}, 144107\relax
\mciteBstWouldAddEndPuncttrue
\mciteSetBstMidEndSepPunct{\mcitedefaultmidpunct}
{\mcitedefaultendpunct}{\mcitedefaultseppunct}\relax
\EndOfBibitem
\bibitem[Dellis and Samios(2010)Dellis, and Samios]{samios:2010}
Dellis,~D.; Samios,~J. Molecular force field investigation for Sulfur
  Hexafluoride: A computer simulation study. \emph{Fluid Phase Equilib.}
  \textbf{2010}, \emph{291}, 81--89\relax
\mciteBstWouldAddEndPuncttrue
\mciteSetBstMidEndSepPunct{\mcitedefaultmidpunct}
{\mcitedefaultendpunct}{\mcitedefaultseppunct}\relax
\EndOfBibitem
\bibitem[Stephenson \latin{et~al.}(1987)Stephenson, Malanowski, and
  Ambrose]{ambrose:1987}
Stephenson,~R.~M.; Malanowski,~S.; Ambrose,~D. \emph{Handbook of the
  thermodynamics of organic compounds; section on vapor-liquid critical
  constants of fluids}; Elsevier, 1987\relax
\mciteBstWouldAddEndPuncttrue
\mciteSetBstMidEndSepPunct{\mcitedefaultmidpunct}
{\mcitedefaultendpunct}{\mcitedefaultseppunct}\relax
\EndOfBibitem
\bibitem[Haynes(2014)]{haynes:2014crc}
Haynes,~W. \emph{CRC Handbook of Chemistry and Physics}; CRC Handbook of
  Chemistry and Physics; CRC Press, 2014; p. 6-85\relax
\mciteBstWouldAddEndPuncttrue
\mciteSetBstMidEndSepPunct{\mcitedefaultmidpunct}
{\mcitedefaultendpunct}{\mcitedefaultseppunct}\relax
\EndOfBibitem
\bibitem[Michels \latin{et~al.}(1954)Michels, Wassenaar, and
  Louwerse]{louwerse:1954xenon}
Michels,~A.; Wassenaar,~T.; Louwerse,~P. Isotherms of xenon at temperatures
  between 0°C and 150°C and at densities up to 515 Amagats (pressures up to
  2800 atmospheres). \emph{Physica} \textbf{1954}, \emph{20}, 99--106\relax
\mciteBstWouldAddEndPuncttrue
\mciteSetBstMidEndSepPunct{\mcitedefaultmidpunct}
{\mcitedefaultendpunct}{\mcitedefaultseppunct}\relax
\EndOfBibitem
\bibitem[Nelder and Mead(1965)Nelder, and Mead]{neldermead:1965}
Nelder,~J.~A.; Mead,~R. A Simplex Method for Function Minimization.
  \emph{Comput. J.} \textbf{1965}, \emph{7}, 308--313\relax
\mciteBstWouldAddEndPuncttrue
\mciteSetBstMidEndSepPunct{\mcitedefaultmidpunct}
{\mcitedefaultendpunct}{\mcitedefaultseppunct}\relax
\EndOfBibitem
\bibitem[Devereux \latin{et~al.}(2014)Devereux, Raghunathan, Fedorov, and
  Meuwly]{MM.dcm:2014}
Devereux,~M.; Raghunathan,~S.; Fedorov,~D.~G.; Meuwly,~M. A Novel,
  computationally efficient multipolar model employing distributed charges for
  molecular dynamics simulations. \emph{J. Chem. Theo. Comp.} \textbf{2014},
  \emph{10}, 4229--4241\relax
\mciteBstWouldAddEndPuncttrue
\mciteSetBstMidEndSepPunct{\mcitedefaultmidpunct}
{\mcitedefaultendpunct}{\mcitedefaultseppunct}\relax
\EndOfBibitem
\bibitem[Unke \latin{et~al.}(2017)Unke, Devereux, and Meuwly]{MM.mdcm:2017}
Unke,~O.~T.; Devereux,~M.; Meuwly,~M. Minimal distributed charges: Multipolar
  quality at the cost of point charge elect rostatics. \emph{J. Chem. Phys.}
  \textbf{2017}, \emph{147}, 161712\relax
\mciteBstWouldAddEndPuncttrue
\mciteSetBstMidEndSepPunct{\mcitedefaultmidpunct}
{\mcitedefaultendpunct}{\mcitedefaultseppunct}\relax
\EndOfBibitem
\bibitem[Devereux \latin{et~al.}(2020)Devereux, Pezzella, Raghunathan, and
  Meuwly]{MM.mdcm:2020}
Devereux,~M.; Pezzella,~M.; Raghunathan,~S.; Meuwly,~M. Polarizable Multipolar
  Molecular Dynamics Using Distributed Point Charges. \emph{J. Chem. Theo.
  Comp.} \textbf{2020}, \emph{16}, 7267--7280\relax
\mciteBstWouldAddEndPuncttrue
\mciteSetBstMidEndSepPunct{\mcitedefaultmidpunct}
{\mcitedefaultendpunct}{\mcitedefaultseppunct}\relax
\EndOfBibitem
\bibitem[Frisch \latin{et~al.}(2016)Frisch, Trucks, Schlegel, Scuseria, Robb,
  Cheeseman, Scalmani, Barone, Petersson, Nakatsuji, Li, Caricato, Marenich,
  Bloino, Janesko, Gomperts, Mennucci, Hratchian, Ortiz, Izmaylov, Sonnenberg,
  Williams-Young, Ding, Lipparini, Egidi, Goings, Peng, Petrone, Henderson,
  Ranasinghe, Zakrzewski, Gao, Rega, Zheng, Liang, Hada, Ehara, Toyota, Fukuda,
  Hasegawa, Ishida, Nakajima, Honda, Kitao, Nakai, Vreven, Throssell,
  Montgomery, Peralta, Ogliaro, Bearpark, Heyd, Brothers, Kudin, Staroverov,
  Keith, Kobayashi, Normand, Raghavachari, Rendell, Burant, Iyengar, Tomasi,
  Cossi, Millam, Klene, Adamo, Cammi, Ochterski, Martin, Morokuma, Farkas,
  Foresman, and Fox]{gaussian16}
Frisch,~M.~J.; Trucks,~G.~W.; Schlegel,~H.~B.; Scuseria,~G.~E.; Robb,~M.~A.;
  Cheeseman,~J.~R.; Scalmani,~G.; Barone,~V.; Petersson,~G.~A.; Nakatsuji,~H.
  \latin{et~al.}  Gaussian˜16 {R}evision {C}.01. 2016; Gaussian Inc.
  Wallingford CT\relax
\mciteBstWouldAddEndPuncttrue
\mciteSetBstMidEndSepPunct{\mcitedefaultmidpunct}
{\mcitedefaultendpunct}{\mcitedefaultseppunct}\relax
\EndOfBibitem
\bibitem[Olney \latin{et~al.}(1997)Olney, Cann, Cooper, and Brion]{olney:1997}
Olney,~T.~N.; Cann,~N.; Cooper,~G.; Brion,~C. Absolute scale determination for
  photoabsorption spectra and the calculation of molecular properties using
  dipole sum-rules. \emph{Chem. Phys.} \textbf{1997}, \emph{223}, 59--98\relax
\mciteBstWouldAddEndPuncttrue
\mciteSetBstMidEndSepPunct{\mcitedefaultmidpunct}
{\mcitedefaultendpunct}{\mcitedefaultseppunct}\relax
\EndOfBibitem
\bibitem[Gussoni \latin{et~al.}(1998)Gussoni, Rui, and Zerbi]{gussoni:1998}
Gussoni,~M.; Rui,~M.; Zerbi,~G. Electronic and relaxation contribution to
  linear molecular polarizability. An analysis of the experimental values.
  \emph{J. Mol. Struct.} \textbf{1998}, \emph{447}, 163--215\relax
\mciteBstWouldAddEndPuncttrue
\mciteSetBstMidEndSepPunct{\mcitedefaultmidpunct}
{\mcitedefaultendpunct}{\mcitedefaultseppunct}\relax
\EndOfBibitem
\bibitem[Branch \latin{et~al.}(1999)Branch, Coleman, and Li]{trf:1999}
Branch,~M.~A.; Coleman,~T.~F.; Li,~Y. A Subspace, Interior, and Conjugate
  Gradient Method for Large-Scale Bound-Constrained Minimization Problems.
  \emph{SIAM J Sci Comput} \textbf{1999}, \emph{21}, 1--23\relax
\mciteBstWouldAddEndPuncttrue
\mciteSetBstMidEndSepPunct{\mcitedefaultmidpunct}
{\mcitedefaultendpunct}{\mcitedefaultseppunct}\relax
\EndOfBibitem
\bibitem[Boys and Bernardi(1970)Boys, and Bernardi]{cpc}
Boys,~S.; Bernardi,~F. The calculation of small molecular interactions by the
  differences of separate total energies. Some procedures with reduced errors.
  \emph{Molecular Physics} \textbf{1970}, \emph{19}, 553--566\relax
\mciteBstWouldAddEndPuncttrue
\mciteSetBstMidEndSepPunct{\mcitedefaultmidpunct}
{\mcitedefaultendpunct}{\mcitedefaultseppunct}\relax
\EndOfBibitem
\bibitem[Grimme \latin{et~al.}(2011)Grimme, Antony, Ehrlich, and
  Krieg]{grimmeD3:2011}
Grimme,~S.; Antony,~J.; Ehrlich,~S.; Krieg,~H. A consistent and accurate ab
  initio parametrization of density functional dispersion correction (DFT-D)
  for the 94 elements H-Pu. \emph{J. Chem. Phys.} \textbf{2011}, \emph{132},
  154104\relax
\mciteBstWouldAddEndPuncttrue
\mciteSetBstMidEndSepPunct{\mcitedefaultmidpunct}
{\mcitedefaultendpunct}{\mcitedefaultseppunct}\relax
\EndOfBibitem
\bibitem[Brooks \latin{et~al.}(2009)Brooks, Brooks~III, MacKerell~Jr., Nilsson,
  Petrella, Roux, Won, Archontis, Bartels, Boresch, Caflisch, Caves, Cui,
  Dinner, Feig, Fischer, Gao, Hodoscek, Im, Kuczera, Lazaridis, Ma,
  Ovchinnikov, Paci, Pastor, Post, Schaefer, Tidor, Venable, Woodcock, Wu,
  Yang, York, and Karplus]{Charmm-Brooks-2009}
Brooks,~B.~R.; Brooks~III,~C.~L.; MacKerell~Jr.,~A.~D.; Nilsson,~L.;
  Petrella,~R.~J.; Roux,~B.; Won,~Y.; Archontis,~G.; Bartels,~C.; Boresch,~S.
  \latin{et~al.}  CHARMM: The Biomolecular Simulation Program. \emph{J. Comp.
  Chem.} \textbf{2009}, \emph{30}, 1545--1614\relax
\mciteBstWouldAddEndPuncttrue
\mciteSetBstMidEndSepPunct{\mcitedefaultmidpunct}
{\mcitedefaultendpunct}{\mcitedefaultseppunct}\relax
\EndOfBibitem
\bibitem[Mart\'{i}nez \latin{et~al.}(2009)Mart\'{i}nez, Andrade, Birgin, and
  Mart\'{i}nez]{martinez:2009}
Mart\'{i}nez,~L.; Andrade,~R.; Birgin,~E.~G.; Mart\'{i}nez,~J.~M. Packmol: A
  Package for Building Initial Configurations for Molecular Dynamics
  Simulations. \emph{J. Chem. Theo. Comp.} \textbf{2009}, \emph{30},
  2157--2164\relax
\mciteBstWouldAddEndPuncttrue
\mciteSetBstMidEndSepPunct{\mcitedefaultmidpunct}
{\mcitedefaultendpunct}{\mcitedefaultseppunct}\relax
\EndOfBibitem
\bibitem[Whitnell \latin{et~al.}(1992)Whitnell, Wilson, and Hynes]{Hynes:1992}
Whitnell,~R.~M.; Wilson,~K.~R.; Hynes,~J.~T. Vibrational relaxation of a
  dipolar molecule in water. \emph{J. Chem. Phys.} \textbf{1992}, \emph{96},
  5354--5369\relax
\mciteBstWouldAddEndPuncttrue
\mciteSetBstMidEndSepPunct{\mcitedefaultmidpunct}
{\mcitedefaultendpunct}{\mcitedefaultseppunct}\relax
\EndOfBibitem
\bibitem[Rey and Hynes(1998)Rey, and Hynes]{Hynes:1998}
Rey,~R.; Hynes,~J.~T. Vibrational phase and energy relaxation of CN$^-$ in
  water. \emph{J. Chem. Phys.} \textbf{1998}, \emph{108}, 142--153\relax
\mciteBstWouldAddEndPuncttrue
\mciteSetBstMidEndSepPunct{\mcitedefaultmidpunct}
{\mcitedefaultendpunct}{\mcitedefaultseppunct}\relax
\EndOfBibitem
\bibitem[Morita and Kato(1998)Morita, and Kato]{kato:1998}
Morita,~A.; Kato,~S. Vibrational relaxation of azide ion in water: The role of
  intramolecular charge fluctuation and solvent-induced vibrational coupling.
  \emph{J. Chem. Phys.} \textbf{1998}, \emph{109}, 5511--5523\relax
\mciteBstWouldAddEndPuncttrue
\mciteSetBstMidEndSepPunct{\mcitedefaultmidpunct}
{\mcitedefaultendpunct}{\mcitedefaultseppunct}\relax
\EndOfBibitem
\bibitem[Sagnella \latin{et~al.}(1999)Sagnella, Straub, Jackson, Lim, and
  Anfinrud]{anfinrud:1999}
Sagnella,~D.~E.; Straub,~J.~E.; Jackson,~T.~A.; Lim,~M.; Anfinrud,~P.~A.
  Vibrational population relaxation of carbon monoxide in the heme pocket of
  photolyzed carbonmonoxy myoglobin: Comparison of time-resolved mid-IR
  absorbance experiments and molecular dynamics simulations. \emph{Proc. Natl.
  Acad. Sci.} \textbf{1999}, \emph{96}, 14324--14329\relax
\mciteBstWouldAddEndPuncttrue
\mciteSetBstMidEndSepPunct{\mcitedefaultmidpunct}
{\mcitedefaultendpunct}{\mcitedefaultseppunct}\relax
\EndOfBibitem
\bibitem[Egorov and Skinner(1996)Egorov, and Skinner]{skinner:1996}
Egorov,~S.~A.; Skinner,~J.~L. {A theory of vibrational energy relaxation in
  liquids}. \emph{J. Chem. Phys.} \textbf{1996}, \emph{105}, 7047--7058\relax
\mciteBstWouldAddEndPuncttrue
\mciteSetBstMidEndSepPunct{\mcitedefaultmidpunct}
{\mcitedefaultendpunct}{\mcitedefaultseppunct}\relax
\EndOfBibitem
\bibitem[Everitt \latin{et~al.}(1998)Everitt, Egorov, and
  Skinner]{skinner:1998}
Everitt,~K.; Egorov,~S.; Skinner,~J. Vibrational energy relaxation in liquid
  oxygen. \emph{Chem. Phys.} \textbf{1998}, \emph{235}, 115--122\relax
\mciteBstWouldAddEndPuncttrue
\mciteSetBstMidEndSepPunct{\mcitedefaultmidpunct}
{\mcitedefaultendpunct}{\mcitedefaultseppunct}\relax
\EndOfBibitem
\bibitem[Everitt and Skinner(1999)Everitt, and Skinner]{skinner:1999}
Everitt,~K.~F.; Skinner,~J.~L. Vibrational energy relaxation of oxygen in
  liquid mixtures with argon. \emph{J. Chem. Phys.} \textbf{1999}, \emph{110},
  4467--4470\relax
\mciteBstWouldAddEndPuncttrue
\mciteSetBstMidEndSepPunct{\mcitedefaultmidpunct}
{\mcitedefaultendpunct}{\mcitedefaultseppunct}\relax
\EndOfBibitem
\bibitem[Bader and Berne(1994)Bader, and Berne]{berne:1994}
Bader,~J.~S.; Berne,~B.~J. Quantum and classical relaxation rates from
  classical simulations. \emph{J. Chem. Phys.} \textbf{1994}, \emph{100},
  8359--8366\relax
\mciteBstWouldAddEndPuncttrue
\mciteSetBstMidEndSepPunct{\mcitedefaultmidpunct}
{\mcitedefaultendpunct}{\mcitedefaultseppunct}\relax
\EndOfBibitem
\bibitem[Skinner and Park(2001)Skinner, and Park]{skinner:2001}
Skinner,~J.~L.; Park,~K. Calculating Vibrational Energy Relaxation Rates from
  Classical Molecular Dynamics Simulations: Quantum Correction Factors for
  Processes Involving Vibration-Vibration Energy Transfer. \emph{J. Phys. Chem.
  B} \textbf{2001}, \emph{105}, 6716--6721\relax
\mciteBstWouldAddEndPuncttrue
\mciteSetBstMidEndSepPunct{\mcitedefaultmidpunct}
{\mcitedefaultendpunct}{\mcitedefaultseppunct}\relax
\EndOfBibitem
\bibitem[Tennyson \latin{et~al.}(2004)Tennyson, Kostin, Barletta, Harris,
  Polyansky, Ramanlal, and Zobov]{dvr3d:2004}
Tennyson,~J.; Kostin,~M.~A.; Barletta,~P.; Harris,~G.~J.; Polyansky,~O.~L.;
  Ramanlal,~J.; Zobov,~N.~F. DVR3D: a program suite for the calculation of
  rotation–vibration spectra of triatomic molecules. \emph{Comput. Phys.
  Commun.} \textbf{2004}, \emph{163}, 85--116\relax
\mciteBstWouldAddEndPuncttrue
\mciteSetBstMidEndSepPunct{\mcitedefaultmidpunct}
{\mcitedefaultendpunct}{\mcitedefaultseppunct}\relax
\EndOfBibitem
\bibitem[Herzberg(1945)]{herzberg:1945}
Herzberg,~G. \emph{Infrared and Raman Spectra of Polyatomic Molecules}; D. Van
  Nostrand Company, Inc., New York, 1945\relax
\mciteBstWouldAddEndPuncttrue
\mciteSetBstMidEndSepPunct{\mcitedefaultmidpunct}
{\mcitedefaultendpunct}{\mcitedefaultseppunct}\relax
\EndOfBibitem
\bibitem[Herzberg and Herzberg(1950)Herzberg, and Herzberg]{herzberg:1950}
Herzberg,~G.; Herzberg,~L. Rotation‐Vibration Spectra of Diatomic and Simple
  Polyatomic Molecules with Long Absorbing Paths VI. The Spectrum of Nitrous
  Oxide (N$_2$O) below 1.2$\mu$. \emph{J. Chem. Phys.} \textbf{1950},
  \emph{18}, 1551--1561\relax
\mciteBstWouldAddEndPuncttrue
\mciteSetBstMidEndSepPunct{\mcitedefaultmidpunct}
{\mcitedefaultendpunct}{\mcitedefaultseppunct}\relax
\EndOfBibitem
\bibitem[Kagann(1982)]{kagann:1982}
Kagann,~R.~H. Infrared absorption intensities for N$_2$O. \emph{J. Mol.
  Spectrosc.} \textbf{1982}, \emph{95}, 297--305\relax
\mciteBstWouldAddEndPuncttrue
\mciteSetBstMidEndSepPunct{\mcitedefaultmidpunct}
{\mcitedefaultendpunct}{\mcitedefaultseppunct}\relax
\EndOfBibitem
\bibitem[Aziz and Slaman(1986)Aziz, and Slaman]{aziz:1986}
Aziz,~R.~A.; Slaman,~M. On the Xe-Xe potential energy curve and related
  properties. \emph{Mol. Phys.} \textbf{1986}, \emph{57}, 825--840\relax
\mciteBstWouldAddEndPuncttrue
\mciteSetBstMidEndSepPunct{\mcitedefaultmidpunct}
{\mcitedefaultendpunct}{\mcitedefaultseppunct}\relax
\EndOfBibitem
\bibitem[Hellmann \latin{et~al.}(2017)Hellmann, J\"ager, and Bich]{bich:2017}
Hellmann,~R.; J\"ager,~B.; Bich,~E. State-of-the-art ab initio potential energy
  curve for the xenon atom pair and related spectroscopic and thermophysical
  properties. \emph{J. Chem. Phys.} \textbf{2017}, \emph{147}, 034304\relax
\mciteBstWouldAddEndPuncttrue
\mciteSetBstMidEndSepPunct{\mcitedefaultmidpunct}
{\mcitedefaultendpunct}{\mcitedefaultseppunct}\relax
\EndOfBibitem
\bibitem[Freeman \latin{et~al.}(1974)Freeman, Yoshino, and Tanaka]{tanaka:1974}
Freeman,~D.~E.; Yoshino,~K.; Tanaka,~Y. Vacuum ultraviolet absorption spectrum
  of the van der Waals molecule Xe$_2$. I. Ground state vibrational structure,
  potential well depth, and shape. \emph{J. Chem. Phys.} \textbf{1974},
  \emph{61}, 4880--4889\relax
\mciteBstWouldAddEndPuncttrue
\mciteSetBstMidEndSepPunct{\mcitedefaultmidpunct}
{\mcitedefaultendpunct}{\mcitedefaultseppunct}\relax
\EndOfBibitem
\bibitem[Maddox \latin{et~al.}(2000)Maddox, Goodyear, and Tucker]{tucker:2000a}
Maddox,~M.~W.; Goodyear,~G.; Tucker,~S.~C. Origins of Atom-Centered Local
  Density Enhancements in Compressible Supercritical Fluids. \emph{J. Phys.
  Chem. B} \textbf{2000}, \emph{104}, 6248--6257\relax
\mciteBstWouldAddEndPuncttrue
\mciteSetBstMidEndSepPunct{\mcitedefaultmidpunct}
{\mcitedefaultendpunct}{\mcitedefaultseppunct}\relax
\EndOfBibitem
\bibitem[Maddox \latin{et~al.}(2000)Maddox, Goodyear, and Tucker]{tucker:2000b}
Maddox,~M.~W.; Goodyear,~G.; Tucker,~S.~C. Effect of Critical Slowing Down on
  Local-Density Dynamics. \emph{J. Phys. Chem. B} \textbf{2000}, \emph{104},
  6266--6270\relax
\mciteBstWouldAddEndPuncttrue
\mciteSetBstMidEndSepPunct{\mcitedefaultmidpunct}
{\mcitedefaultendpunct}{\mcitedefaultseppunct}\relax
\EndOfBibitem
\bibitem[Berne and Pecora(2000)Berne, and Pecora]{berne:2000}
Berne,~B.~J.; Pecora,~R. \emph{Dynamic light scattering: with applications to
  chemistry, biology, and physics}; Courier Corporation, 2000\relax
\mciteBstWouldAddEndPuncttrue
\mciteSetBstMidEndSepPunct{\mcitedefaultmidpunct}
{\mcitedefaultendpunct}{\mcitedefaultseppunct}\relax
\EndOfBibitem
\bibitem[Bondi(1964)]{vdwradii:1964}
Bondi,~A. van der Waals Volumes and Radii. \emph{J. Phys. Chem.} \textbf{1964},
  \emph{68}, 441--451\relax
\mciteBstWouldAddEndPuncttrue
\mciteSetBstMidEndSepPunct{\mcitedefaultmidpunct}
{\mcitedefaultendpunct}{\mcitedefaultseppunct}\relax
\EndOfBibitem
\bibitem[Yardley(1980)]{yardley1980}
Yardley,~J. \emph{Introduction to Molecular Energy Transfer}; Academic Press,
  New York, 1980\relax
\mciteBstWouldAddEndPuncttrue
\mciteSetBstMidEndSepPunct{\mcitedefaultmidpunct}
{\mcitedefaultendpunct}{\mcitedefaultseppunct}\relax
\EndOfBibitem
\end{mcitethebibliography}


\providecommand{\latin}[1]{#1}
\makeatletter
\providecommand{\doi}
  {\begingroup\let\do\@makeother\dospecials
  \catcode`\{=1 \catcode`\}=2 \doi@aux}
\providecommand{\doi@aux}[1]{\endgroup\texttt{#1}}
\makeatother
\providecommand*\mcitethebibliography{\thebibliography}
\csname @ifundefined\endcsname{endmcitethebibliography}
  {\let\endmcitethebibliography\endthebibliography}{}
\begin{mcitethebibliography}{6}
\providecommand*\natexlab[1]{#1}
\providecommand*\mciteSetBstSublistMode[1]{}
\providecommand*\mciteSetBstMaxWidthForm[2]{}
\providecommand*\mciteBstWouldAddEndPuncttrue
  {\def\EndOfBibitem{\unskip.}}
\providecommand*\mciteBstWouldAddEndPunctfalse
  {\let\EndOfBibitem\relax}
\providecommand*\mciteSetBstMidEndSepPunct[3]{}
\providecommand*\mciteSetBstSublistLabelBeginEnd[3]{}
\providecommand*\EndOfBibitem{}
\mciteSetBstSublistMode{f}
\mciteSetBstMaxWidthForm{subitem}{(\alph{mcitesubitemcount})}
\mciteSetBstSublistLabelBeginEnd
  {\mcitemaxwidthsubitemform\space}
  {\relax}
  {\relax}

\bibitem[Koner \latin{et~al.}(2020)Koner, San Vicente~Veliz, Bemish, and
  Meuwly]{koner:2020}
Koner,~D.; San Vicente~Veliz,~J.~C.; Bemish,~R.~J.; Meuwly,~M. Accurate
  reproducing kernel-based potential energy surfaces for the triplet ground
  states of N$_2$O and dynamics for the N + NO $\leftrightarrow$ O + N$_2$ and
  N$_2$ + O $\rightarrow$ 2N + O reactions. \emph{Phys. Chem. Chem. Phys.}
  \textbf{2020}, \emph{22}, 18488--18498\relax
\mciteBstWouldAddEndPuncttrue
\mciteSetBstMidEndSepPunct{\mcitedefaultmidpunct}
{\mcitedefaultendpunct}{\mcitedefaultseppunct}\relax
\EndOfBibitem
\bibitem[Devereux \latin{et~al.}(2020)Devereux, Pezzella, Raghunathan, and
  Meuwly]{MM.mdcm:2020}
Devereux,~M.; Pezzella,~M.; Raghunathan,~S.; Meuwly,~M. Polarizable Multipolar
  Molecular Dynamics Using Distributed Point Charges. \emph{J. Chem. Theo.
  Comp.} \textbf{2020}, \emph{16}, 7267--7280\relax
\mciteBstWouldAddEndPuncttrue
\mciteSetBstMidEndSepPunct{\mcitedefaultmidpunct}
{\mcitedefaultendpunct}{\mcitedefaultseppunct}\relax
\EndOfBibitem
\bibitem[Dellis and Samios(2010)Dellis, and Samios]{samios:2010}
Dellis,~D.; Samios,~J. Molecular force field investigation for Sulfur
  Hexafluoride: A computer simulation study. \emph{Fluid Phase Equilib.}
  \textbf{2010}, \emph{291}, 81--89\relax
\mciteBstWouldAddEndPuncttrue
\mciteSetBstMidEndSepPunct{\mcitedefaultmidpunct}
{\mcitedefaultendpunct}{\mcitedefaultseppunct}\relax
\EndOfBibitem
\bibitem[Aziz and Slaman(1986)Aziz, and Slaman]{aziz:1986}
Aziz,~R.~A.; Slaman,~M. On the Xe-Xe potential energy curve and related
  properties. \emph{Mol. Phys.} \textbf{1986}, \emph{57}, 825--840\relax
\mciteBstWouldAddEndPuncttrue
\mciteSetBstMidEndSepPunct{\mcitedefaultmidpunct}
{\mcitedefaultendpunct}{\mcitedefaultseppunct}\relax
\EndOfBibitem
\bibitem[Rotondaro \latin{et~al.}(2022)Rotondaro, Jain, Erramilli, and
  Ziegler]{ziegler:2022}
Rotondaro,~M.~C.; Jain,~A.; Erramilli,~S.; Ziegler,~L.~D. Ultrafast 2DIR
  comparison of rotational energy transfer, isolated binary collision
  breakdown, and near critical fluctuations in Xe and SF$_6$ solutions.
  \emph{J. Chem. Phys.} \textbf{2022}, \emph{157}, 174305\relax
\mciteBstWouldAddEndPuncttrue
\mciteSetBstMidEndSepPunct{\mcitedefaultmidpunct}
{\mcitedefaultendpunct}{\mcitedefaultseppunct}\relax
\EndOfBibitem
\end{mcitethebibliography}

\end{document}


\section{System Setup}

\begin{table}
\caption{System setup of N$_2$O concentration $c(\mathrm{N_2O})$,
  molar volumes $V_m$ and critical density ratio $\rho^*$ of one
  N$_2$O in 343 SF$_6$ molecules and one N$_2$O in 600 Xe atoms.}
  \label{sitab_conc}
\begin{tabular}{ccc||ccc}
\hline\hline
\multicolumn{3}{c||}{N$_2$O/SF$_6$} & \multicolumn{3}{c}{N$_2$O/Xe} \\
\hline
$c(\mathrm{N_2O})$ (mol/l) & $V_m(\mathrm{SF_6})$ (cm$^3$/mol) & $\rho^*$ &
$c(\mathrm{N_2O})$ (mol/l) & $V_m(\mathrm{Xe})$ (cm$^3$/mol) & $\rho^*$ \\
\hline
0.20 & 4934 & 0.04  & 0.34  & 2984 & 0.04  \\
0.51 & 1974 & 0.10  & 0.84  & 1194 & 0.10  \\
0.81 & 1234 & 0.16  & 1.26  & 796  & 0.15  \\
1.52 & 658  & 0.30  & 3.10  & 323  & 0.37  \\
3.39 & 295  & 0.67  & 4.11  & 244  & 0.49  \\
4.36 & 229  & 0.86  & 5.19  & 193  & 0.62  \\
5.02 & 199  & 0.99  & 5.53  & 181  & 0.66  \\
5.93 & 169  & 1.17  & 6.28  & 159  & 0.75  \\
6.89 & 145  & 1.36  & 7.29  & 137  & 0.87  \\
7.65 & 131  & 1.51  & 7.79  & 128  & 0.93  \\
     &      &       & 11.14 & 90   & 1.33  \\
     &      &       & 12.99 & 77   & 1.55  \\
\hline \hline
\end{tabular}
\end{table}

\begin{table}
\caption{Bonded and non-bonded parameters. Larger sets of 
parameters are made available on a github as full parameter files.}
\label{sitab_tab2}
\begin{tabular}{c|cc}
\hline\hline
\textbf{Residues} & \multicolumn{2}{c}{Parameters} \\
\hline \hline 
\textbf{N$_2$O} & \multicolumn{2}{c}{} \\
\hline
Intramolecular Potential & \\
RKHS\cite{koner:2020} & \multicolumn{2}{c}{\url{https://github.com/MMunibas/SCF_N2O/}} \\
 & \multicolumn{2}{c}{in, e.g., \url{N2O_Xe/source/pes1_rRz.csv}} \\
 & \multicolumn{2}{c}{or \url{N2O_SF6/source/pes1_rRz.csv}} \\
\hline 
Electrostatic Potential & \\
polarizable MDCM\cite{MM.mdcm:2020} & \multicolumn{2}{c}{\url{https://github.com/MMunibas/SCF_N2O/}} \\
 & \multicolumn{2}{c}{in, e.g., \url{N2O_Xe/source/n2o_xe.dcm}} \\
 & \multicolumn{2}{c}{or \url{N2O_SF6/source/n2o_sf6.dcm}} \\
\hline 
Polarizabilities &  $\alpha$ (1/\AA$^3$) &  \\
N1 & $0.94$ &  \\
N2 & $0.94$ &  \\
O & $0.94$ &  \\
\hline 
Non-bonded in SF$_6$ & $\epsilon_i$ (kcal/mol)& $R_{\mathrm{min},i}/2$ (\AA) \\
N1 & $0.2592$ & $1.716$ \\
N2 & $0.1542$ & $1.618$ \\
O  & $0.2089$ & $1.555$ \\
\hline 
Non-bonded in Xe & $\epsilon_i$ (kcal/mol)& $R_{\mathrm{min},i}/2$ (\AA) \\
N1 & $0.3775$ & $1.637$ \\
N2 & $0.0002$ & $2.565$ \\
O  & $0.3684$ & $1.497$ \\
\hline \hline
\textbf{SF$_6$} & \multicolumn{2}{c}{Samios and coworker\cite{samios:2010}} \\
\hline 
Bonds & $k_b$ (kcal/mol/\AA$^2$) & $r_\mathrm{min}$ (\AA) \\
S-F & $165.746$ & $1.565$ \\
\hline
Angles & $k_\Theta$ (kcal/mol/rad$^2$) & $\Theta_\mathrm{min}$ ($^\circ$) \\
F-S-F & $73.461$ & $90.0$ \\
\hline 
Electrostatic Potential & \\
polarizable MDCM\cite{MM.mdcm:2020} * & \multicolumn{2}{c}{\url{https://github.com/MMunibas/SCF_N2O/}} \\
 & \multicolumn{2}{c}{in, e.g., \url{N2O_Xe/source/n2o_xe.dcm}} \\
 & \multicolumn{2}{c}{or \url{N2O_SF6/source/n2o_sf6.dcm}} \\
\hline 
Non-bonded & $\epsilon_i$ (kcal/mol)& $R_{\mathrm{min},i}/2$ (\AA) \\
S & $0.3257$ & $1.822$ \\
F & $0.0541$ & $1.659$ \\
\hline \hline
\textbf{Xe} & \multicolumn{2}{c}{Aziz and coworker\cite{aziz:1986}} \\
\hline 
Electrostatic Potential & $q$ (e) & $\alpha$ (1/\AA$^3$) \\
polarizable MDCM\cite{MM.mdcm:2020} ** & $0.0$ & $2.964$ \\
\hline 
Non-bonded & $\epsilon_i$ (kcal/mol)& $R_{\mathrm{min},i}/2$ (\AA) \\
Xe & $0.5610$ & $2.181$ \\
\hline \hline
\multicolumn{3}{l}{* Contributing only to N$_2$O-SF$_6$ interaction} \\
\multicolumn{3}{l}{** Contributing effectively only to N$_2$O-Xe interaction}
\end{tabular}
\end{table}

\clearpage

\begin{figure}[htb!]
\begin{center}
\includegraphics[width=0.80\textwidth]{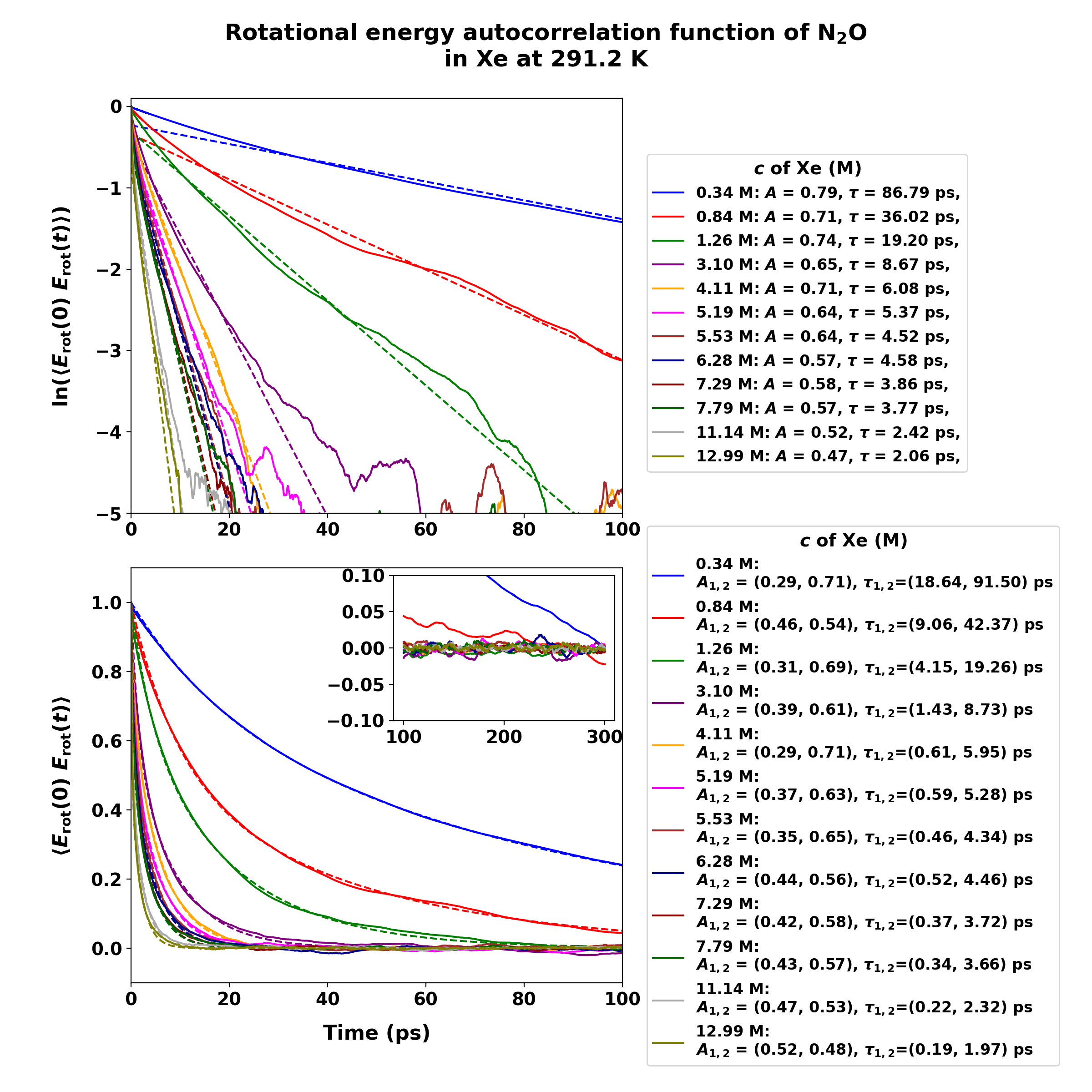}
\caption{
Rotational energy correlation function of N$_2$O in Xe at
different solvent concentrations as solid lines and optimized fits of a 
the single-exponential function (top) to the logarithm and of a 
bi-exponential function (bottom) as dashed lines.}
\label{sifig_erot_xe}
\end{center}
\end{figure}

\begin{figure}[htb!]
\begin{center}
\includegraphics[width=0.80\textwidth]{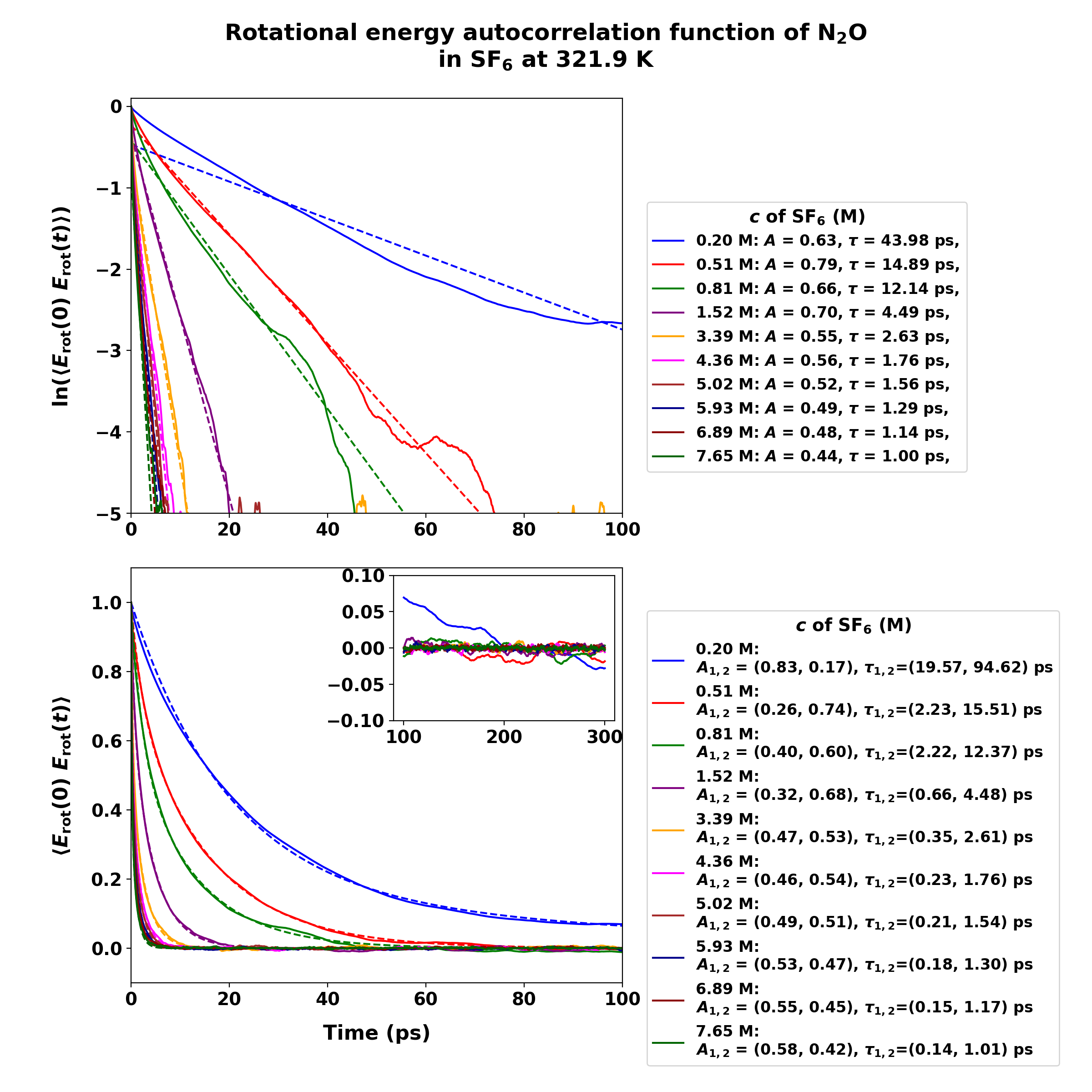}
\caption{
Rotational energy correlation function of N$_2$O in SF$_6$ at
different solvent concentrations as solid lines and optimized fits of a 
the single-exponential function (top) to the logarithm and of a 
bi-exponential function (bottom) as dashed lines.}
\label{sifig_erot_sf6}
\end{center}
\end{figure}

\clearpage

\begin{figure}[htb!]
\begin{center}
\includegraphics[width=0.90\textwidth]{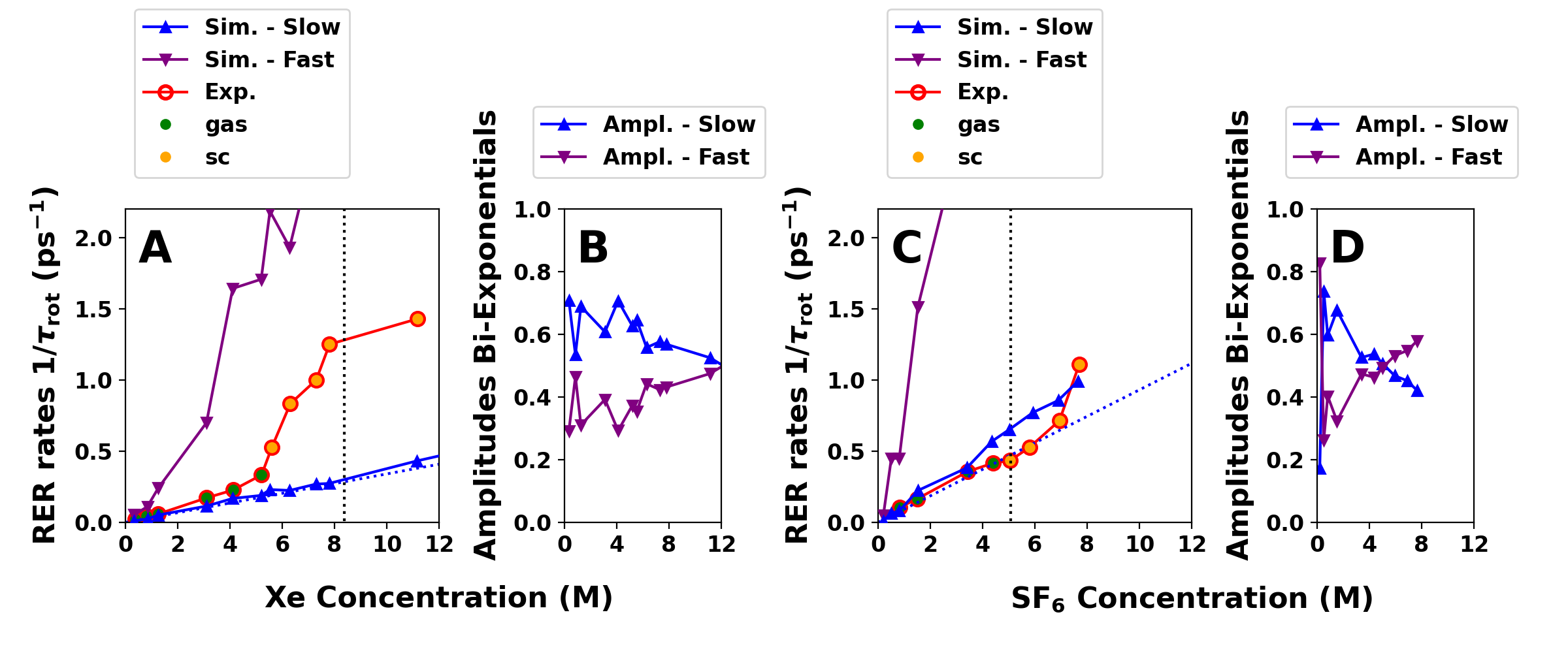}
\caption{
  RER rates from the RER times $\tau_\mathrm{rot,slow}$ 
  and $\tau_\mathrm{rot,fast}$ 
  of a fitted bi-exponential function 
  $C_2(t) = A \mathrm{e}^{-t/\tau_\mathrm{fast}} 
  + (1-A)\mathrm{e}^{-t/\tau_\mathrm{slow}} + \Delta$
  to correlation function
  of the rotational energy $E_\mathrm{rot}(t)$ of N$_2$O in (A) xenon
  and (C) SF$_6$ at different solvent concentrations (solid blue lines).
  The solid red line with colored circle markers shows the experimentally
  measured RER rates and indicate the solvent state gaseous or
  supercritical (sc).\cite{ziegler:2022}
  The dotted blue line is an linear extrapolation of the first 3 
  simulated RER rates. The vertical dotted lines
  mark the experimentally observed solvent concentration at the
  respective critical density of xenon and SF$_6$, respectively.
  Panel B and D shows the amplitudes of the bi-exponential fit function
  for which the RER rates are shown in panel A and C, respectively.}
\label{sifig_rotrates_bi}
\end{center}
\end{figure}

\begin{figure}[htb!]
\begin{center}
\includegraphics[width=0.90\textwidth]{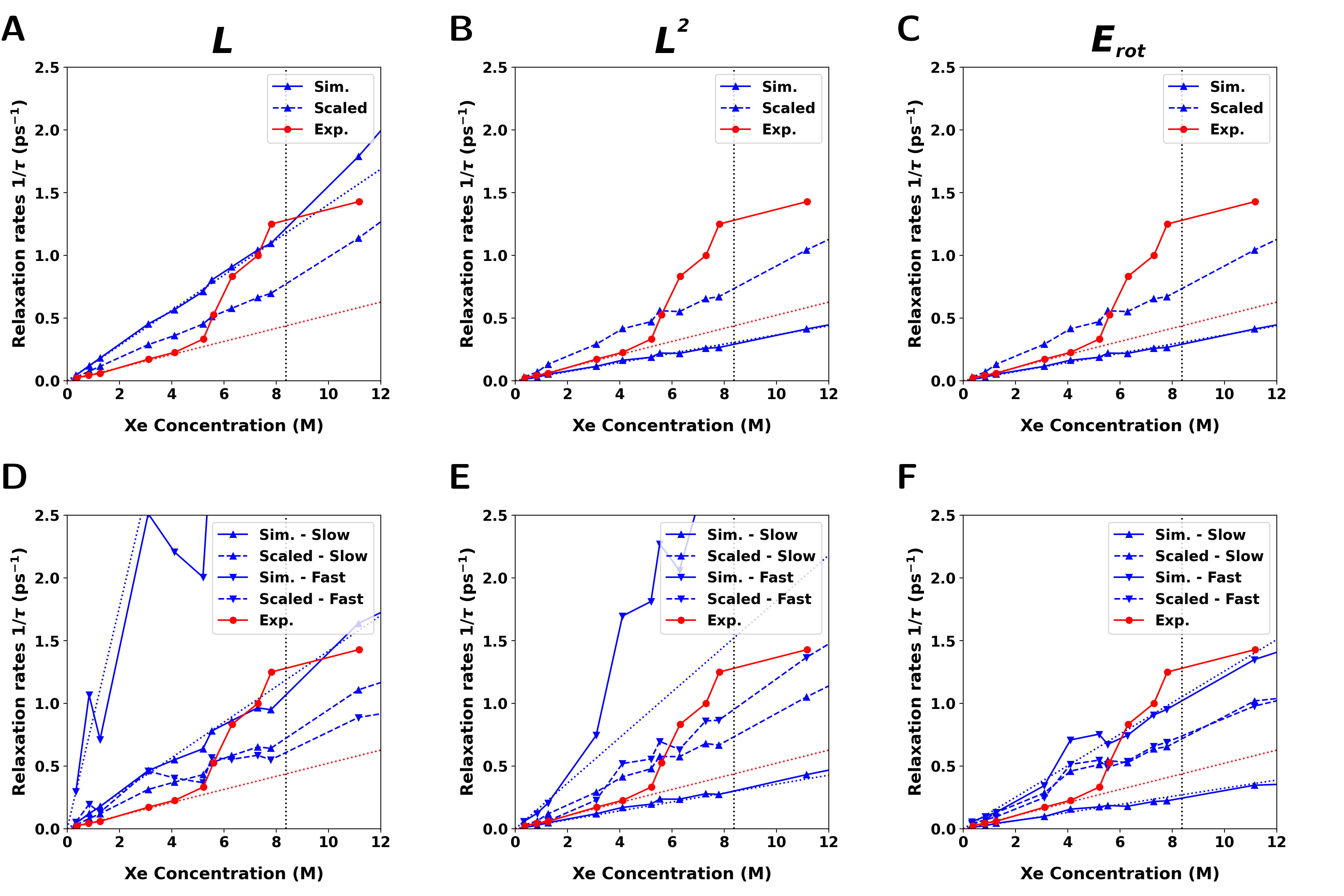}
\caption{Rotational relaxation rates from the lifetimes of a fitted
single- (\textbf{A}-\textbf{C}) and bi-exponential function (\textbf{D}-\textbf{F}) 
to correlation function of the  angular momentum (\textbf{A}, \textbf{D}), 
squared angular momentum (\textbf{B}, \textbf{E}) or rotational energy 
(\textbf{C}, \textbf{F}) of N$_2$O in Xe and at different solvent concentrations. 
The solid red line with circle markers shows the experimentally 
measured rotational relaxation rates,\cite{ziegler:2022}
and the dashed blue lines are the fitted rotational relaxation rates scaled to 
minimizing deviation from experiments.
The vertical dashed lines mark the experimentally observed solvent concentration
at the respective critical density xenon.}
\label{sifig_rotrates_xe}
\end{center}
\end{figure}

\clearpage

\begin{figure}[htb!]
\begin{center}
\includegraphics[width=0.90\textwidth]{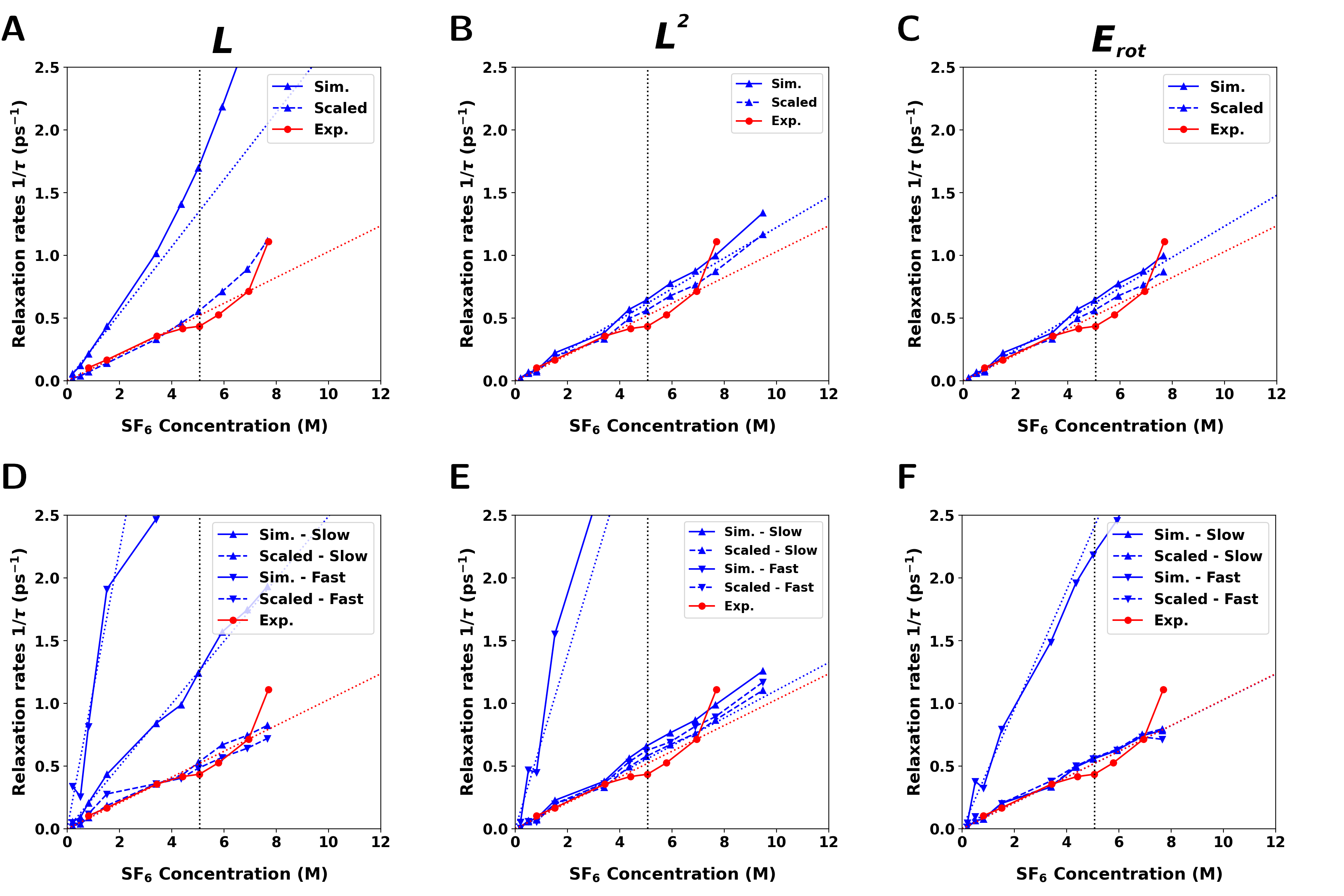}
\caption{Rotational relaxation rates from the lifetimes of a fitted
single- (\textbf{A}-\textbf{C}) and bi-exponential function (\textbf{D}-\textbf{F}) 
to correlation function of the  angular momentum (\textbf{A}, \textbf{D}), 
squared angular momentum (\textbf{B}, \textbf{E}) or rotational energy 
(\textbf{C}, \textbf{F}) of N$_2$O in SF$_6$ and at different solvent concentrations. 
The solid red line with circle markers shows the experimentally 
measured rotational relaxation rates,\cite{ziegler:2022}
and the dashed blue lines are the fitted rotational relaxation rates scaled to 
minimizing deviation from experiments.
The vertical dashed lines mark the experimentally observed solvent concentration
at the respective critical density SF$_6$.}
\label{sifig_rotrates_sf6}
\end{center}
\end{figure}

\clearpage

\begin{figure}[htb!]
\begin{center}
\includegraphics[width=0.80\textwidth]{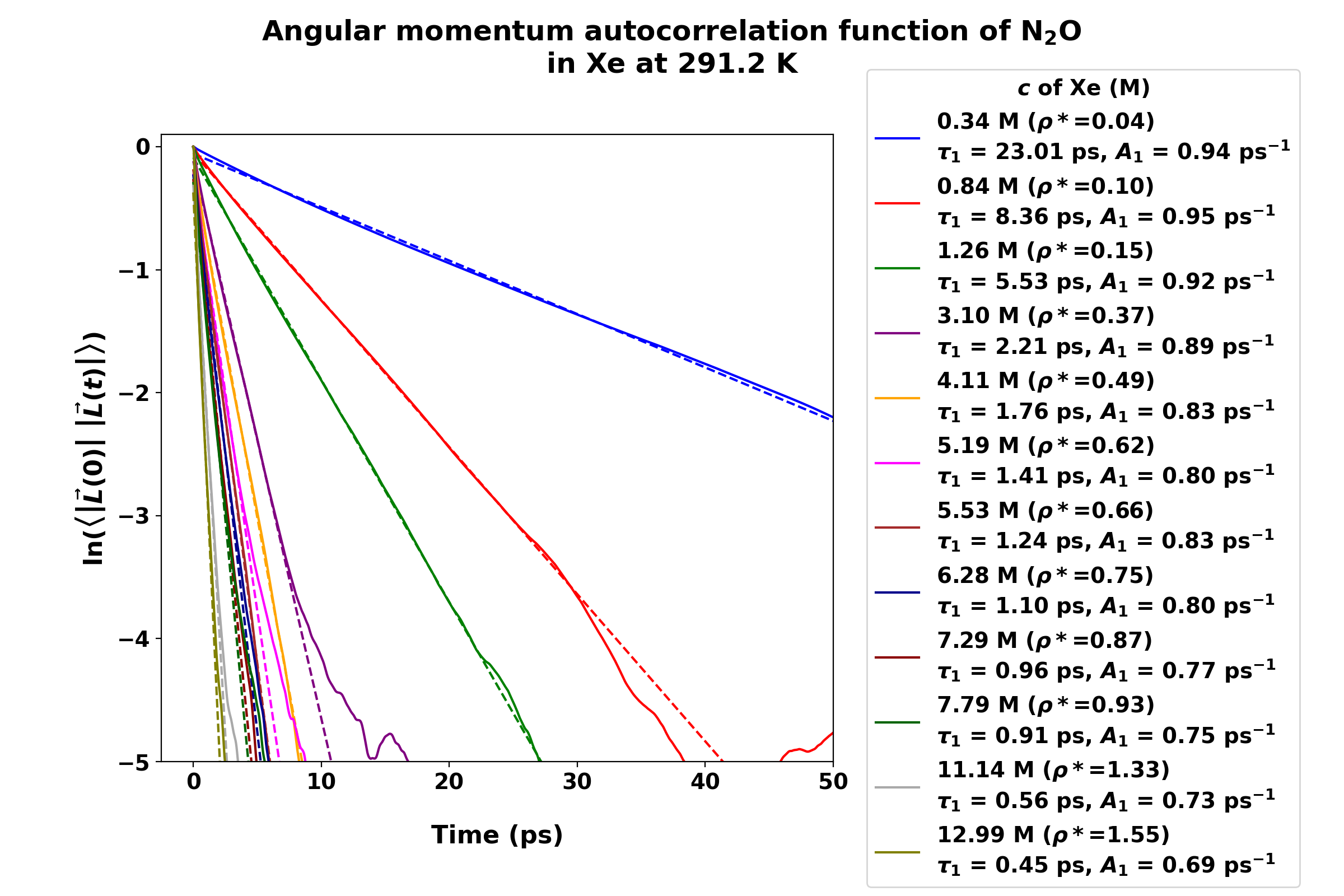}
\includegraphics[width=0.80\textwidth]{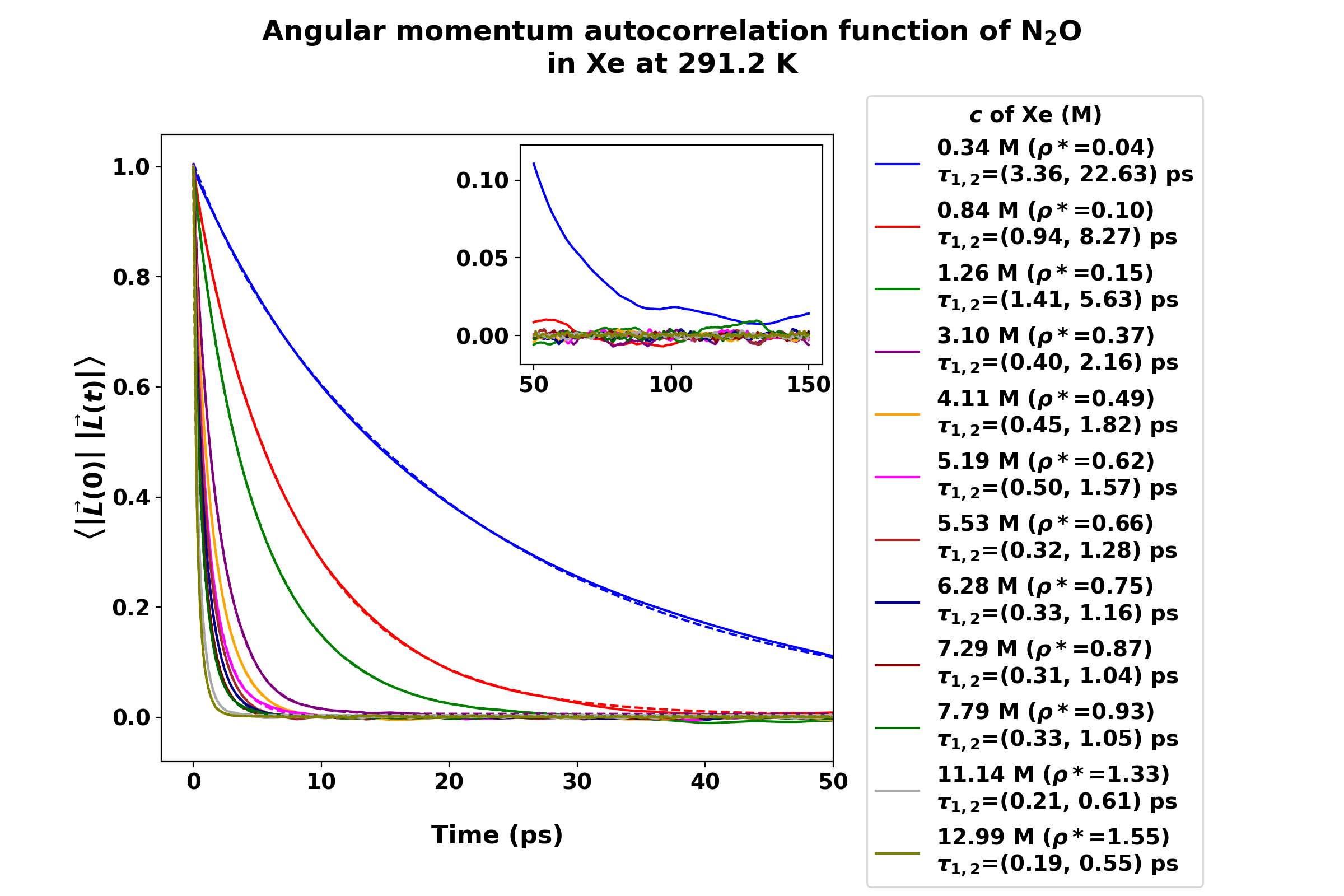}
\caption{
Angular momentum correlation function of N$_2$O in Xe at
different solvent concentrations as solid lines and optimized fits of a 
the single-exponential function (top) to the logarithm and of a 
bi-exponential function (bottom) as dashed lines.}
\label{sifig_lxyz_xe}
\end{center}
\end{figure}

\begin{figure}[htb!]
\begin{center}
\includegraphics[width=0.80\textwidth]{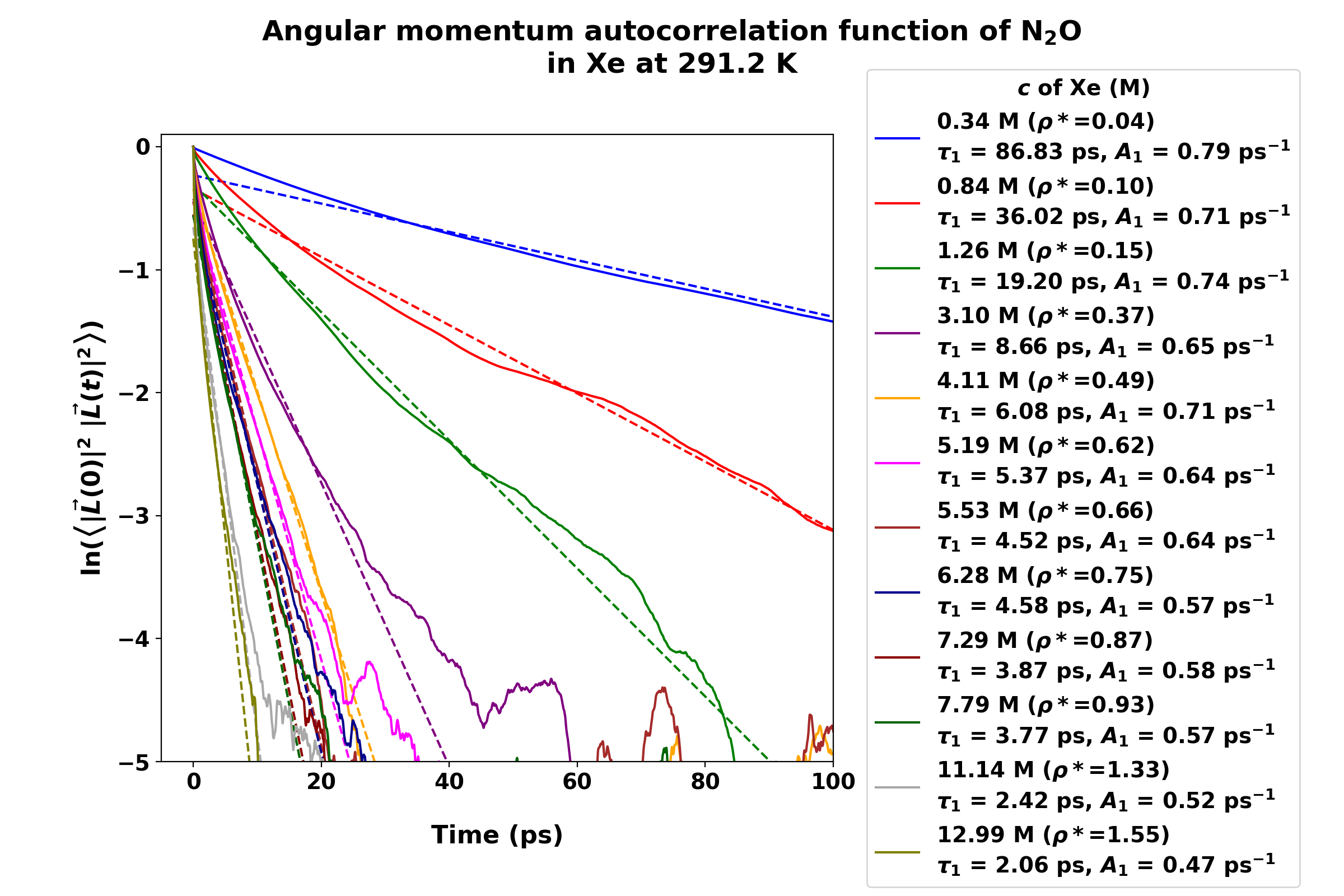}
\includegraphics[width=0.80\textwidth]{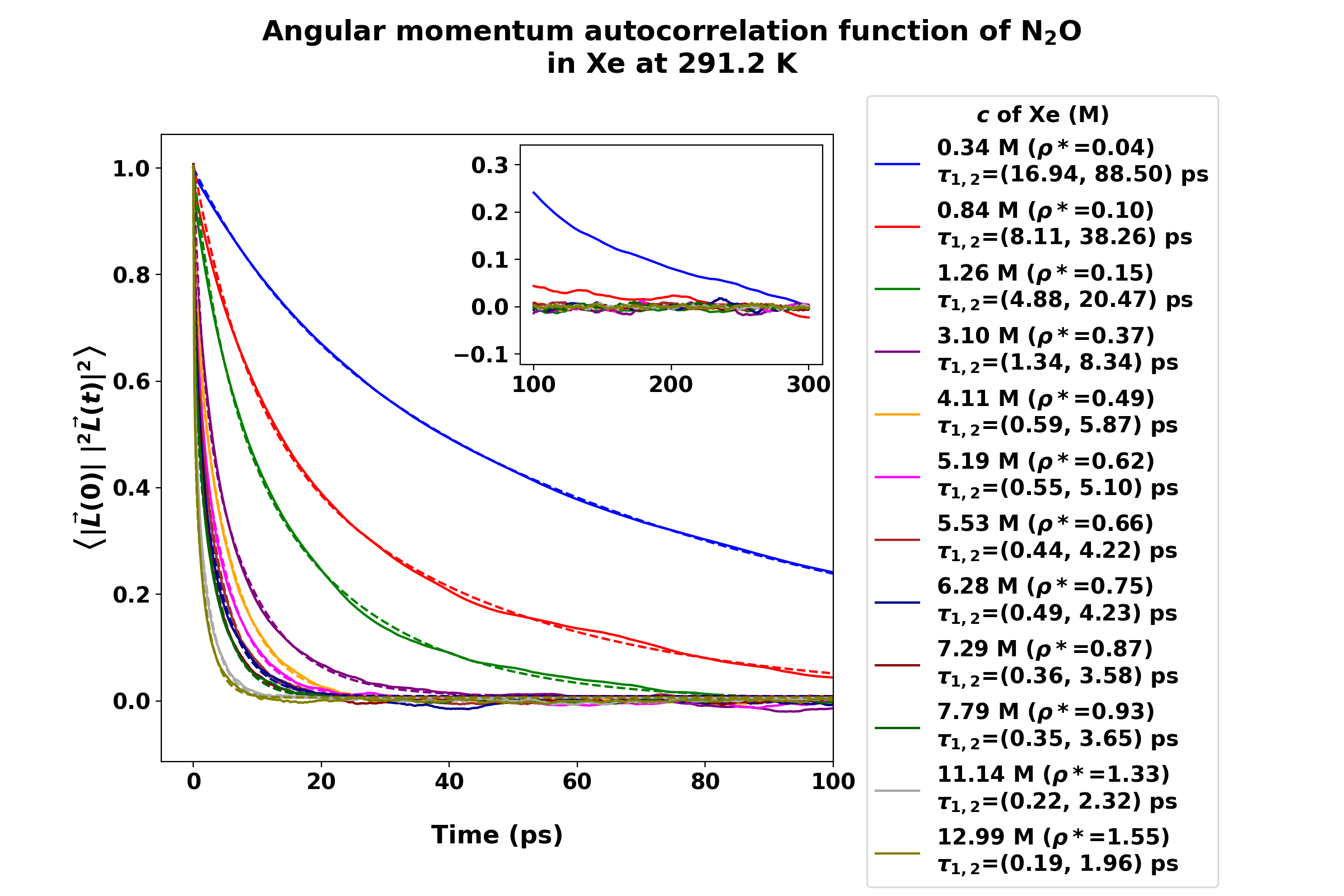}
\caption{
Squared angular momentum correlation function of N$_2$O in Xe at
different solvent concentrations as solid lines and optimized fits of a 
the single-exponential function (top) to the logarithm and of a 
bi-exponential function (bottom) as dashed lines.}
\label{sifig_l2_xe}
\end{center}
\end{figure}

\begin{figure}[htb!]
\begin{center}
\includegraphics[width=0.80\textwidth]{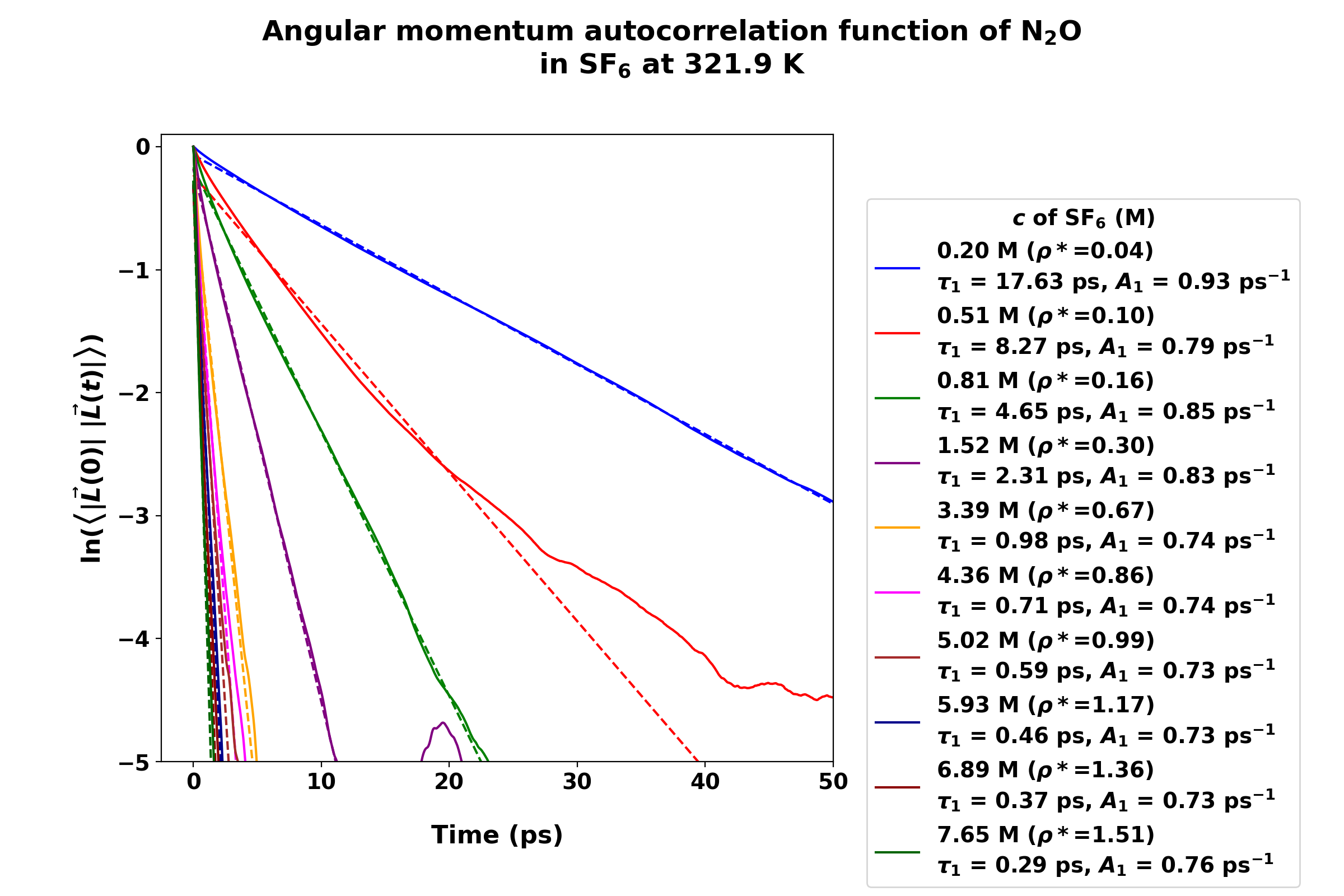}
\includegraphics[width=0.80\textwidth]{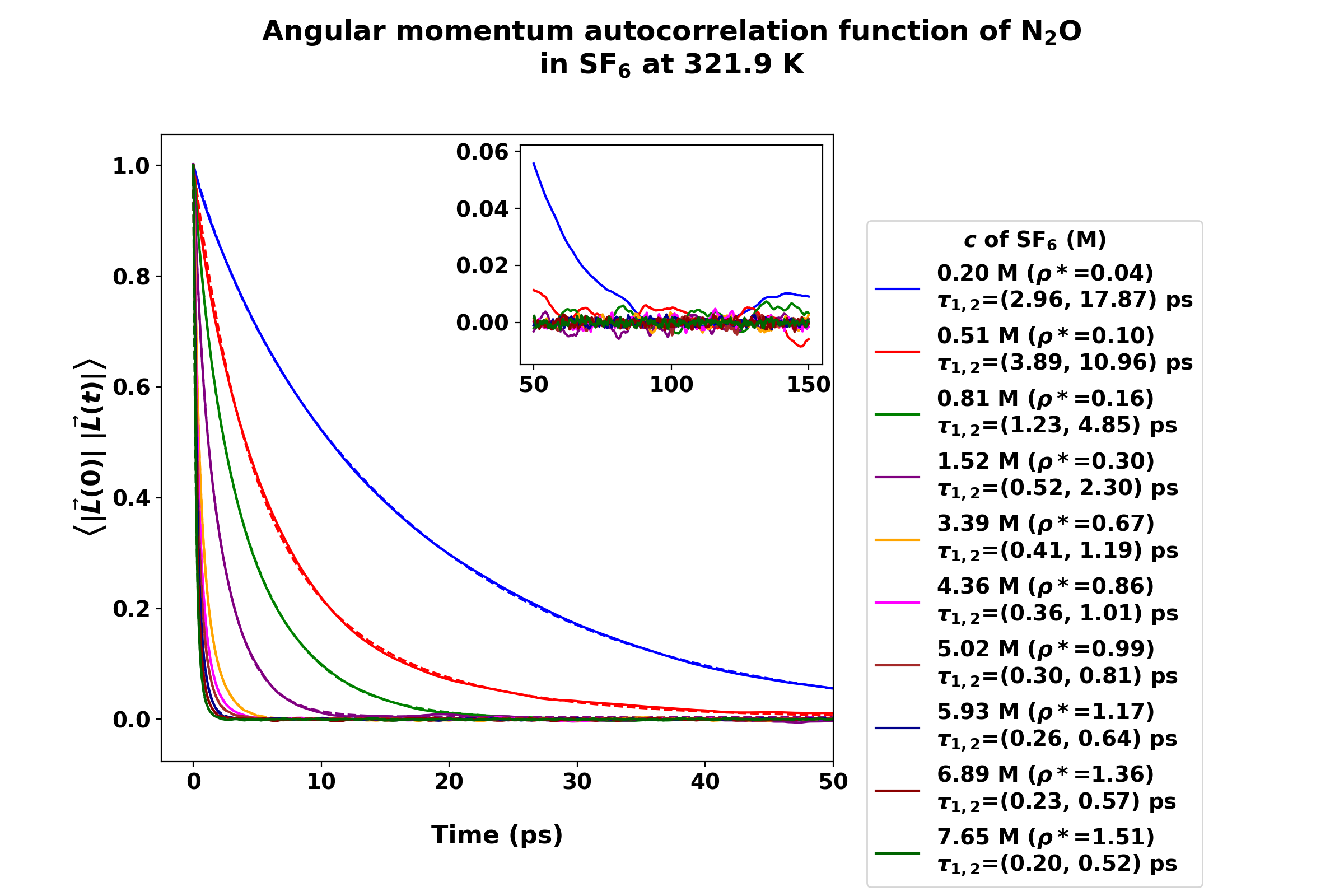}
\caption{
Angular momentum correlation function of N$_2$O in SF$_6$ at
different solvent concentrations as solid lines and optimized fits of a 
the single-exponential function (top) to the logarithm and of a 
bi-exponential function (bottom) as dashed lines.}
\label{sifig_lxyz_sf6}
\end{center}
\end{figure}

\begin{figure}[htb!]
\begin{center}
\includegraphics[width=0.80\textwidth]{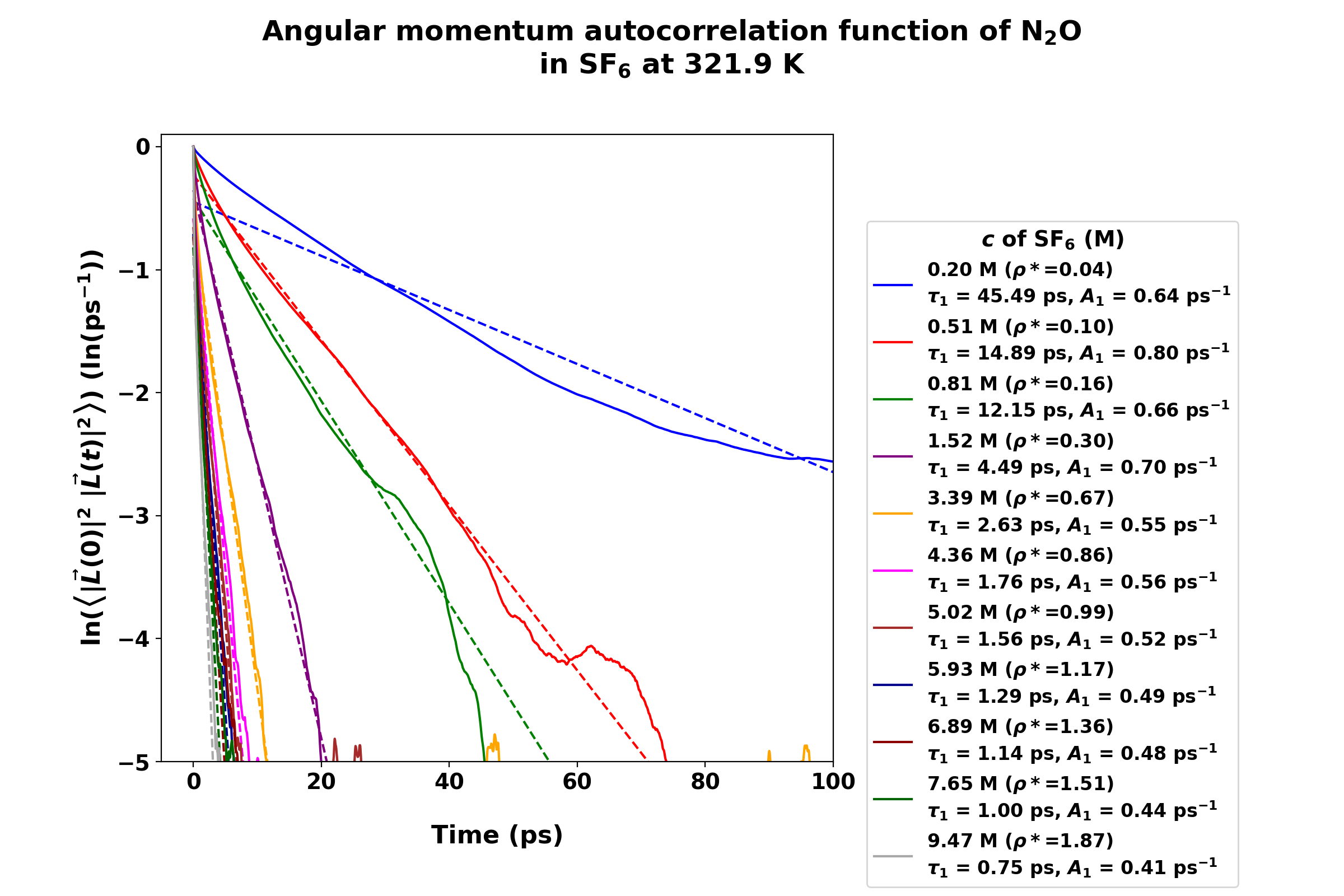}
\includegraphics[width=0.80\textwidth]{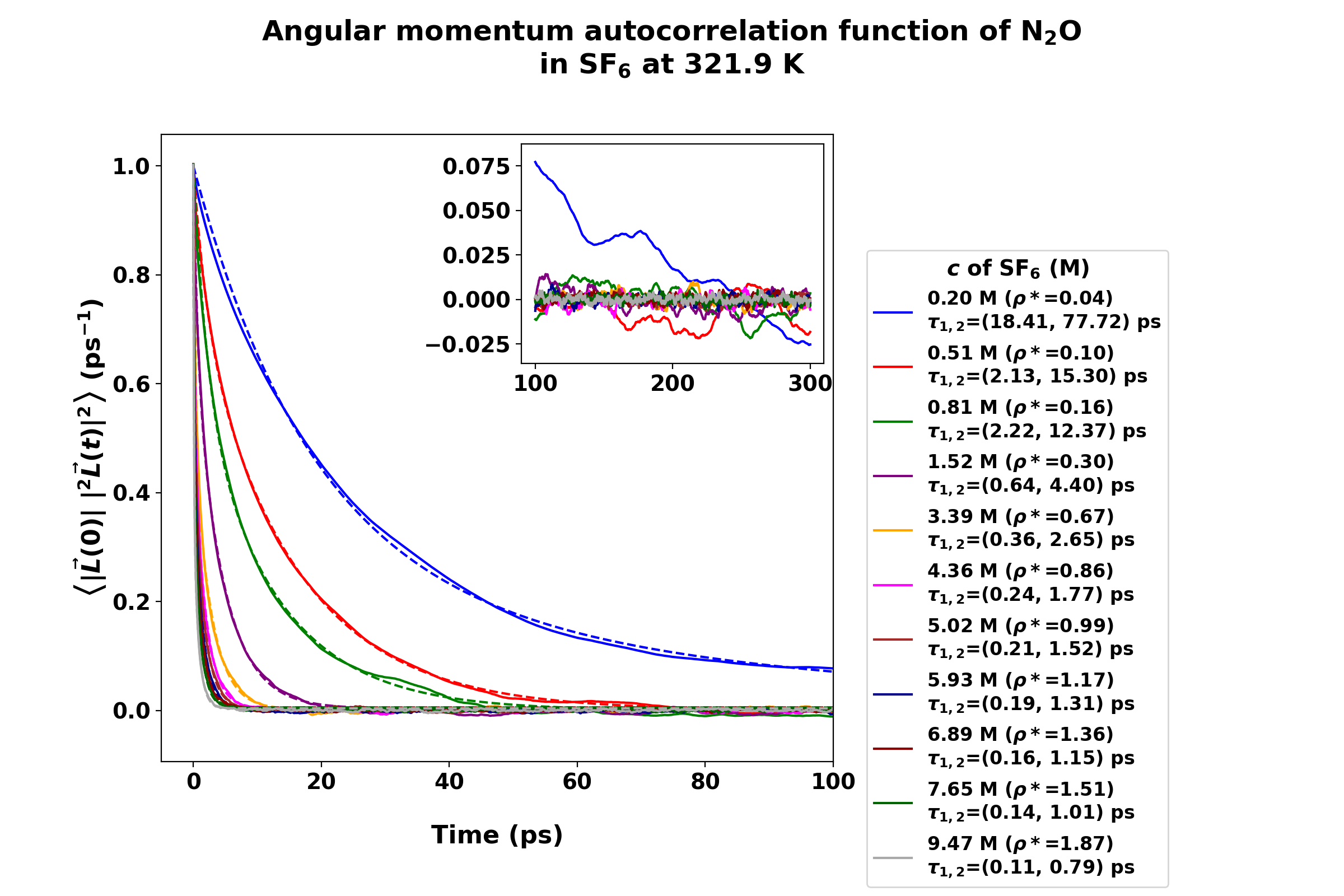}
\caption{
Squared angular momentum correlation function of N$_2$O in SF$_6$ at
different solvent concentrations as solid lines and optimized fits of a 
the single-exponential function to the logarithm (top) and of a 
bi-exponential function (bottom) as dashed lines.}
\label{sifig_l2_sf6}
\end{center}
\end{figure}

\begin{figure}[htb!]
\begin{center}
\includegraphics[width=0.90\textwidth]{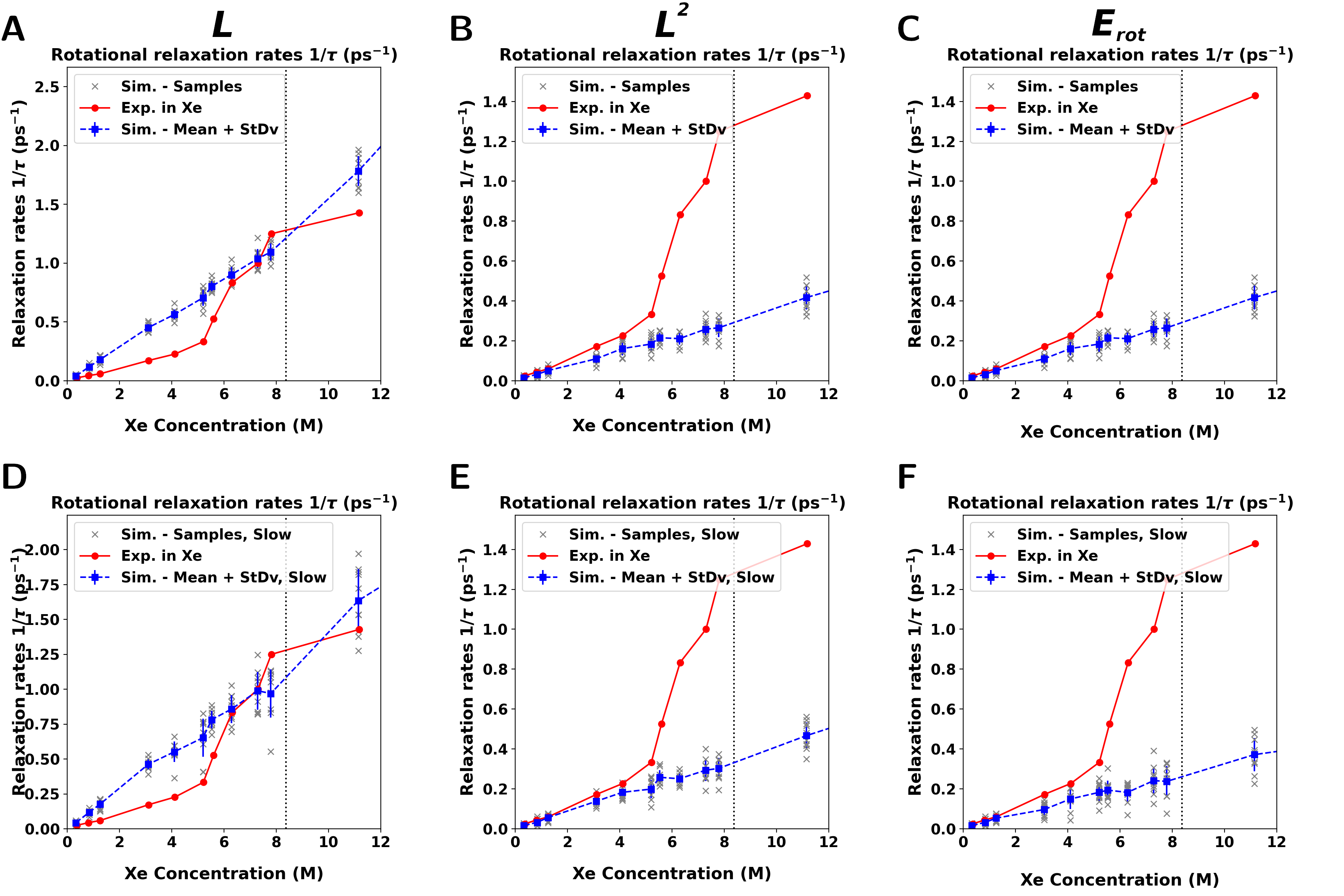}
\caption{Rotational relaxation rates from the lifetimes of a fitted
single- (\textbf{A}-\textbf{C}) and bi-exponential function (\textbf{D}-\textbf{F}) 
to correlation function of the  angular momentum (\textbf{A}, \textbf{D}), 
squared angular momentum (\textbf{B}, \textbf{E}) or rotational energy 
(\textbf{C}, \textbf{F}) of N$_2$O in Xe and at different solvent concentrations. 
The solid red line with circle markers shows the experimentally 
measured rotational relaxation rates,\cite{ziegler:2022}
and the dashed blue lines are the fitted rotational relaxation rates scaled to 
minimizing deviation from experiments.
The vertical dashed lines mark the experimentally observed solvent concentration
at the respective critical density xenon.}
\label{sifig_sample_rotrates_xe}
\end{center}
\end{figure}

\begin{figure}[htb!]
\begin{center}
\includegraphics[width=0.90\textwidth]{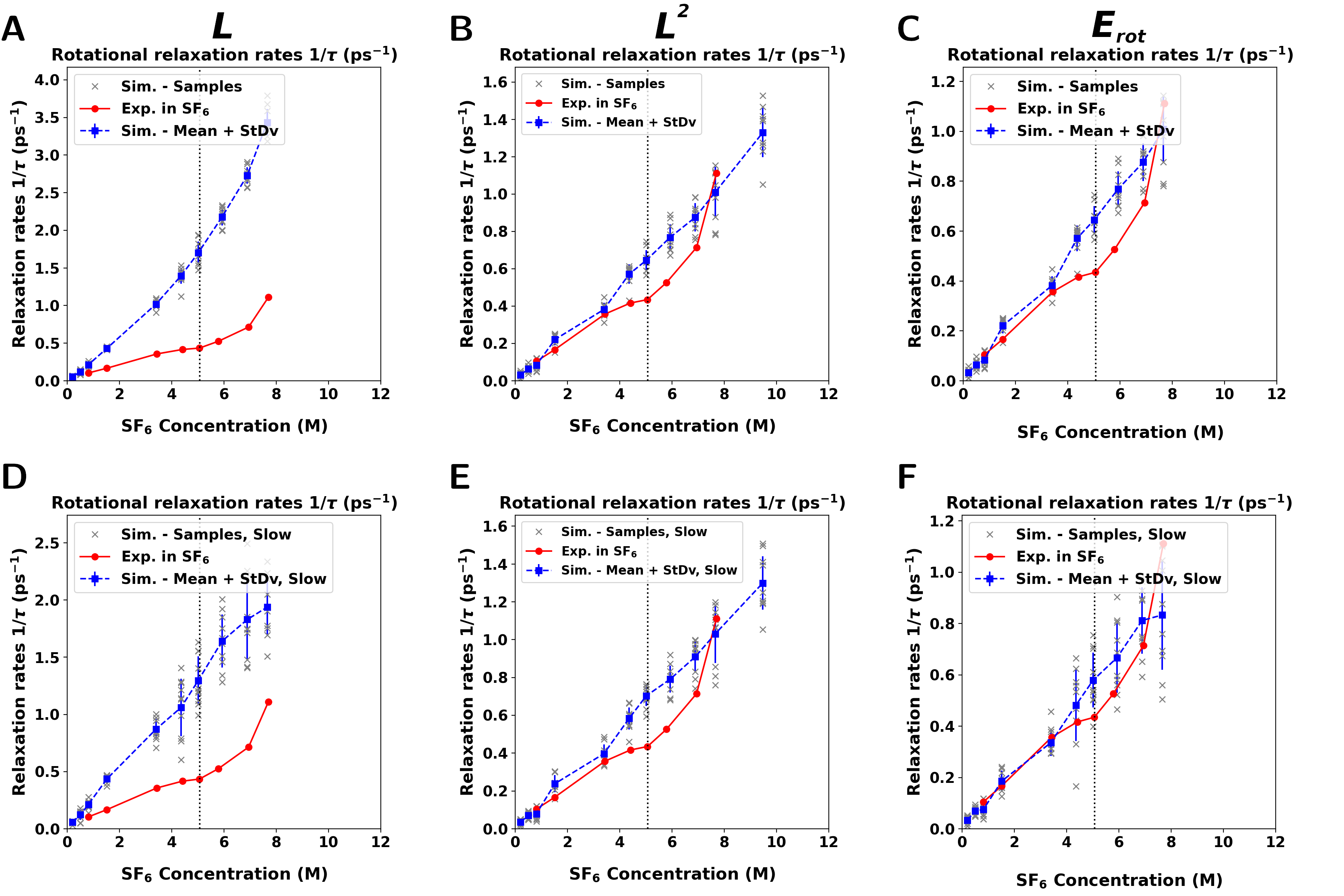}
\caption{Average of obtained rotational relaxation rates from the lifetimes of a fitted
single- (\textbf{A}-\textbf{C}) and bi-exponential function (\textbf{D}-\textbf{F}) 
to the single correlation functions of the angular momentum (\textbf{A}, \textbf{D}), 
squared angular momentum (\textbf{B}, \textbf{E}) or rotational energy 
(\textbf{C}, \textbf{F}) of each sample run of N$_2$O in SF$_6$ and 
at different solvent concentrations (solid blue line). 
The grey markers .
The solid red line with circle markers shows the experimentally 
measured rotational relaxation rates,\cite{ziegler:2022}
and the dashed blue lines are the fitted rotational relaxation rates scaled to 
minimizing deviation from experiments.}
\label{sifig_sample_rotrates_sf6}
\end{center}
\end{figure}

\begin{figure}[htb!]
\begin{center}
\includegraphics[width=0.80\textwidth]{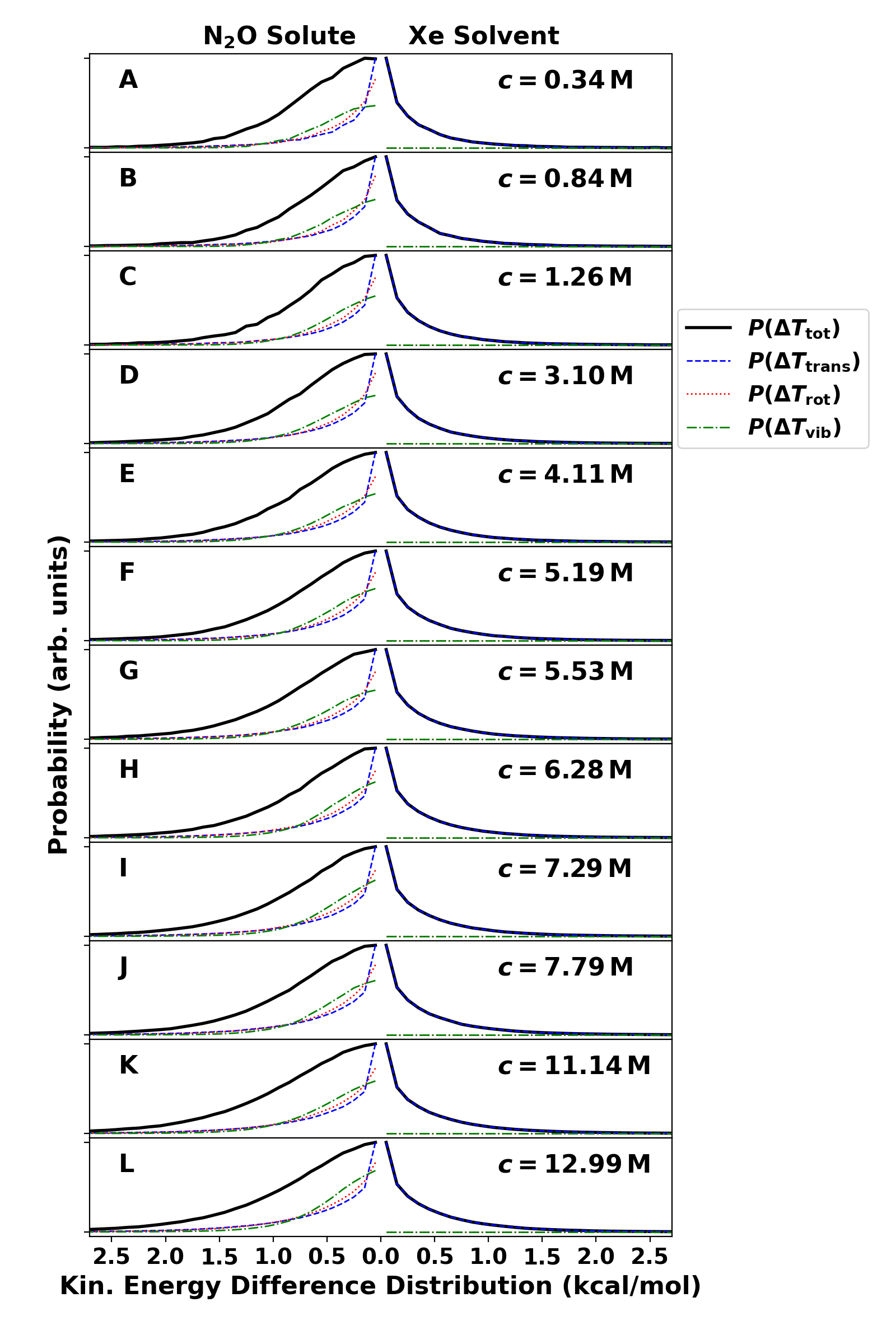}
\caption{Absolute Kinetic energy difference probability distribution
  $P(\Delta T)$ of the total, translational, rotational and
  vibrational kinetic energy in N$_2$O after an collision event with a
  xenon solvent atom for different solvent concentrations. Left part
  of the axis shows the $P(\Delta \bar{T})$ for the N$_2$O solute and
  right part for the xenon solvent.}
\label{sifig_ekin_dist_xe}
\end{center}
\end{figure}

\begin{figure}[htb!]
\begin{center}
\includegraphics[width=0.80\textwidth]{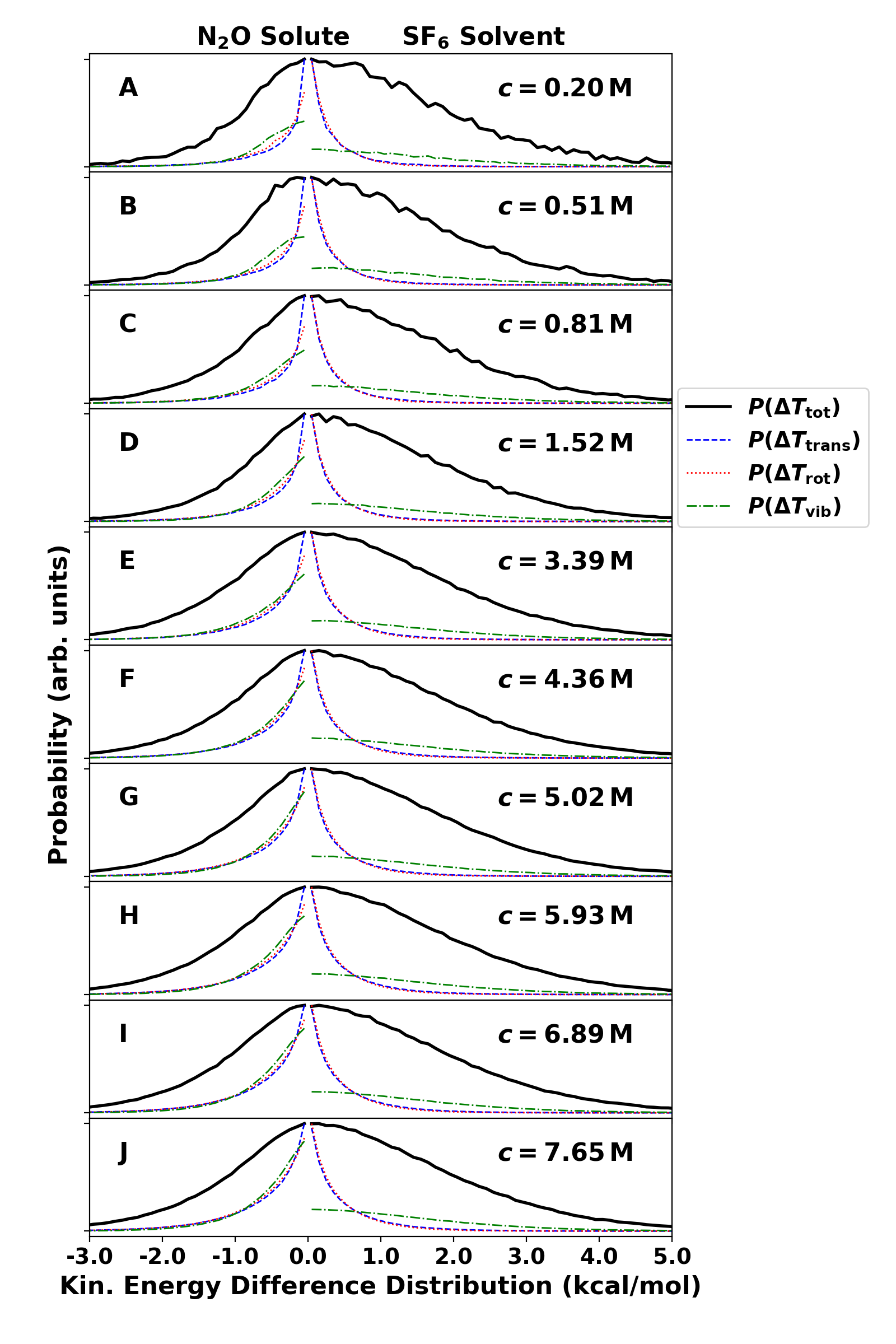}
\caption{Absolute Kinetic energy difference probability distribution
  $P(\Delta T)$ of the total, translational, rotational and
  vibrational kinetic energy in N$_2$O after an collision event with a
  SF$_6$ solvent molecule for different solvent concentrations. Left
  part of the axis shows the $P(\Delta \bar{T})$ for the N$_2$O solute
  and right part for the SF$_6$ solvent.}
\label{sifig_ekin_dist_sf6}
\end{center}
\end{figure}

\begin{figure}[htb!]
\begin{center}
\includegraphics[width=0.80\textwidth]{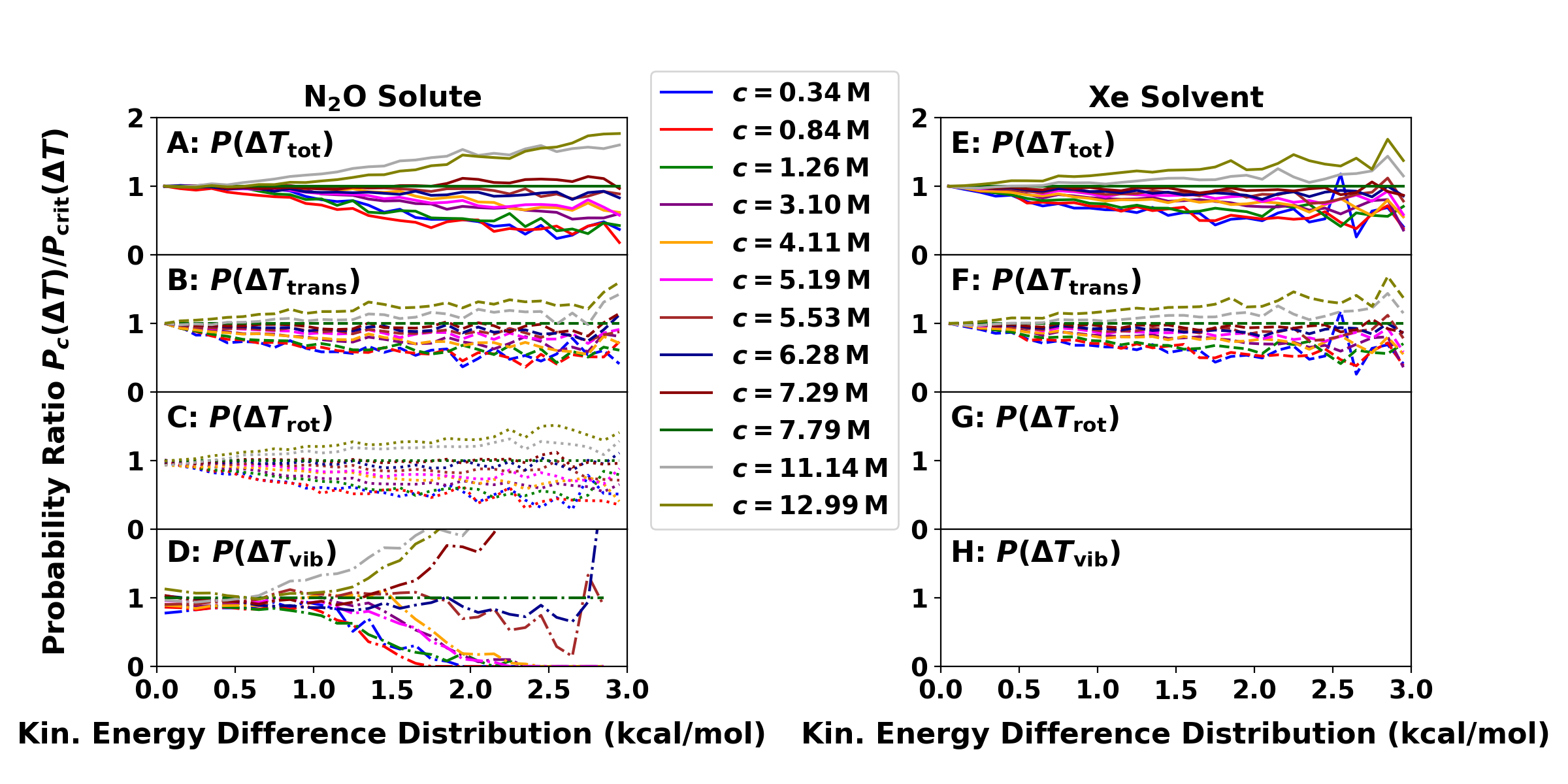}
\caption{Kinetic energy difference probability distribution 
ratio $P_c(\Delta T)/P_\mathrm{crit}(\Delta T)$ between the 
absolute kinetic energy difference probability distribution $P_c(\Delta T)$
at solvent concentration $c$ and the respective distribution closest to the
critical distribution $P_\mathrm{crit}(\Delta T)$ at 
$c_\mathrm{crit} \approx 7.79$\,M.}
\label{sifig_ekin_reldist_xe}
\end{center}
\end{figure}

\begin{figure}[htb!]
\begin{center}
\includegraphics[width=0.80\textwidth]{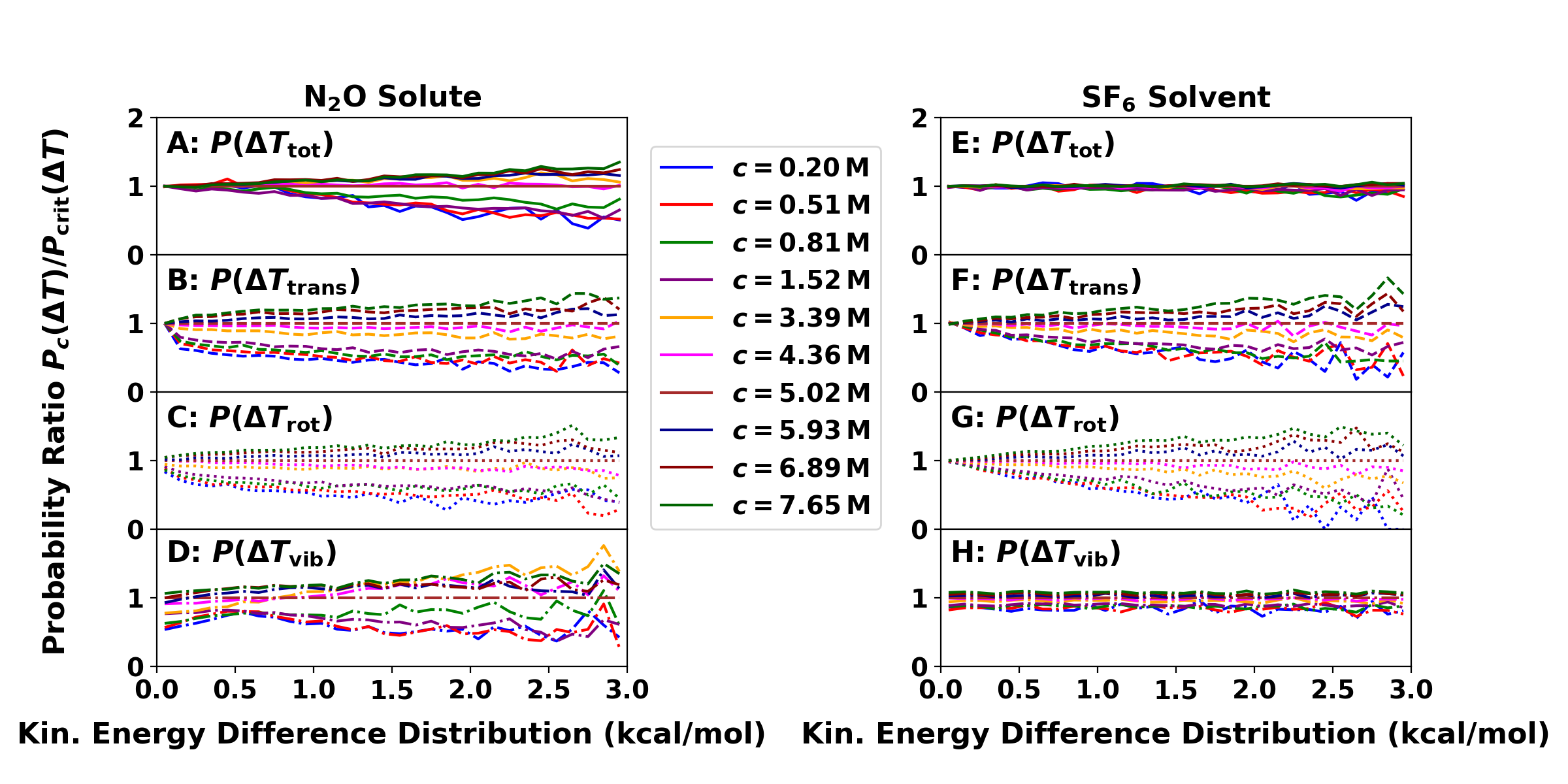}
\caption{Kinetic energy difference probability distribution 
ratio $P_c(\Delta T)/P_\mathrm{crit}(\Delta T)$ between the 
absolute kinetic energy difference probability distribution $P_c(\Delta T)$
at solvent concentration $c$ and the respective distribution at the
critical distribution $P_\mathrm{crit}(\Delta T)$ at 
$c_\mathrm{crit} \approx 5.02$\,M.}
\label{sifig_ekin_reldist_sf6}
\end{center}
\end{figure}

\begin{figure}[htb!]
\begin{center}
\includegraphics[width=0.90\textwidth]{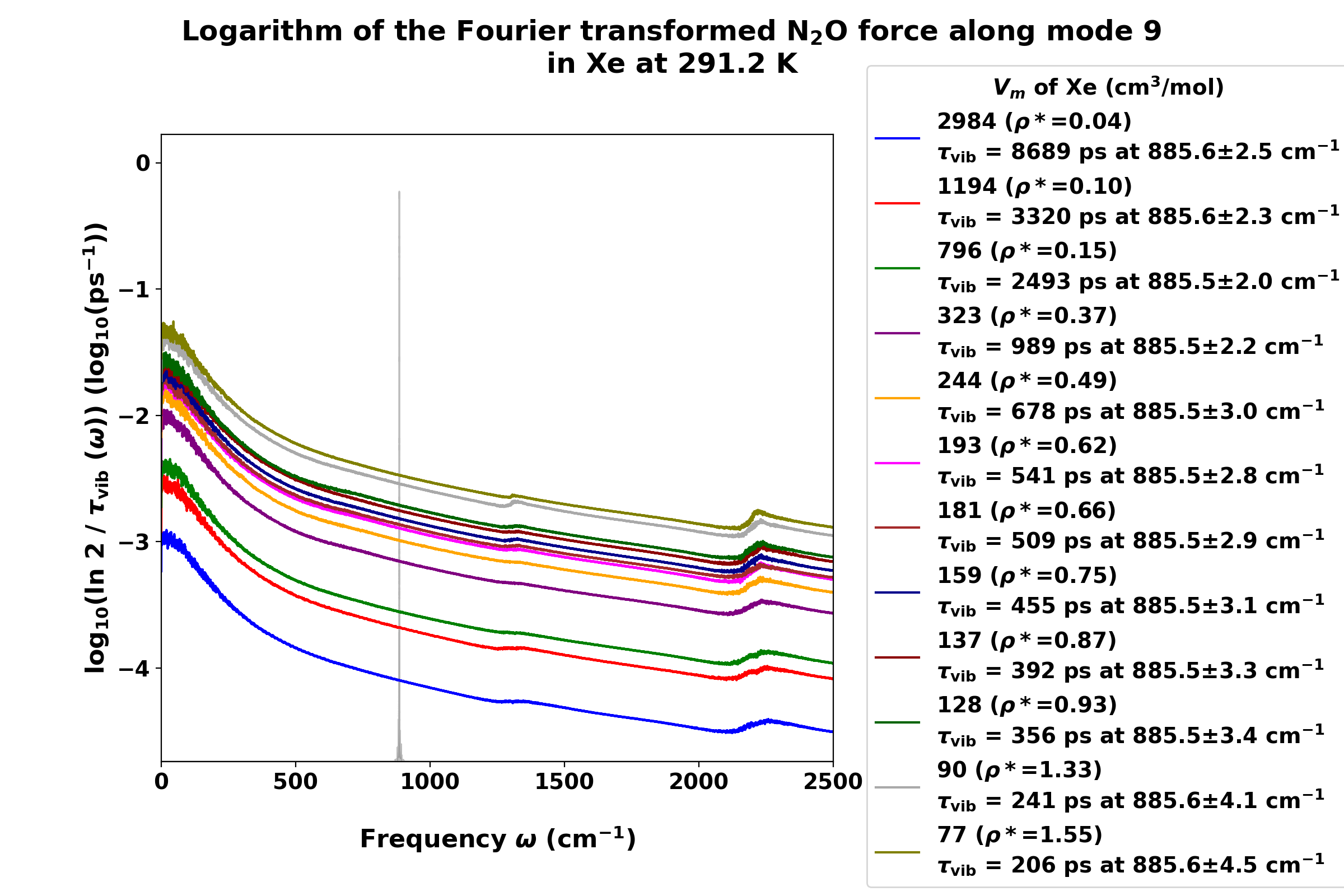}
\caption{Frequency dependent friction function for the vibrational transition
$\nu_\mathrm{s}, \nu_\mathrm{as} = 0, 1$ to $\nu_\mathrm{s}, \nu_\mathrm{as} =1, 0$
of N$_2$O in xenon at different solvent concentration. The gray histogram shows the
average transition frequency for all solvents and the transition frequency mean and 
standard deviation is given in the legend with the respective vibrational relaxation
value $\tau_\mathrm{vib}$.}
\label{sifig_vibtimes_xe}
\end{center}
\end{figure}

\begin{figure}[htb!]
\begin{center}
\includegraphics[width=0.90\textwidth]{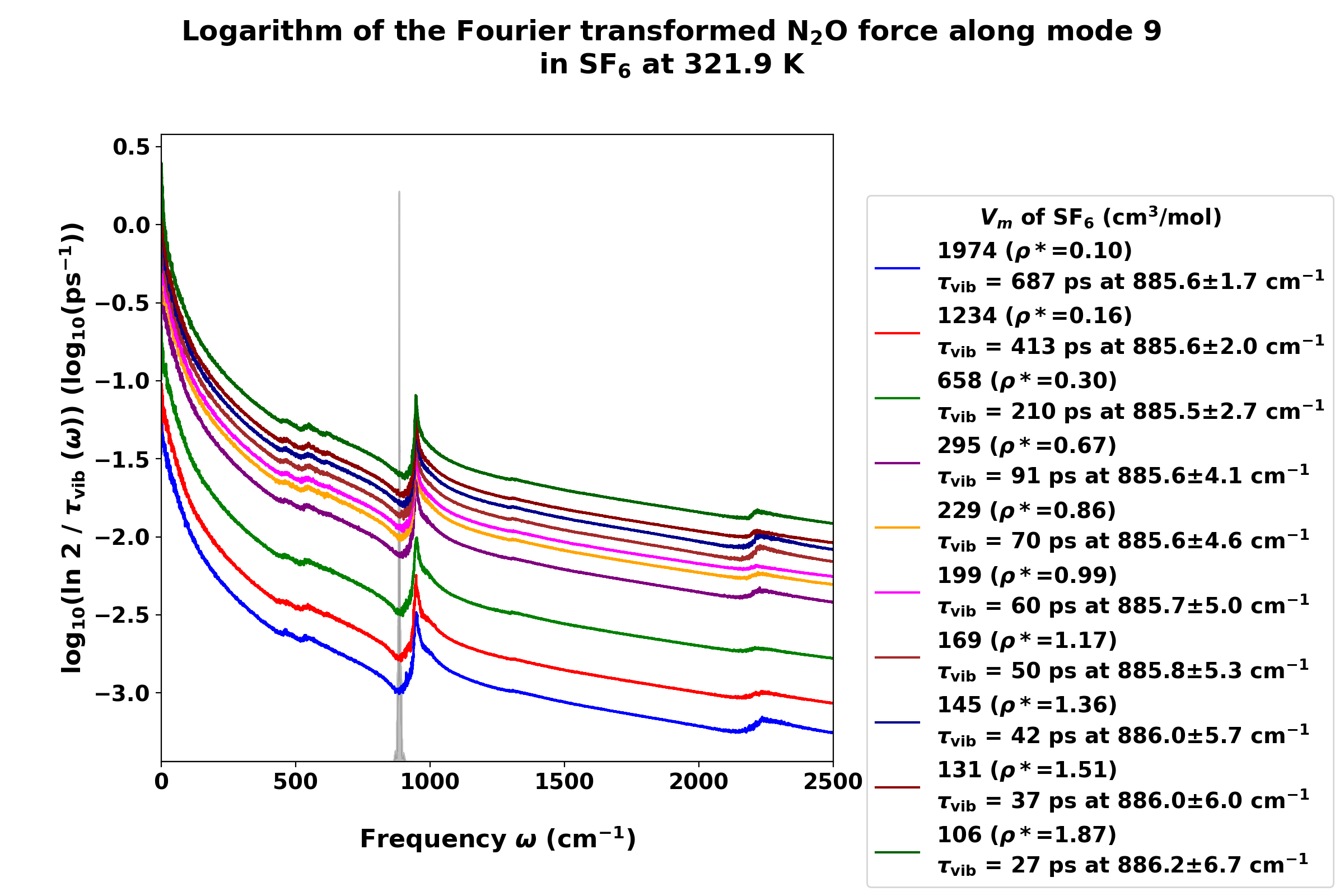}
\caption{Frequency dependent friction function for the vibrational
  transition $\nu_\mathrm{s}, \nu_\mathrm{as} = 0, 1$ to
  $\nu_\mathrm{s}, \nu_\mathrm{as} =1, 0$ of N$_2$O in SF$_6$ at
  different solvent concentration. The gray histogram shows the
  average transition frequency for all solvents and the transition
  frequency mean and standard deviation is given in the legend with
  the respective vibrational relaxation value $\tau_\mathrm{vib}$.
  The vertically dotted magenta line show the position of the SF$_6$
  stretch frequency at 762\,$cm^{-1}$. The dashed green line next to
  it is shifted by $-8$\,cm$^{-1}$ which is the experimentally
  measured frequency difference between the SF$_6$ stretch and the
  transition frequency between the N2O stretch vibrations.}
\label{sifig_vibtimes_sf6}
\end{center}
\end{figure}

\clearpage

\bibliography{refs}